\documentclass[prd,aps,floats,floatfix,eqsecnum,nofootinbib]{revtex4}
\usepackage{verbatim,graphicx,amssymb,amsbsy,bm,amsmath,rotating,epsfig}
\usepackage{hyperref}
\usepackage{fontenc}
\usepackage{graphics}
\usepackage{rotate,psfrag}
\usepackage{color}
\usepackage{amsthm,amssymb}
\usepackage{slashed}
\usepackage{sidecap}
\usepackage{multirow}
\newcommand{\be}{\begin{equation}}
\newcommand{\ee}{\end{equation}}
\newcommand{\bea}{\begin{eqnarray}}
\newcommand{\eea}{\end{eqnarray}}
\newcommand{\bp}{\ensuremath{\mathbf{p}}}

\newcommand{\br}{\ensuremath{\mathbf{r}}}
\begin{document}

\title{\Large Towards the Chalonge Meudon Workshop 2013 \\ 
HIGHLIGHTS and CONCLUSIONS of the Chalonge 
CIAS Meudon Workshop 2012: \\
WARM DARK MATTER GALAXY FORMATION IN AGREEMENT WITH OBSERVATIONS:

\medskip

Ecole Internationale d'Astrophysique Daniel Chalonge

Meudon campus of Observatoire de Paris

in the historic Ch\^ateau, 6-8 June 2012.}

\author{\Large \bf  P.L. Biermann$^{(a)}$,  H.J. de Vega$^{(b,c)}$,    N.G. Sanchez$^{(c)}$}

\date{\today}

\affiliation{$^{(a)}$ MPI-Bonn, Germany \& Univ of Alabama, Tuscaloosa, USA.\\
$^{(b)}$ LPTHE, Universit\'e
Pierre et Marie Curie (Paris VI) et Denis Diderot (Paris VII),
Laboratoire Associ\'e au CNRS UMR 7589, Tour 24, 5\`eme. \'etage, 
Boite 126, 4, Place Jussieu, 75252 Paris, Cedex 05, France. \\
$^{(c)}$ Observatoire de Paris,
LERMA. Laboratoire Associ\'e au CNRS UMR 8112.
 \\61, Avenue de l'Observatoire, 75014 Paris, France.}

\begin{abstract}
The turning point operated recently in the Dark Matter research,
WDM emerged impressively over Cold Dark Matter (CDM) as the leading Dark Matter candidate, 
considerably clarifies and simplifies galaxies and galaxy formation in agreement with 
observations and naturally re-inserts galaxies in cosmology ($\Lambda$WDM). 
Warm Dark Matter (WDM) research is progressing fastly, the subject is new and WDM 
essentially {\it works}, naturally reproducing the astronomical observations over all 
the scales, small (galactic) as well as large and cosmological scales ($\Lambda$WDM). 
Evidence that Cold Dark Matter ($\Lambda$CDM), CDM+ baryons and its proposed tailored cures do not work in galaxies is staggering, and the CDM wimps (DM particles heavier than 1 GeV) are strongly disfavoured combining theory with galaxy astronomical observations. In the tradition of this series, the Chalonge School Meudon Workshop 2012 approached DM in a fourfold coherent way: astronomical observations of DM structures (galaxy and cluster properties, haloes, rotation curves, density profiles, surface density and scaling laws), $\Lambda$WDM N-body simulations in agreement with observations, WDM theoretical astrophysics and cosmology (kinetic theory, Boltzmann-Vlasov evolution, the mass halo function, halo models, improved perturbative approachs), the quantum pressure of WDM fermions forming the observed cores and their sizes, WDM particle and nuclear physics (sterile neutrinos) and its experimental search. Nicola Amorisco, Peter Biermann, Subinoy Das, Hector J. de Vega, Ayuki Kamada, Elena Ferri on behalf 
of the MARE collaboration, Igor D. Karanchetsev, Wei Liao, Marc Lovell, Manolis Papastergis, Norma G. Sanchez, Patrick Valageas, Casey Watson, Jesus Zavala, He Zhang present here their highlights of the Workshop. 
Cored (non cusped) DM halos and WDM (keV scale mass) are clearly determined from theory and 
astronomical observations, they naturally produce the observed structures at all scales; 
keV sterile neutrinos are the most suitable candidates, they naturally appear in minimal extensions of the standard 
model of particle physics. $\Lambda$WDM simulations with keV particles remarkably 
reproduce the observations, the small and large structures, sizes of local minivoids and velocity functions.  
Inside galaxy cores, below $ \sim 100$ pc, $N$-body classical physics simulations 
are incorrect for WDM because at such scales quantum effects are important for WDM.
Quantum calculations (Thomas-Fermi approach) for the WDM fermions provide galaxy cores, 
galaxy masses, velocity dispersions and density profiles in remarkable agreement with the observations.
All evidences point to a dark matter particle mass around 2 keV.
Baryons, which represent 16\% of DM, are expected to give a correction to pure WDM results. 
The summary and conclusions by H. J. de Vega and N. G. Sanchez stress among other points the 
impressive evidence that DM particles 
have a mass in the keV scale and that those keV scale 
particles naturally produce the small scale structures observed in galaxies and the cored density profiles 
with their observed sizes.  keV scale sterile neutrinos are the most serious DM candidates and deserve 
dedicated study and experimental searchs. Astrophysical constraints put the mass in the range $ 1 < m < 4 $ keV.  
Peter Biermann presents his live minutes of the Workshop and concludes that a right-handed sterile neutrino of 
mass of a few keV is the most serious DM candidate. Interestingly enough, 
MARE -and hopefully an adapted KATRIN- experiment could provide a sterile neutrino signal. It will be 
a fantastic discovery to detect dark matter in a beta decay. 
There is a formidable WDM work to perform ahead of us, these highlights point some of the 
directions where it is worthwhile to put the effort. Photos of the Workshop are included. 
\end{abstract}

\maketitle

\tableofcontents

\begin{figure}[htbp]
\epsfig{file=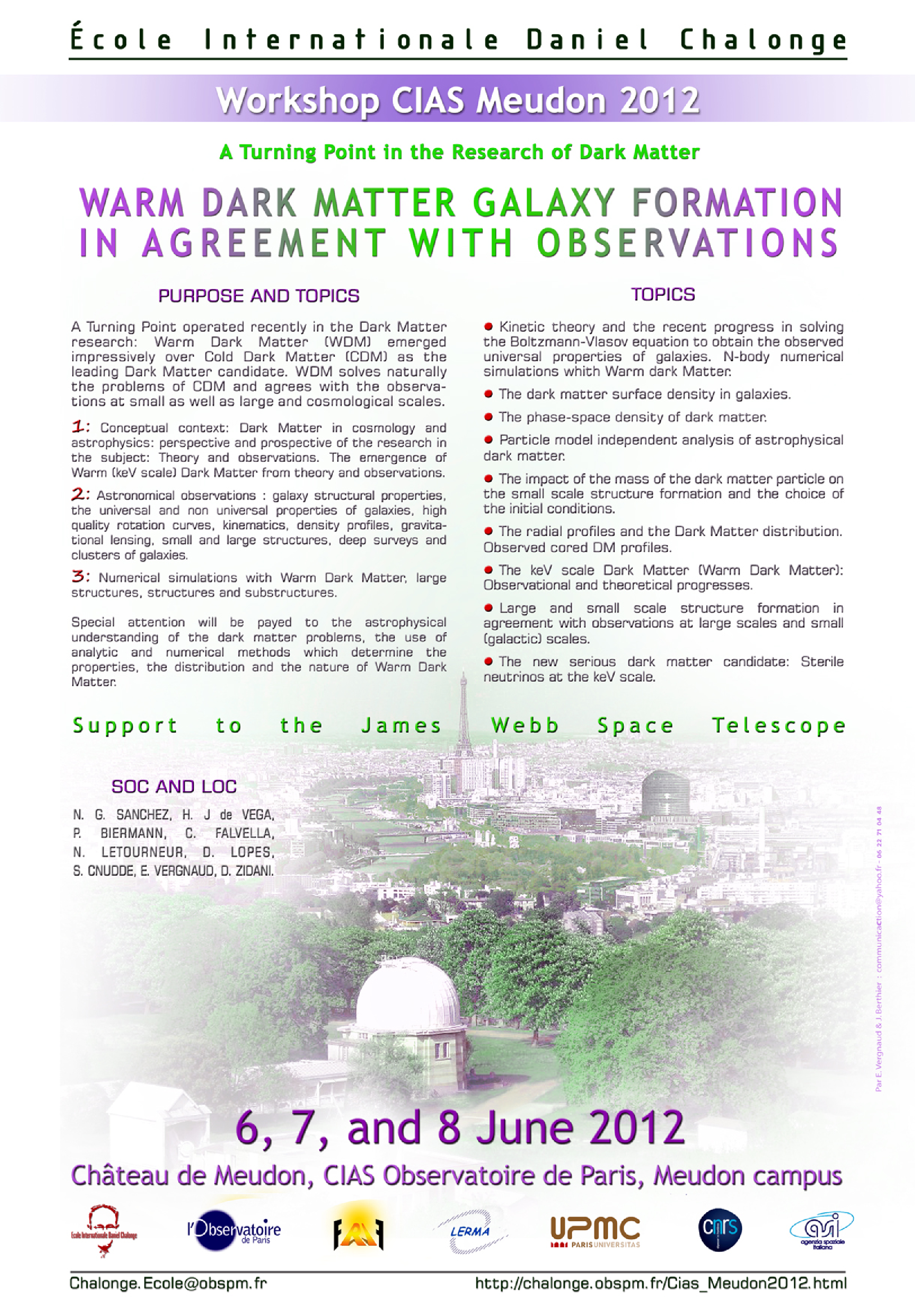,width=14cm,height=18cm}
\caption{Poster of the Workshop}
\end{figure}

\begin{figure}[h]
\includegraphics[width=150mm,height=100mm]{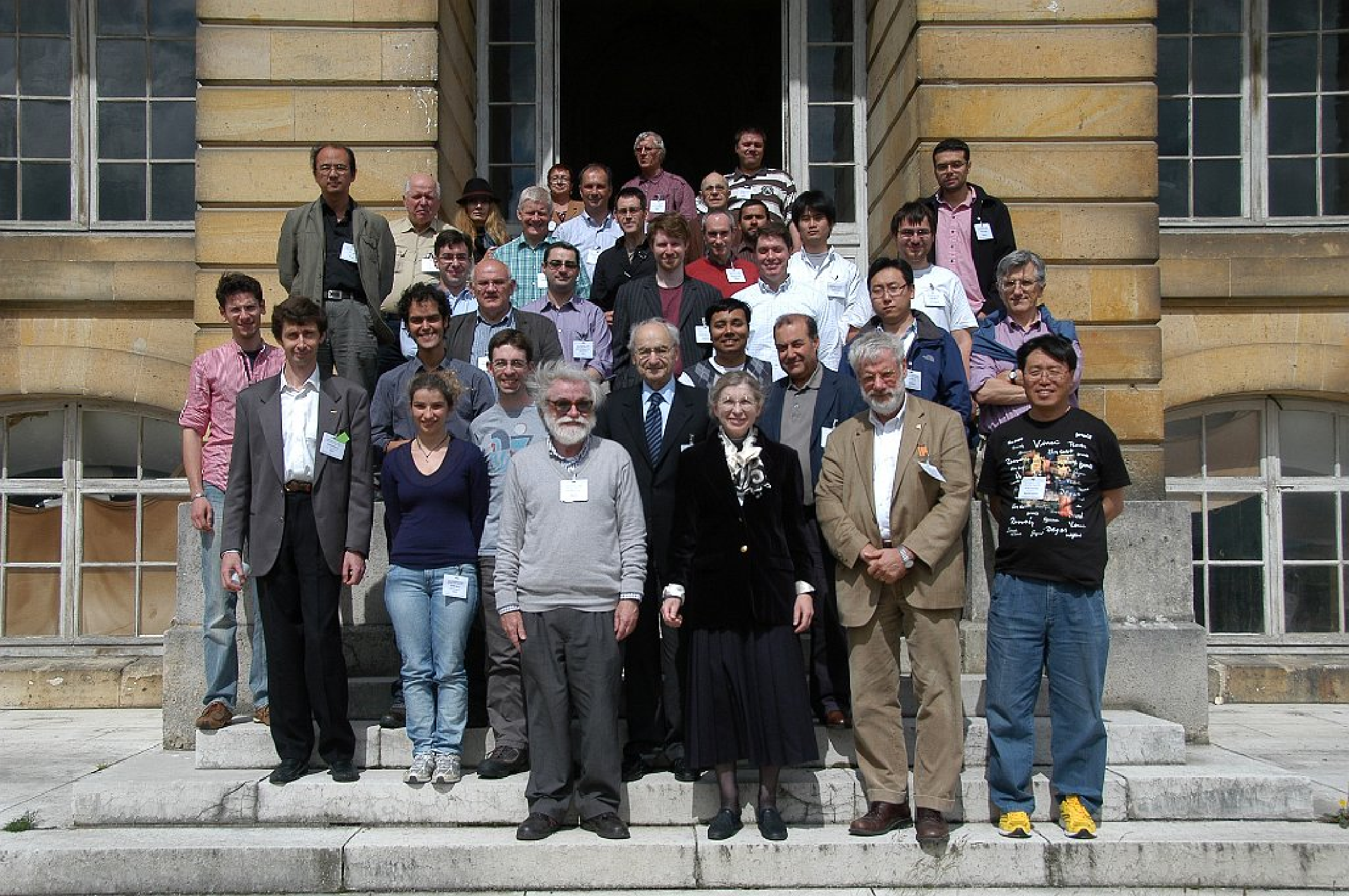}
\caption{Photo of the Group}
\end{figure}

\section{Purpose of the Workshop, Context and Introduction}

This Workshop addresses the turning point in the research of Dark Matter represented by 
Warm Dark Matter (WDM) putting together astrophysical, cosmological and particle WDM,
astronomical observations, theory and WDM numerical simulations which naturally reproduce the observations, 
as well as the experimental search for the WDM particle candidates (sterile neutrinos).

\medskip

Recently, $\Lambda$WDM  emerged impressively over $\Lambda$CDM
($\Lambda$-Cold Dark Matter) whose small -galactic- scale problems are staggering. 
$\Lambda$WDM solves naturally the problems of $\Lambda$CDM and {\it agrees} with the observations 
at small as well as large and cosmological scales.

\medskip

This Workshop is the third of a new series dedicated to Dark Matter, producing fast and growing new 
progresses in the subject.

\medskip

The first Workshop of this series in the Meudon Castle CIAS in June 2010 allowed to identify  and understand 
the issues of the serious problems faced by Cold Dark Matter (CDM) to reproduce the galactic 
(and even cluster of galaxies) observations. \\

The 2010  and 2011 Workshops served as well to verify and better understand the always growing 
amount of confusion in the 
CDM research (CDM + baryons), namely the increasing number and cyclic change of arguments, counter-arguments and ad-hoc 
mechanisms introduced in the CDM simulations over most of twenty years, in trying to deal with the CDM small 
scale crisis: Cusped profiles and overabundance of substructures. Too many satelites are predicted by CDM 
simulations while cored profiles and no such overabundant substructures are seen by astronomical observations. 
A host of ad-hoc mechanisms are proposed and advocated to cure the CDM problems. `Baryon and supernovae feedbacks',
non circular motions, triaxiality, mergers, `cusps hidden in cores', 
`strippings' are some of such mechanisms tailored or exagerated for the purpose of obtaining the desired result
without having a well established physical basis. For example, the strong "baryon and 
supernovae feedback" introduced to transform the CDM cusps into cores in baryon+CDM simulations
corresponds to a large star formation rate contradicting the observations.

\medskip

On the CDM particle physics side, 
the problems are no less critical: So far, all the {\it dedicated} experimental searches after most of 
twenty five years to find the theoretically proposed CDM particle candidate (WIMP) have {\bf failed}. The
CDM indirect searches (invoking CDM annihilation) to explain cosmic ray positron excesses, are in crisis as 
well, as wimp annihilation models are plagued with growing tailoring or fine tuning, and in any case, 
such cosmic rays excesses are well explained and reproduced by natural astrophysical process and sources, 
as non-linear acceleration, shocks and magnetic winds around massive explosion stars, quasars, or the 
interaction of the primary cosmic-ray protons with the interstellar medium producing positrons. The so-called 
and repeatedealy invoked `wimp miracle' is nothing but been able to solve one equation with three unknowns  
(mass, decoupling temperature, and annhiliation cross section) within WIMP models
theoretically motivated by SUSY model building twenty years ago
(SUSY was very fashionable at that time and believed a popular motivation for many proposals).

\medskip

After more than twenty five years -and as often in big-sized science-, CDM research has by now its own internal 
inertia: growing simulations involve large super-computers and large number of people working with, CDM particle 
wimp search involve large and longtime planned experiments, huge number of people, (and huge budgets); one should not be surprised in 
principle, if a fast strategic change would not yet operate in the CDM and wimp research, although
they would progressively decline. Similar situation happens in the CDM+baryon (super)computer simulations, 
in which tailored models fail to reproduce observations and ever increasing `baryon complexity' 
is advocated as the reason of such a failure. Moreover, the contradiction appears: 
if all the cures to the drastic failure of CDM in galaxies is assumed to be produced by baryons: Why then use CDM ?

\bigskip

In contrast to the CDM situation, the WDM research situation is progressing fast, the subject is new and WDM essentially {\it works}, 
naturally reproducing the observations over all the scales, small as well as large and cosmological scales ($\Lambda$WDM). $N$-body CDM 
simulations {\bf fail} to produce the observed structures for {\bf small} scales less than some kpc.
Both $N$-body WDM and CDM simulations yield {\bf identical and correct} structures 
for scales larger than some kpc.
At intermediate scales WDM give the {\bf correct abundance} of substructures .

{\vskip 0.2cm} 

Inside galaxy cores, below  $ \sim 100$ pc, $N$-body classical physics simulations 
are incorrect for WDM because quantum effects are important in WDM at these scales.
WDM predicts correct structures for small scales (below kpc) when its {\bf quantum} nature is
taken into account.

\bigskip

This 2012 Workshop addressed the last progresses made in WARM DARK MATTER GALAXY FORMATION IN AGREEMENT WITH OBSERVATIONS. In the tradition of the Chalonge School, an effort of clarification and synthesis is  made by combining in a conceptual framework, theory, analytical, observational and numerical simulation results which reproduce observations.

\bigskip

\begin{center} 

{\bf The subject have been approached in a fourfold coherent way:}

\bigskip

(I) Conceptual context 

\bigskip

(II) Astronomical observations linked to the galaxy structural properties and to structure 
formation at different large and small (galactic) scales.

\bigskip

(III) WDM Numerical simulations which reproduce observations at large and small (galactic) scales.

\medskip

WDM analytical developments (Thomas-Fermi approach) which includes quantum mechanical tratement of WDM
fermions, reproduce galaxy masses, sizes, velocity dispersions , galaxy cores and their correct sizes,
from the compact dwarfs to the larger spirals and ellipticals.

\bigskip

(IV) WDM particle candidates, keV sterile neutrinos: particle models, laboratory constraints from beta decay 
and electron capture and from X-ray astronomical observations.

\end{center}

\bigskip

\begin{center}

{\bf The Topics covered included:}

\end{center}

\medskip

Recent progress in solving the Boltzmann-Vlasov equation to obtain the observed properties of galaxies (and clusters of galaxies), linear and non-linear WDM analytical approaches.
N-body numerical simulations with Warm Dark Matter; the surface density; scaling laws, universality and Larson laws;  the phase-space density. Particle model independent analysis of astrophysical dark matter. The impact of the mass of the dark matter particle on the small scale structure formation.
The quantum mechanical effects of WDM (in the Thomas-Fermi approach), essential to describe the observed cores and their sizes. reproduce:  the observed cores and their sizes, the observed galaxy masses and their sizes, the phase space density and velocity dispersions from the dwarfs to the larger galaxies (elliptical, spirals). dwarf galaxies are quantum macroscopic WDM objects, 

\medskip

The radial profiles and the Dark Matter distribution. Observed cored density profiles. The ever increasing problems 
of $\Lambda$CDM at small scales. The keV scale Dark Matter (Warm Dark Matter): Observational and theoretical progresses. 
Perturbative  approachs and the halo model. Large and small scale structure formation in agreement with 
observations at large and small (galactic) scales. The new serious dark matter candidate: Sterile neutrinos 
at the keV scale:  Particle physics models of sterile neutrinos, astrophysical constraints (Lyman alpha, 
Supernovae, weak lensing surveys),  experimental searches of keV sterile neutrinos

\medskip

Nicola Amorisco, Peter Biermann, Subinoy Das, 
Hector J. de Vega, Elena Ferri on behalf of the MARE collaboration, Ayuki Kamada, 
Igor Karanchetsev, Wei Liao, Marco Lombardi, 
Marc Lovell, Manolis Papastergis, 
Norma G. Sanchez, Patrick Valageas, Casey Watson, Jesus Zavala and He Zhang
present here their highlights of the Workshop. 

\medskip

\begin{center}

{\bf The Summary and Conclusions:}

\end{center}

Summary and conclusions are presented at the end by H. J. de Vega and N. G. Sanchez in three subsections: 

\medskip

A. General view and clarifying remarks. 

B. Conclusions. 

C. The present context and future in the DM research.

\medskip

The conclusions stress among other points the 
impressive evidence that DM particles have a mass in the keV scale and that those keV scale particles naturally 
produce the small scale structures observed in galaxies. The quantum mechanical effects of WDM (in the Thomas-Fermi approach), remarkably reproduce:  the observed cores and their sizes, the observed galaxy masses and their sizes, the phase space density and velocity dispersions from the compact dwarfs to the larger galaxies (elliptical, spirals). Dwarf galaxies are natural quantum macroscopic objects for WDM, 
while larger galaxies are naturally WDM dilute semiclassical objects. Adding baryons will not change these results. Baryons are a correction to WDM effects. Wimps (DM particles heavier than 1 GeV) are strongly 
disfavoured combining theory with galaxy astronomical observations. keV scale sterile neutrinos are the most 
serious DM candidates and deserve dedicated 
experimental searchs and simulations. Astrophysical constraints including Lyman alpha bounds put the 
sterile neutrinos mass in 
the range $ 1 $ keV  $< m_{DM} < 4$ keV. The most complete recent study (with the quantum WDM Thomas-Fermi approach) confronted to observations, points to a fermion of 2 keV. MARE -and hopefully 
an adapted KATRIN- experiment could provide a sterile neutrino signal. The experimental search for these 
serious DM candidates appears urgent and important: It will be a a fantastic discovery to 
detect dark matter (the keV sterile neutrino) in a beta decay. 

\medskip

\begin{center}

{\bf Summary of the Workshop by Peter Biermann}: 

\end{center}

\medskip

First of all very many arguments based on observations and simulations were presented, that CDM + baryons is not sufficient to explain a plethora of data (discussed by Hector de Vega, Karachentsev, Kamada, Das, Papastergis, Amorisco, Penarrubia, Lovell, Zavala, Valageas, Corasaniti, and Norma Sanchez).  Particle physics and new ground experiments were also presented (talks by Liao, Ferri, and Zhang).  

\medskip

It is striking, how data on cores of galaxies, as well as the large scales of the dark matter distribution in galaxies, the free-streaming length, and the galaxy distribution on larger scales all consistently point to  keV warm dark matter.

\medskip

It may be helpful to note that warm dark matter allows star formation to begin very early in the universe (discussed by Marco Lombardi); since warm dark matter strongly influences early star formation in the universe, some observational limits like the Lyman-alpha bound may be strongly influenced, and so may not be reliable to set limits on the mass of the dark matter particle.  

\medskip

Dark matter decay gives an X-ray emission line, here extensively discussed by Casey Watson.  Second, at this meeting we had both model-dependent and model-independent quantitative bounds on the mass of a sterile neutrino as a dark matter candidate:  It appears that we need a more careful analysis of the Lyman-$\alpha$ data, in order to obtain additional constraints on the various models.  Assuming certain models like Dodelson \& Widrow and Asaka et al. 2010, one can obtain more restrictive bounds, like $m_{DM} \; > \; 2.08 \; {\rm keV}$ (Popa \& Ana Vasile 2007, 2008; Ayuki Kamada 2012 here) from WMAP data, and $m_{DM} \; < \; 2.2 \; {\rm keV}$ (C. Watson et al. 2012); restricting oneself to the same model, as used by these  teams, would allow a well determined mass already.   

\medskip

Bounds and determination of the mass scale of the dark matter particle derived from phase-space arguments are model independent, since they just depend on the character of the particle as a Fermion, and such bounds can be derived from observed small galaxies.  These data suggest a mass range of 2 - 4 keV (Hector de Vega \& Norma Sanchez, as well as various other groups with similar limits). Strigari et al. 2006 give $ m < 2.2$ keV from dwarf ellipticals, but using X-ray limits and so this limit depends on a specific model for the particle.   Since the cores of modest galaxies are observed to host either a nuclear star cluster or a central black hole, the observed masses of the smallest observed black holes at the centers of galaxies can be used with various phase-space limits to also estimate the dark matter particle mass (Munyaneza et al. 1998 - 2006, and in a much more detailed consideration of the inner structure of galaxies described by the quantum warm dark matter fermions Destri, de Vega \& Sanchez 2012); this suggests a few keV.  

\medskip

A positive detection could be a small finite distributed Thomson depth in the cosmic radiation all the way back to recombination, with no spatial variation, which in the Biermann \& Kusenko (2006) approach could be attributed to the decay of a sterile neutrino; if such a Thomson depth is measured, its strength is a direct measure of the sterile neutrino mass.  Better would be if we had a positive observation, an X-ray line in emission, consistently detected in several extragalactic fields, or best perhaps, a positive detection in an Earth-based experiment, such as planned and built up in various laboratories such as in Milano.  Thus, many model-dependent arguments as well as model-independent arguments suggest a dark matter particle mass somewhere close to 2 keV, with fairly reliable limits  $0.4$ keV $ < m_{DM} < 4$ keV.

\medskip

There is an encouraging and formidable WDM work to perform ahead us, these highlights point some of the directions
where to put the effort.

\bigskip

Sessions lasted for three full days in the beautiful Meudon campus of Observatoire de Paris, where CIAS 
`Centre International d'Ateliers Scientifiques' is located. All sessions took place in the historic 
Chateau building, (built in 1706 by the great architect Jules-Hardouin Mansart in orders by King Louis XIV 
for his son the Grand Dauphin).

\bigskip

The meeting is open to all scientists interested in the subject. 
All sessions are plenary followed by discussions. 
The format of the Meeting is intended to allow easy and 
fruitful mutual contact and communication. Large time is devoted  
to discussions. All informations about the meeting, are displayed at

\begin{center}

{\bf http://chalonge.obspm.fr/Cias\_Meudon2012.html}

\end{center}

The presentations by the lecturers are available on line 
(in .pdf format) in `Programme and Lecturers' in the above link, 
as well as the Album of photos of the Workshop: 

\begin{center}

{\bf http://chalonge.obspm.fr/Programme\_CIAS2012.html}

\bigskip

{\bf http://chalonge.obspm.fr/albumCIAS2012/index.html}

\end{center}

We thank all again, both lecturers and participants, 
for having contributed so much to the great success of this 
Workshop and look forward to seeing you again in the next 
Workshop of this series.     

\medskip

We thank the Observatoire de Paris, CIAS for their support; secretariat, logistics and technical
assistance,  Nicole Letourneur, Sylvain Cnudde, Djilali Zidani; 
DIM ACAV; LERMA, LPTHE and all those who contributed so efficiently to the successful organization of this Workshop.

\medskip
      
\begin{center}
                                   
With compliments and kind regards,

\bigskip  
                                            
Peter L. Biermann, Hector J de Vega, Norma G Sanchez

\end{center}

\newpage

\section{Programme and Lecturers}

\bigskip

\begin{itemize}

\item{{\bf Nicola C. AMORISCO} (Institute of Astronomy, Univ of Cambridge, Cambridge, UK)
Dark matter cores and cusps: the case of multiple stellar populations in dwarf spheroidals.}

\medskip

\item{{\bf Peter BIERMANN} (MPI-Bonn, Germany \& Univ of Alabama, Tuscaloosa, USA)
Warm Dark Matter overview: cosmological and astrophysical signatures.}

\medskip

\item{{\bf Pier Stefano CORASANITI} (CNRS LUTH Observatoire de Paris, Meudon, France)
The mass halo function and halo models}

\medskip

\item{{\bf Subinoy DAS} (Inst. für Theoretische Teilchenphysik und Kosmologie RWTH, Aachen Univ., Aachen, Germany)
Cosmological Limits on Hidden Warm Dark Matter}

\medskip

\item{{\bf Hector J. DE VEGA} (CNRS LPTHE Univ de Paris VI, France)
Warm Dark Matter from theory and galaxy observations}

\medskip

\item{{\bf Elena FERRI} (INFN Universi\`a di Milano Biccoca, Italy)
The MARE experiment to measure the mass of light (active) and heavy (sterile) neutrinos}

\medskip

\item{{\bf Ayuki KAMADA} (IPMU Inst. Physics \& Mathematics of the Universe, Univ of Tokyo, Japan)
Structure Formation in Warm Dark Matter Models.}

\medskip

\item{{\bf Igor D. KARACHENTSEV} (SAO-Special Astrophysical Observatory, Nizhnii Arkhyz, Russia)
Cosmography of the Local Universe.}

\medskip

\item{{\bf Wei LIAO} (Inst.Modern Physics, East China Univ.Science \& Technology, Shanghai, P. R. China)
On the detection of keV scale neutrino Warm Dark Matter in Beta decay experiments.}

\medskip

\item{{\bf Marco LOMBARDI} (University of Milano, Italy)
Star Formation Rates in Molecular Clouds and the Nature of the Extragalactic Scaling Relations}

\medskip

\item{{\bf Marc LOVELL} (Inst. Comput. Cosmology, Univ Durham, Durham, UK)
Numerical simulations of WDM halos}

\medskip

\item{{\bf Manolis PAPASTERGIS} (Centre Radiophysics. \& Space Research, Cornell Univ, Ithaca NY, USA)
The velocity width function of galaxies from the ALFALFA survey: Dark matter implications.}

\medskip

\item{{\bf Jorge PENARRUBIA} (Instituto Astrof\'{\i}sica Andalucía - IAA-CSIC, Granada, Spain)
Measuring the Dark Mass Profiles of Dwarf Spheroidal Galaxies.}

\medskip

\item{{\bf Norma G. SANCHEZ} (CNRS LERMA Observatoire de Paris, Paris, France)
Warm Dark Matter Galaxy Formation in Agreement with Observations}

\medskip

\item{{\bf Patrick VALAGEAS} (Inst. de Physique Th\'eorique, Orme de Merisiers, CEA-Saclay, Gif-sur-Yvette, France)
Perturbation approaches and halo models for large-scale structures and related issues}

\medskip

\item{{\bf Casey WATSON} (Millikin Univ, Dept Physics \& Astron., Decatur, Illinois, USA)
Using X-ray Observations to Constrain Sterile Neutrino Warm Dark Matter}

\medskip

\item{{\bf Jesus ZAVALA} (Univ Waterloo, Dept Phys \& Astr, Ontario, Canada)
The velocity function of galaxies in the local environment from Cold and Warm Dark Matter simulations} 

\medskip

\item{{\bf He ZHANG} (Max-Planck-Inst. f\"r Kernphysik, Heidelberg, Germany)
Sterile Neutrinos for Warm Dark Matter in Flavor Symmetry Models.}

\medskip

\end{itemize}

\newpage

\section{Highlights by the Lecturers}

\subsection{Nicola C. AMORISCO and N. Wyn EVANS}

\vskip  0.0cm 

\begin{center}

N.C.A. and N.W.E.: Institute of Astronomy, University of Cambridge, Cambridge, UK 

\bigskip

{\bf Dark Matter Cores and
  Cusps: The Case of Multiple Stellar Populations in Dwarf Spheroidals} 

\end{center}

\medskip

The population of dwarf spheroidal satellites (dSphs) of the Galaxy
furnishes an unprecedented opportunity to test the physics underlying
the most extreme dark matter dominated systems. In particular, the
prediction of a cusped density profile for cold dark matter haloes has 
naturally focused attention on the structure of the innermost parts. 
Although progress on this line of enquiry has been hampered by the
difficulties posed by the mass-anisotropy degeneracy, the subject has 
received new impetus by the growing realization that the stellar
content of dSphs is complex. Carina, Fornax, Sculptor and Sextans all display
evidence for the co-existence of at least two stellar populations with
different metallicities -- usually an older, more spatially extended
and metal-poor population and a younger, more spatially concentrated,
metal-rich population. The different stellar populations clearly have 
different kinematics: in Sculptor, for example, while the older component displays a fairly 
flattish velocity dispersion profile, the dispersion profile of the younger component often shows signs of a quite sharp decline with radius. 

\bigskip

The coexistence of two different stellar populations in the same
potential provides a stronger instrument to probe the properties of
the dark matter halo in dSphs. A first approach is already provided in 
the Jeans analysis of the two populations in the Sculptor dSph [1]. They found that, 
though a Navarro-Frenk-White (NFW) halo is still statistically compatible with the kinematical properties of the two stellar populations of the Sculptor dSph, a cored halo is a better
fit in the context of the Jeans equations. Recently, a statistical
method for assigning probabilities as to whether dSph stars belong to
metal-rich and metal-poor sub-populations has been developed [2]. 
The output from this statistical analysis was used to conclude that
a cusped NFW profile is ruled out in both the Sculptor and Fornax dSphs.
Interestingly, it has been prooved that such a result is a direct consequence
of the tensor virial theorem [3]. Also, the consequences of the hypothesis
of spherical symmetry have been investigated in this work, 
and it was found that a core of at least 120 pc is in fact 
necessary in Sculptor independently of the geometry of the system.

\bigskip

In [4], we re-analyze the data for the two stellar populations in the Sculptor
dSph by using phase space models, instead of the less complete Jeans
analysis, adopting the general family of Michie-King distribution
functions (King 1962, Michie 1963). This represents a fairly flexible family of centrally
isotropic distribution functions with an adjustable radial anisotropy in the outer regions. 
The two stellar populations are embedded in a dark halo whose density profile can be shaped at
will, cored or cusped. Our approach can significantly strengthen
the analysis by requiring the models to reproduce at the same time the
luminosity and kinematic profiles for both populations. 
This enables us to determine the mass profile of the Sculptor
dSph with unprecedented accuracy.  

\bigskip

We show that cored halo models are preferred over cusped halo
models. Although both can fit the data, cored dark matter haloes provide a 
significantly better description, with a likelihood ratio test rejecting NFW models at any 
  significance level higher than $0.05\%$. There is clear
evidence of discrepancies in the surface brightness profile of the
metal-rich stars near the centre for cusped haloes. It is worth to highlight that it is
the combination of surface photometry with the stellar kinematics --
rather than the kinematics alone -- that is providing the decisive
evidence against cusped haloes. We conclude 
  that the kinematics of multiple populations in
  dSphs provides a substantial new challenge for theories of galaxy
  formation, with the weight of available evidence strongly against
  dark matter cusps at the centre.

\bigskip

With an eye to challenge that future kinematic dataset present to
dynamical modelling, in [5] we have devised an efficient method to extract the shape information
for line profiles of discrete kinematic data. The Gauss-Hermite expansion, which is
widely used to describe the line of sight velocity distributions
extracted from absorption spectra of elliptical galaxies, is not
readily applicable to samples of discrete stellar velocity
measurements, accompanied by individual measurement errors and
probabilities of membership. These include datasets on the kinematics
of globular clusters and planetary nebulae in the outer parts of
elliptical galaxies, as well as giant stars in the Local Group
galaxies and the stellar populations of the Milky Way. We introduce
two parameter families of probability distributions describing
symmetric and asymmetric distortions of the line profiles from
Gaussianity. These are used as the basis of a maximum likelihood
estimator to quantify the shape of the line profiles. Tests show that
the method outperforms a Gauss-Hermite expansion for discrete data,
with a lower limit for the relative gain of $\approx 2$ for sample
sizes $N \approx 800$. To ensure that our methods can give reliable
descriptions of the shape, we develop an efficient test to assess
the statistical quality of the obtained fit.

\bigskip

As an application, we turn our attention to the discrete velocity
datasets of the dwarf spheroidals (dSphs) of the Milky Way. Sculptor
and Fornax have datasets of $\gtrsim 1000$ line of sight velocities of
probable member stars. In Sculptor, the symmetric deviations are
everywhere consistent with velocity distributions more peaked than
Gaussian. In Fornax, instead, there is an evolution in the symmetric deviations of
the line profile from a peakier to more flat-topped distribution on
moving outwards. Although the datasets for Carina and Sextans are
smaller, they still comprise several hundreds of stars. Our methods
are sensitive enough to detect evidence for velocity distributions
more peaked than Gaussian. These results suggest a radially biased orbital structure
for the outer parts of Sculptor, Carina and Sextans. On the other hand, tangential 
anisotropy is favoured in Fornax. This is all consistent with a
picture in which Fornax may have had a different evolutionary history
to Sculptor, Carina and Sextans.\\

\vspace{1cm}
{\bf References}

\begin{description}

\item[1] Battaglia G., Helmi A., Tolstoy E., Irwin M., Hill V., \& Jablonka P.\ 2008a, ApJL, 681, L13, [arXiv:0802.4220]
  
\item[2] Walker M.~G. \&  Pe{\~n}arrubia J., 2011, ApJ, 742, 20, [arXiv:1108.2404]

\item[3] Agnello A. \& N.W. Evans 2012, ApJL, in press, [arXiv:1205.6673]

\item[4] Amorisco N.C. \& N.W. Evans 2010, MNRAS, 419, 184, [arXiv:1106.1062]

\item[5]  Amorisco N.C. \& N.W. Evans 2010, MNRAS, in press, [arXiv:1204.5181]

\end{description}

\newpage

\subsection{Peter L. BIERMANN}

\vskip -0.3cm

\begin{center}

MPI-Bonn, Germany \& Univ of Alabama, Tuscaloosa, USA

\bigskip

{\bf Warm Dark Matter and its cosmological and astrophysical signatures}

\end{center}

\medskip

Dark matter has been first detected 1933 (Zwicky) and basically behaves like a non-EM-interacting gravitational gas of particles.  From particle physics Supersymmetry suggests that there should be a lightest supersymmetric particle, which is a dark matter candidate, possibly visible via decay in odd properties of energetic particles and photons.

\medskip

Discoveries were made: i) An upturn in the CR-positron fraction. ii) An upturn in the CR-electron spectrum. iii) A cosmic ray anisotropy in data from Tibet, SuperK, Milagro, and now at two energies, in arriving cosmic rays at 20 TeV and 400 TeV with IceCube. iv) A flat radio emission component near the Galactic Center (WMAP haze). v) A corresponding IC component in gamma rays (Fermi haze and Fermi bubble), a flat $\gamma$-spectrum at the Galactic Center (Fermi). vi) The 511 keV annihilation line also near the Galactic Center. vii) An upturn in the CR-spectra of all elements from Helium, with a hint of an upturn for Hydrogen. viii) A derivation of the interaction grammages separately for CR-protons and CR-heavy nuclei, and ix) Have the complete cosmic ray spectrum with a steep powerlaw leading to a dip near $3 \, 10^{18}$ eV in terms of $E^{3} \, (d N/d E)$, then a broad bump near $5 \, 10^{19}$ eV, turning down towards $10^{21}$ eV (KASCADE, IceTop, KASCADE-Grande, Auger).  

\medskip

All these features can be quantitatively explained building on the action of cosmic rays accelerated in the magnetic winds of very massive stars, when they explode (Biermann et al. 2009 - 2011): this work is based on predictions from 1993 (Biermann 1993, Biermann \& Cassinelli 1993, Biermann \& Strom 1993, Stanev et al 1993; review at ICRC Calgary 1993); this approach is older and simpler than adding WR-star supernova CR-contributions with pulsar wind nebula CR-contributions, is also simpler than using magnetic field enhancement in ISM-shocks, and also simpler than using the decay of a postulated particle.  This concept gives an explanation for the cosmic ray spectrum as Galactic plus one extragalactic source, Cen A (Gopal-Krishna et al. 2010, Biermann \& de Souza 2012).  The data do not require any extra source population below the MWBG induced turnoff - commonly referred to as the GZK-limit: Greisen (1966), Zatsepin \& Kuzmin (1966).  This is possible, since the magnetic horizon appears to be quite small (consistent with the cosmological MHD simulations of Ryu et al. 2008).  It also entails that Cen A is our highest energy physics laboratory accessible to direct observations of charged particles.  

\medskip

All this allows to go back to galaxy data to derive the key properties of the dark matter particle: Work by Hogan \& Dalcanton (2000), Gilmore et al. (from 2006, 2009), Strigari et al. (2008), Boyanovsky et al. (2008), Gentile et al. (2009) and de Vega \& Sanchez (2010) clearly points to a keV particle.  A right-handed neutrino is a Fermion candidate to be this particle (e.g. Kusenko \& Segre 1997; Fuller et al. 2003; Kusenko 2004; also see Kusenko et al. 2010, and Lindner et al. 2010; for a review see Kusenko 2009; Biermann \& Kusenko 2006; Stasielak et al. 2007; Loewenstein et al. 2009). 

\medskip

The keV WDM particle has the advantage to allow star formation very early, near redshift 80, and so also allows the formation of supermassive black holes: they possibly formed out of agglomerating massive stars, in the gravitational potential well of the first DM clumps, whose mass in turn is determined by the properties of the DM particle.  The supermassive star gives rise to a large HII region, possibly dominating the Thomson depth observed.  This black hole formation can be thought of as leading to a highly energetic supernova remnant, a Hyper Nova Remnant (HNR).  Black holes in turn also merge, but in this manner start their mergers at masses of a few million solar masses, about ten percent of the baryonic mass inside the initial dark matter clumps.  This readily explains the supermassive black hole mass function (Caramete \& Biermann 2010).  

\medskip

The action of the formation of the first super-massive black holes allows a possible path to determine the dark matter particle mass, under the proviso that it is a right-handed neutrino, as advocated by some (e.g., Kusenko 2009):  a)  Determine the Galactic radio background spectrum, and check for residual all-sky emission. b)  Determine the extragalactic radio background spectrum, if possible (Kogut et al. 2011). c)  Match it with various models, such a the Hyper Nova Remnants radio emission, d) Such a match implies angular fine structure of this emission on the sky, which may be detectable. e)  Determine the Thomson depth through to recombination, match it with the HII regions, HNRs or any other model, and determine, if possible its angular structure on the sky. f)  If there is a residual Thomson depth which is not structured, then all the normal mechanisms fail due to their spatial distribution, including the HII regions and HNRs, and only a very distributed source of ionization could explain it. g) The strength of the residual Thomson depth directly scales with the action of the decay of a dark matter particle such as a right-handed neutrino:  This gives the mass of the particle, given sufficient accuracy.  

\medskip

Our conclusion is that a right-handed neutrino of a mass of a few keV is the most interesting candidate to constitute dark matter.

\newpage

\subsection{Subinoy DAS}

\vskip -0.3cm

\begin{center}
Institute for Theoretical Particle Physics and Cosmology, RWTH Aachen, Germany 

\bigskip

{\bf Model Independent Constraints on Warm Hidden Thermal Dark Matter } 

\end{center}

\medskip

We discuss general hidden warm dark matter candidate with almost no interaction (or very feeble) with standard model particles so that it is not in thermal contact with visible sector but we assume it is thermalised with in a hidden sector due to some interaction. While encompassing the standard cold WIMP scenario, we do not require the freeze-out process to be non-relativistic. Rather, freeze-out may also occur when dark matter particles are semi-relativistic and giving rise to possibility of thermal production of warm dark matter through hidden sector. We also show that if the warm dark matter originates as a thermal relic through annihilation, to produce correct relic density and obey other cosmological constraints, the hidden sector has to be at a lower temperature than visible sector.

We briefly discuss a unified treatment [1] of the freeze-out of thermal relics in hidden sectors for arbitrary cross section $\sigma$, only requiring that constraints from cosmology are satisfied and find viable dark matter that freezes out when it is relativistic, semirelativistic, or nonrelativistic. We find an model independent lower bound for dark matter mass $m_{\chi} \geq 1.5$ \rm{keV}. Our result applies to sterile neutrino warm dark matter which might or might not be in thermal equilibrium  with visible sector. Especially, if the dark matter sterile neutrinos are not in thermal equilibrium with visible sector but somehow has enough interactions in the dark sector to thermalise it, this lower bound of $1.5$ \rm keV is pretty generic and applies in such situation.\\

The second part of this proceeding focuses on an exotic dark matter canditate which arises from the existence of \rm{eV} mass sterile neutrino through a late phase transition. Due to existence of a strong scalar force the  light sterile states get trapped into stable degenerate micro nuggets. We find that its signature in matter power spectra is close to a warm dark matter candidate. 

\bigskip

{\bf Freeze-out in hidden sector and constraints on warm hidden DM:}
For a  temperature-dependent dark matter cross section $\sigma f(x)$, a general ( warm, cold or hot) freeze-out is described by [1]

\begin{equation}
\frac{dy}{dx}=-\frac{1}{x^2}\left[y^2 - y_0^2(x;\xi)\right] + y \, \frac{\, \,d\ln{f(x)}}{dx} \, .
\label{eq:fulleq}
\end{equation}
where $y(x;\xi)$ is the scaled yeild of dark matter and  $f(x)= \left(\frac{3}{x^2} + \frac{\frac{5}{4}+x }{1+x}\right) \,$ with $x= m/T$. Numerically we solve this equation for different regimes of freeze-out and find the hidden dark matter relic density which depends on dark matter mass ($m_{\chi}$), cross-section ($\sigma (x)$) and hidden to visible sector temperature 

After solving the relic abundance for hidden dark matter as a function of 3 parameters -- hidden to visible temperature ratio, dark matter cross-section and its mass -- $(\xi, \sigma, m_{\chi})$, we put the phase space bound on the dark matter mass and  free-streaming  bound from  structure formation. In fig.1 we show the bounds on $(\xi, \sigma)$ parameter space for different dark matter mass and we see that as the mass become smaller free-streaming bound cuts the parameter space to a large extent and for mass $m_{\chi} \geq 1.5$ \rm{keV} there remains no viable region in parameter space. This is a particle model independent lower limit on fermionic warm dark matter mass. 

\bigskip

{\bf Warm dark matter from phase transition}:
In the last part of this proceeding we focus on recent hints by MiniBoone and reactor experiment which along with LSND may point out to the existence of \rm eV scale strile neutrino. We discuss a exotic warm dark matter like scenario where being triggered by a scalar field  such eV sterile states which was highly relativistic deep in radiation dominated era went through a phase transition few efoldings before matter radiation equality and clumped into micro-nuggets. The signatue of the this dark matter are always close to that of warm dark matter particle as far as the matter power spectra is concerned. This is due the fact that light sterile states was behaving like hot dark matter before the phase transition and then transitions inot cold dark matter nuggets. This means we always get suppression in power in smaller scales [2] in the matter power spectra -- so it mimics WDM.

\bigskip

The basic idea behind this sort of warm like dark matter is : at some point in radiation dominated era the scalar force between light fermions  wins over radiation pressure and make bound states of light fermions which behaves like dark matter. The stability of the nugget is achieved when scalar force is balanced by degenerate fermi pressure of the light sterile fermions. 
The size of the nugget depends on the strength and the range of the scalar force. The dark matter nugget profile is solved by solving the Klein-Gordon equation and force balance equation\\

$\phi'' + \frac{2}{r} \phi' = \frac{dV(\phi)}{d\phi} - \frac{dln[m(\phi)]}{d\phi} T_{\mu}^{\mu}$ \,\,

$\frac{dp}{d\phi}= \frac{d[ln(m(\phi))]}{d\phi} ( 3 p -\rho)$\newline

where $m(\phi), p, \rho$  is the scalar profile dependent  mass, pressure and energy density of the dark matter fermion. 
Using Thomas-Fermi approximation, we  numerically solved the profile for the dark matter nugget which has been shown above.
 
\begin{figure*}[t]
\centering
\includegraphics[width=2.2in]{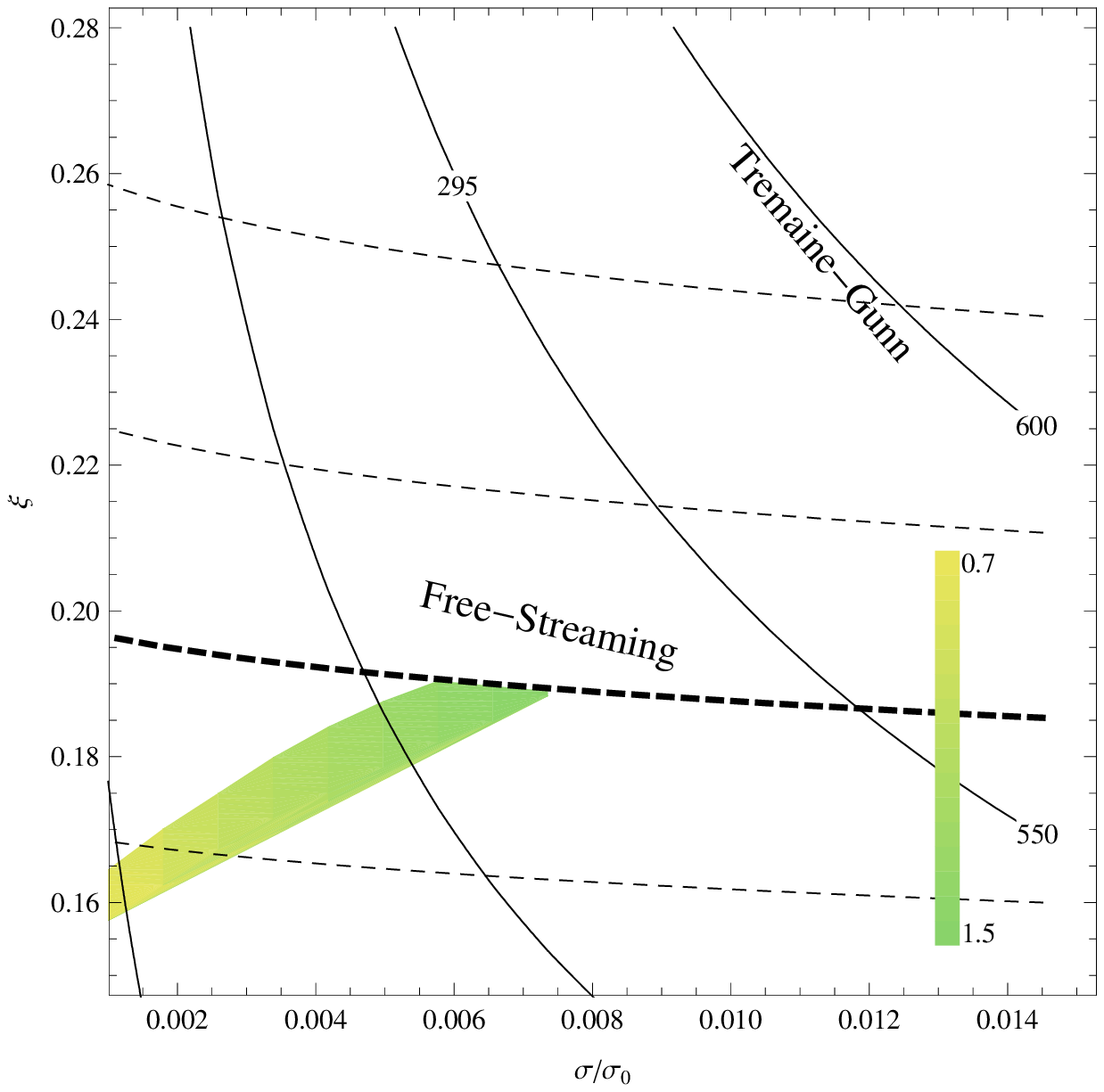}\quad
\includegraphics[width=2.2in]{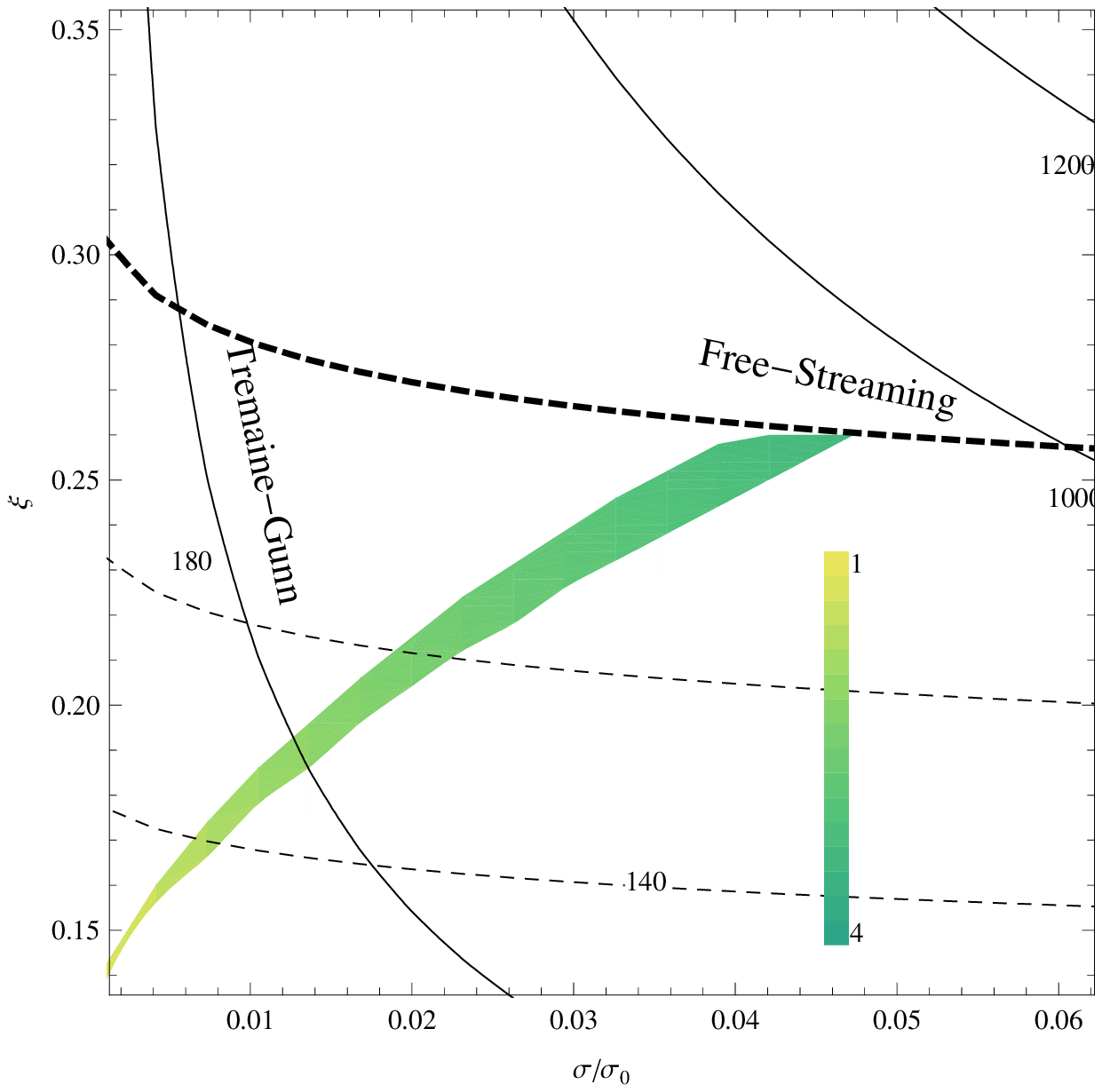}\quad
\includegraphics[width=2.2in]{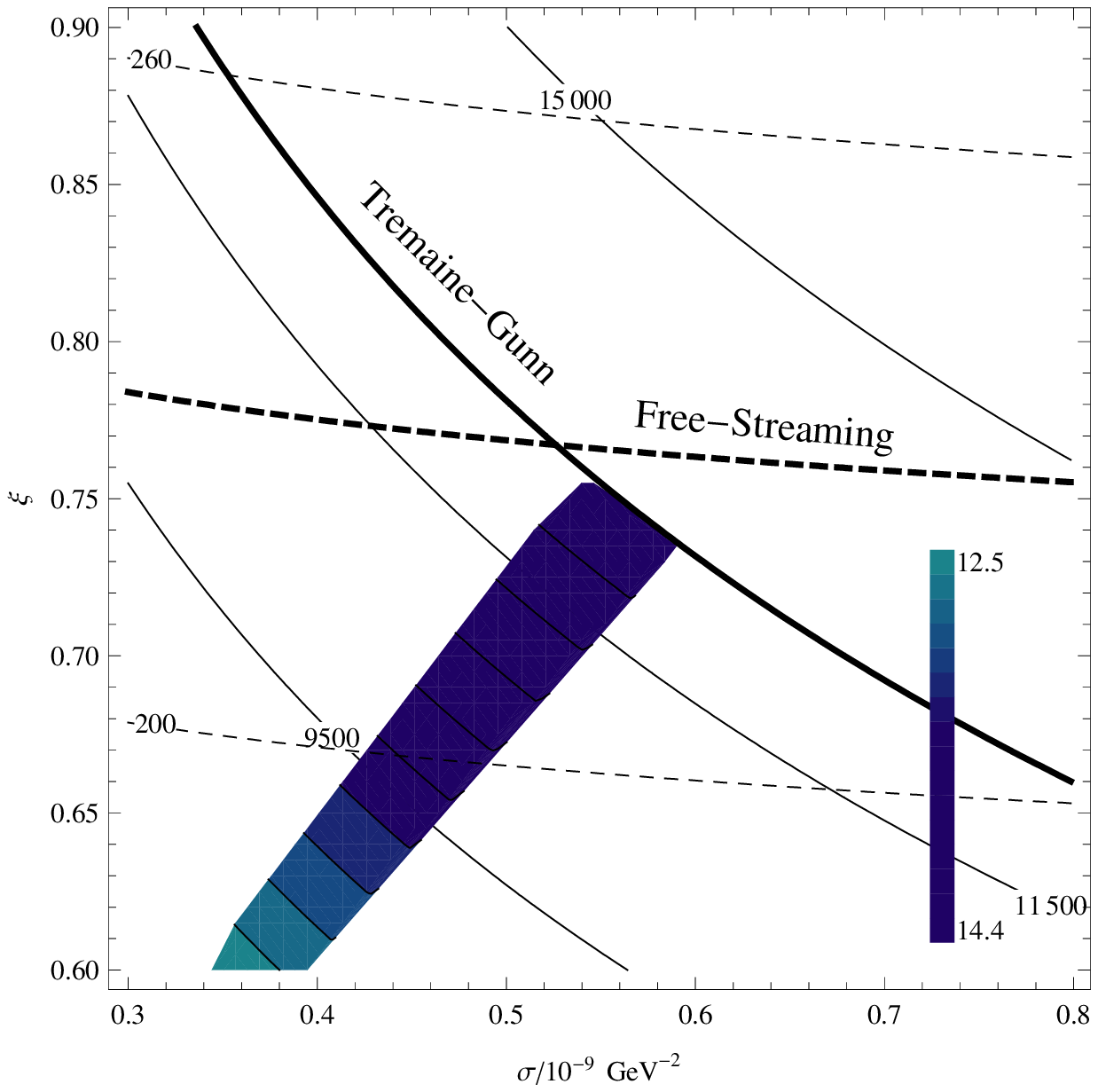}\quad
\caption{Representative constraints in $({\sigma}/{\sigma_0}, \xi)$ plane for \mbox{$m_{\chi}=1900$\,\,{\rm eV}} (left panel), \mbox{3\,\,\rm{keV}} (middle panel), and \mbox{$12.5$\,\,\rm{keV}} (right panel).  Here, $\sigma_0=10^{-9}$ $\rm GeV^{-2}$. The horizontal dashed
 lines are contours of free-streaming length while the bold one corresponds to  $ \lambda^{FSH} \simeq 230\,\,\rm{kpc}$. The solid lines are contours of minimum Tremaine-Gunn mass, $m_{min}^{TG}(\xi,\sigma)$. We see that the free-streaming bound is more stringent for such low dark matter mass. The colored region corresponds to the allowed values of dark matter density ($ 0.1\leq \Omega_{DM} h^2 \leq 0.114$) and that also obey the Free-streaming and Tremaine-Gunn bounds. The corresponding values of hidden-sector 
$x_f^h \equiv m/T_f^h$ are shown in the sidebar to illustrate the nature of decoupling (i.e., whether relativistic, nonrelativistic or semirelativistic). We see that for low mass $x_f^h $ is smaller and decoupling tends to be relativistic, while for higher mass $x_f^h $ is larger with values corresponding to either semirelativistic (warm) or nonrelativistic freeze out. }\label{Contours123}
\end{figure*}
 
 \bigskip
 
 In Fig 2.  we show the bound structure of eV mass neutrino in dark matter nugget, where we see that the potential barrier at the boundary of the nugget makes the dark matter nugget stable. We find the density of sterile fermions as a function of distance and show that it drops to zero at the boundary of the nugget. Also we show a generic warm dark matter like effect in linear matter power spectra where dark matter was free-streaming  like hot neutrinos and then went through a phase transition in RDE  few e-folding matter radiation equality.

\begin{figure*}[]
\centering
\includegraphics[scale=0.55]{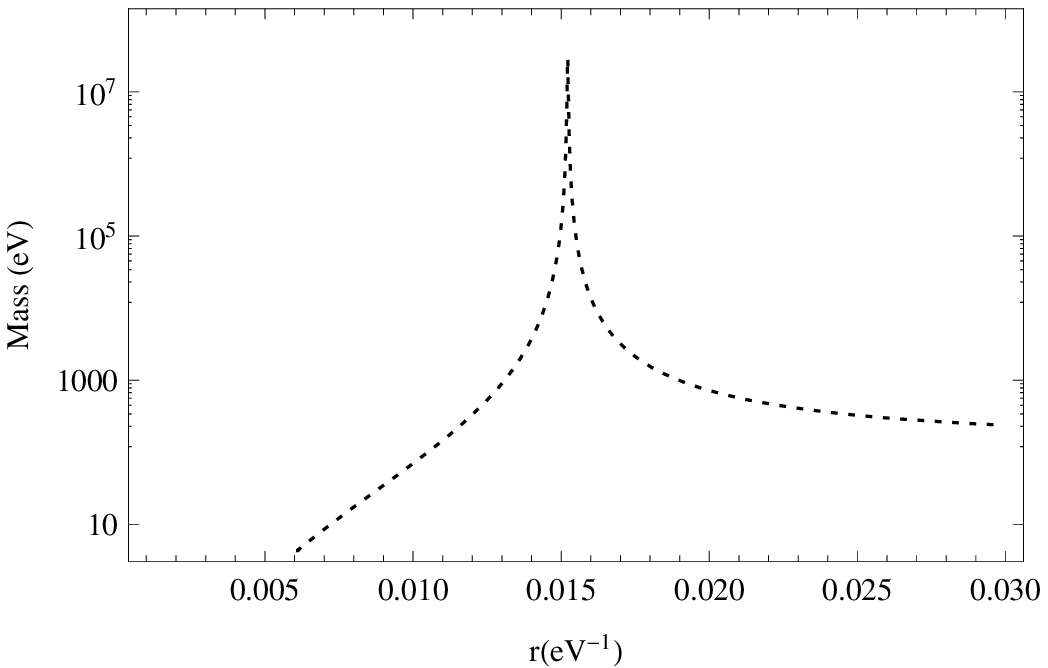}\quad
\includegraphics[scale=0.52]{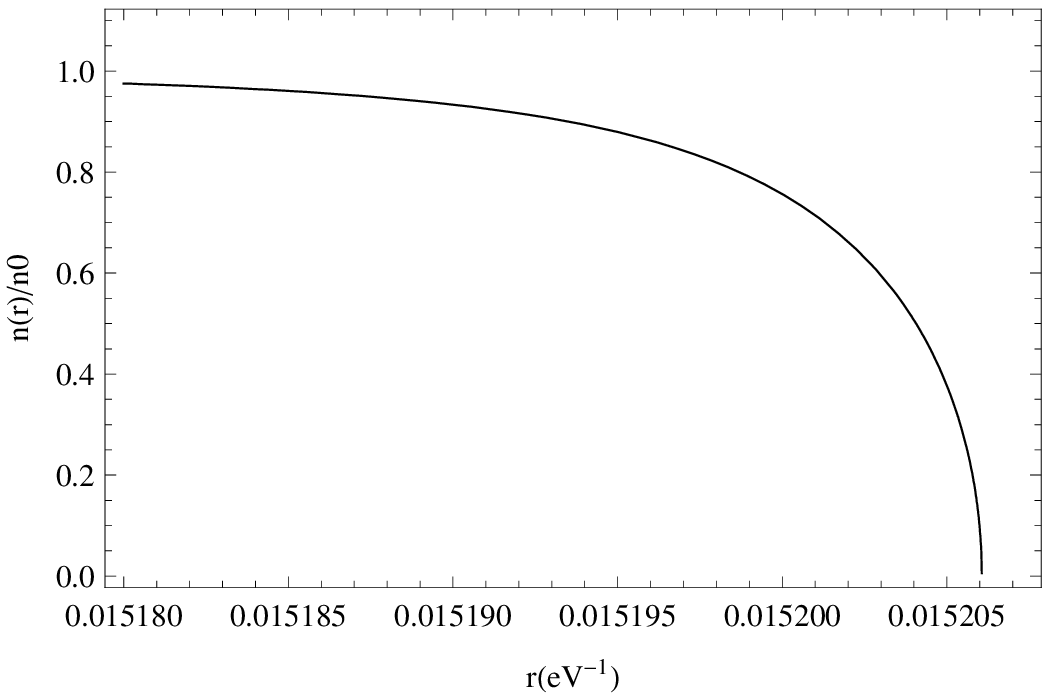}\quad
\includegraphics[scale=0.3]{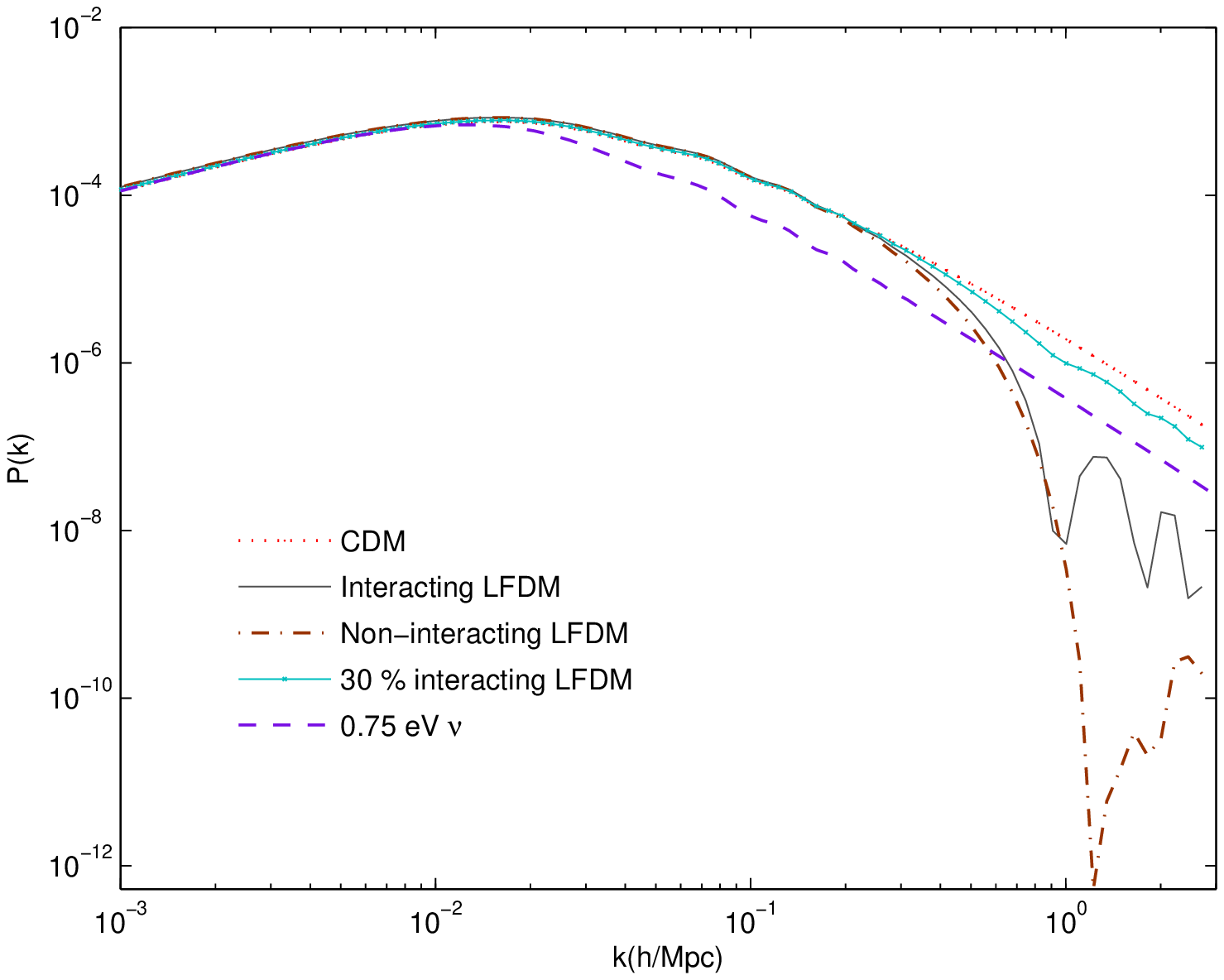}\quad
\caption{ 
 Left panel shows the fermion mass as function of distance r from the center of Dark Matter nugget. The radius is calculated to be  $~ 10^{-7}$ cm. The middle panel shows the fermion number density$\frac{n}{n_0}$ ($n_0$ being the density at the center) as a function of distance. The radius is determined when number density drops to zero. The right panel shows the warm dark matter like effect of such relatively late phase transition to DM nugget. The suppression in power in smaller scale is the signature of free-streaming before the formation of dark matter nugget. }\label{numassr}
\end{figure*}

\bigskip

{\bf References}

\begin{description}

\item[1]   S.~Das and K.~Sigurdson,Phys.\ Rev.\ D {\bf 85}, 063510 (2012)
  [arXiv:1012.4458 [astro-ph.CO]].

\item[2]  S.~Das and N.~ Weiner, Phys.\ Rev.\ D {\bf 84}, 123511 (2011)

\end{description}

\newpage

\subsection{Hector J. DE VEGA and Norma G. SANCHEZ}

\vskip -0.3cm

\begin{center}

HJdV: LPTHE, CNRS/Universit\'e Paris VI-P. \& M. Curie \& Observatoire de Paris.\\
NGS: Observatoire de Paris, LERMA \& CNRS

\bigskip

{\bf Quantum WDM fermions and gravitation determine the observed galaxy structures I} 

\end{center}

Warm dark matter (WDM) means DM particles with mass $ m $ in the keV scale.
For large scales, for structures beyond $ \sim 100$ kpc, WDM and CDM yield identical results 
which agree with observations. For intermediate scales, WDM gives the correct abundance of substructures.
Inside galaxy cores, below $ \sim 100$ pc, $N$-body classical physics simulations 
are incorrect for WDM because at such scales quantum WDM effects are important.
Quantum calculations (Thomas-Fermi approach) for the WDM fermions provide galaxy cores, 
galaxy masses, velocity dispersions and density profiles in agreement with the observations.
All evidences point to a dark matter particle mass around 2 keV.
Baryons, which represent 16\% of DM, are expected to give a correction to pure WDM results.
The detection of the DM particle depends upon the particle physics model.
Sterile neutrinos with keV scale mass (the main WDM candidate) can be detected in 
beta decay for Tritium and Renium and in the electron capture in Holmiun.
The sterile neutrino decay into X rays can be detected observing DM
dominated galaxies and through the distortion of the black-body CMB spectrum.
The effective number of neutrinos, N$_{\rm eff}$ measured by WMAP9 and Planck satellites
is compatible with one or two Majorana sterile neutrinos in the eV mass scale.
The WDM contribution to  N$_{\rm eff}$ is of the order $ \sim 0.01 $ and therefore
too small to be measurable by CMB observations.

So far, {\bf not a single valid} objection arose against WDM.

\vskip 0.1 cm

DM particles decouple due to the universe expansion, their distribution function 
{\bf freezes out} at decoupling. The characteristic length scale after decoupling
is the {\bf free streaming scale (or Jeans' scale)}. Following the DM evolution since
ultrarelativistic decoupling by solving the linear Boltzmann-Vlasov equations
yields (see [1]),
\be\label{fs}
r_{Jeans} = 57.2 \, {\rm kpc}
\; \frac{\rm keV}{m} \; \left(\frac{100}{g_d}\right)^{\! \frac13} \; , 
\ee
where $ g_d $ equals the number of UR degrees of freedom at decoupling. 
DM particles can {\bf freely} propagate over distances of the order of the free streaming scale.
Therefore, structures at scales smaller or of the order of $ r_{Jeans} $ are {\bf erased}
for a given value of $ m $.

\vskip 0.1 cm

The observed size of the DM galaxy substructures is in the 
$ \sim 1 - 100 $ kpc scale. Therefore, eq.(\ref{fs}) indicates that $ m $ 
should be in the keV scale. That is, Warm Dark Matter particles.
This indication is confirmed by the phase-space density observations in galaxies
[2] and further relevant evidence from galaxy observations [5,6].

\vskip 0.1 cm 

CDM particles with $ m \sim 100$ GeV have $ r_{Jeans} \sim 0.1 $ pc.
Hence CDM structures keep forming till scales as small as the solar system.
This result from the linear regime is confirmed  as a {\bf robust result} by
$N$-body CDM simulations. However, it has {\bf never been observed} in the sky. 
Adding baryons to CDM does not cure this serious problem. 
There is {\bf over abundance} of small structures in CDM and in CDM+baryons
(also called the satellite problem). CDM has {\bf many serious} conflicts with observations as:
\begin{itemize}
\item{Galaxies naturally grow through merging in CDM models.
Observations show that galaxy mergers are {\bf rare} ($ <10 \% $).}
\item{Pure-disk galaxies (bulgeless) are observed whose formation 
through CDM is unexplained}.
\item{CDM predicts {\bf cusped} density profiles: $ \rho(r) \sim 1/r $ for small $ r $.
Observations show {\bf cored } profiles: $ \rho(r) $  bounded for small $ r $.
Adding by hand strong enough feedback from baryons in the CDM models
can eliminate cusps but spoils the star formation rate.}
\end{itemize}
$N$-body CDM simulations {\bf fail} to produce the observed structures 
for {\bf small} scales less than some kpc.
Both $N$-body WDM and CDM simulations yield {\bf identical and correct} structures 
for scales larger than some kpc.
At intermediate scales WDM give the {\bf correct abundance} of substructures [7].

{\vskip 0.1cm} 

Inside galaxy cores, below  $ \sim 100$ pc, $N$-body classical physics simulations 
are incorrect for WDM because quantum effects are important in WDM at these scales.
WDM predicts correct structures for small scales (below kpc) when its {\bf quantum} nature is
taken into account [4].

{\vskip 0.1cm} 

All searches of CDM particles (wimps) look for $ m \gtrsim 1 $ GeV [8].
The fact that the DM mass is in the keV scale explains why no detection 
has been reached so far.
Moreover, past, present and future reports of signals of such CDM experiments
{\bf cannot be due to DM detection} because the DM particle mass is in the keV scale.
The inconclusive signals in such experiments should be originated by phenomena
of other kinds. Notice that the supposed wimp detection signals reported in refs. [8]
contradict each other supporting the notion that these  signals are {\bf unrelated to any DM
detection}.

\vskip 0.1cm 

Positron excess in cosmic rays are unrelated to DM physics but to astrophysical
sources and astrophysical mechanisms and can be explained by them [9].

\begin{center}
{\bf Quantum physics in Galaxies}
\end{center}

In order to determine whether a physical system has a classical or quantum nature
one has to compare the average distance between particles with their
de Broglie wavelength. The de Broglie wavelength of DM particles in a galaxy can be expressed as
\be\label{LdB}
\lambda_{dB}  = \frac{\hbar}{m \; v} \; ,
\ee
where $ v $ is the velocity dispersion, while the average interparticle distance $ d $ can be estimated as
\be\label{dis}
d = \left( \frac{m}{\rho_h} \right)^{\! \! \frac13} \; ,
\ee
where $ \rho_h $ is the average density in the  galaxy core.
We can measure the classical or quantum character of the system by considering the ratio
$$ 
{\cal R} \equiv \frac{\lambda_{dB}}{d}
$$
By using the phase-space density, $  Q_h \equiv \frac{\rho_h}{\sigma^3} $
and eqs.(\ref{LdB})-(\ref{dis}), $ \cal R $ can be expressed as [4]
$$
{\cal R} = \hbar \; \left( \frac{Q_h}{m^4}\right)^{\! \! \frac13} \; .
$$
Using now the observed  values of $ Q_h $ from Table \ref{pgal} yields $ \cal R $ in the range
\be\label{quant}
 2 \times 10^{-3}  \; \left( \displaystyle \frac{\rm keV}{m}\right)^{\! \frac43}
< {\cal R} < 1.4 \; \left( \displaystyle \frac{\rm keV}{m}\right)^{\! \frac43}
\ee
The larger value of $ \cal R $ is for ultracompact dwarfs while the smaller value of $ \cal R $ 
is for big spirals.

\vskip 0.1 cm 

The ratio $ \cal R $ around unity clearly implies a macroscopic quantum object.
Notice that $ \cal R $ expresses solely in terms of $ Q $ and hence 
$ (\hbar^3 \; Q/m^4) $ measures how quantum or classical is the system, here, the galaxy. 
Therefore, eq.(\ref{quant}) clearly shows {\bf solely from observations} 
that compact dwarf galaxies are natural macroscopic quantum objects for WDM [4].

\vskip 0.1 cm 

We see from eq.(\ref{quant}) that for CDM, that is for $ m\gtrsim $ GeV,
$ {\cal R}_{CDM} \lesssim 10^{-8} $, and therefore quantum effects are negligible in CDM.

\vskip 0.1 cm 

We estimate the quantum pressure in galaxies using the Pauli principle together
with the Heisenberg relations and show that dwarf galaxies are supported by the
fermionic {\it WDM quantum pressure} [4].

We obtain quantitaive results for all types of galaxies using the Thomas-Fermi approach
in [4] (see also the part II of this contribution in page 29).


\begin{table}
\begin{tabular}{|c|c|c|c|c|c|} \hline  
 & & & & & \\
 Galaxy  & $ \displaystyle \frac{r_h}{\rm pc} $ & $  \displaystyle \frac{v}{\frac{\rm km}{\rm s}} $
& $ \displaystyle  \frac{\hbar^{\frac32} \;\sqrt{Q_h}}{({\rm keV})^2} $ & 
$ \rho(0)/\displaystyle \frac{M_\odot}{({\rm pc})^3} $ & $ \displaystyle \frac{M_h}{10^6 \; M_\odot} $
\\ & & & & & \\ \hline 
Willman 1 & 19 & $ 4 $ & $ 0.85 $ & $ 6.3 $ & $ 0.029 $
\\ \hline  
 Segue 1 & 48 & $ 4 $ & $ 1.3 $ & $ 2.5 $ & $ 1.93 $ \\ \hline  
  Leo IV & 400 & $ 3.3 $ & $ 0.2 $ & $ .19 $ & $ 200 $ \\ \hline  
Canis Venatici II & 245 & $ 4.6 $ & $ 0.2 $   & $ 0.49 $ & $ 4.8 $
\\ \hline  
Coma-Berenices & 123 & $ 4.6 $  & $ 0.42 $   & $ 2.09 $  & $ 0.14 $
\\ \hline  
 Leo II & 320 & $ 6.6 $ & $ 0.093 $  & $ 0.34 $ & $ 36.6 $
\\  \hline  
 Leo T & 170 & $ 7.8 $ &  $ 0.12 $  & $ 0.79 $ & $ 12.9 $
\\ \hline  
 Hercules & 387 & $ 5.1 $ &  $ 0.078 $  & $ 0.1 $ & $ 25.1 $
\\ \hline  
 Carina & 424 & $ 6.4 $ & $ 0.075 $  & $ 0.15 $ & $ 32.2 $
\\ \hline 
 Ursa Major I & 504 & 7.6  &  $ 0.066 $  & $ 0.25 $ & $ 33.2 $
\\ \hline  
 Draco & 305 & $ 10.1 $ &  $ 0.06 $  & $ 0.5 $ & $ 26.5 $
\\ \hline  
 Leo I & 518  & $ 9 $ &  $ 0.048 $  & $ 0.22 $ & $ 96 $
\\ \hline  
 Sculptor & 480  & $ 9 $ & $ 0.05 $  & $ 0.25  $ & $ 78.8 $
\\ \hline 
 Bo\"otes I & 362 & $ 9 $ & $ 0.058 $  & $ 0.38 $ & $ 43.2 $
\\ \hline  
 Canis Venatici I & 1220  & $ 7.6 $ & $ 0.037 $ & $ 0.08 $ & $ 344 $
\\ \hline  
Sextans & 1290 & $ 7.1 $ & $ 0.021 $ & $ 0.02 $ & $ 116 $
\\ \hline 
 Ursa Minor & 750 & $ 11.5 $ & $ 0.028 $  & $ 0.16 $ & $ 193 $
\\ \hline  
 Fornax  & 1730 & $ 10.7 $ & $ 0.016 $  & $ 0.053  $ & $ 1750 $
\\  \hline  
 NGC 185  & 450 & $ 31 $ & $ 0.033 $ & $ 4.09 $ & $ 975 $
\\ \hline  
 NGC 855  & 1063 & $ 58 $ & $ 0.01 $ & $ 2.64 $ & $ 8340 $
\\ \hline  
  Small Spiral  & 5100  & $ 40.7 $ & $ 0.0018 $ & $ 0.029 $ & $ 6900 $
\\ \hline  
NGC 4478 & 1890 & $ 147 $ & $ 0.003 $ & $ 3.7 $ & $ 6.55 \times 10^4 $
\\ \hline  
 Medium Spiral & $ 1.9 \times 10^4 $ & $ 76.2 $ & $ 3.7 \times 10^{-4} $ & $ 0.0076 $ & $ 1.01 \times 10^5 $
\\ \hline  
 NGC 731 & 6160 & $ 163 $ & $ 9.27 \times 10^{-4} $ & $ 0.47 $ & $ 2.87 \times 10^5 $
\\ \hline 
 NGC 3853   & 5220 & $ 198 $ & $ 8.8 \times 10^{-4} $ & $ 0.77 $  
& $ 2.87 \times 10^5 $ \\ \hline 
NGC 499  & 7700 &  $ 274 $ & $ 5.9 \times 10^{-4} $ & 
$ 0.91 $ & $ 1.09 \times 10^6 $ \\   \hline 
Large Spiral & $ 5.9 \times 10^4 $ & $ 125 $ & $ 0.96 \times 10^{-4} $ & $ 2.3 \times 10^{-3} $ & 
$ 1. \times 10^6 $ \\ \hline  
\end{tabular}
\caption{Observed values $ r_h $, velocity dispersion $ v, \;  \sqrt{Q_h}, \; \rho(0)$ 
and $ M_h $ covering from ultracompact galaxies to large spiral galaxies
from refs. [5]. The phase space density is larger for smaller galaxies, both in mass and size.
Notice that the phase space density is obtained
from the stars velocity dispersion which is expected to be smaller than the DM  velocity dispersion.
Therefore, the reported $ Q_h $ are in fact upper bounds to the true values [5].}
\label{pgal}
\end{table}

{\bf References}

\begin{description}

\item[1] D. Boyanovsky,  H J de Vega, N. G. Sanchez, 	
Phys. Rev.  {\bf D 78}, 063546 (2008).

\item[2] H. J. de Vega, N. G. S\'anchez, 
MNRAS {\bf 404}, 885 (2010) and Int. J. Mod. Phys. {\bf A 26}, 1057 (2011).

\item[3] H. J. de Vega, N. G. S\'anchez, Phys. Rev. D85, 043516 (2012)
and  D85, 043517 (2012).

\item[4] C. Destri, H. J. de Vega, N. G. Sanchez, arXiv:1204.3090,
New Astronomy {\bf 22}, 39 (2013) and arXiv:1301.1864, Astroparticle Physics, 46, 14 (2013).

\item[5] G. Gilmore et al., Ap J, 663, 948 (2007).
M. Walker, J. Pe\~narrubia, Ap. J. 742, 20 (2011).
P. Salucci et al., MNRAS, 378, 41 (2007).
H. J. de Vega,  P. Salucci, N. G. Sanchez, 
New Astronomy {\bf 17}, 653 (2012) and references therein.
J. D. Simon, M. Geha,  Ap J, 670, 313 (2007) and references therein.
J. P. Brodie  et al., AJ, 142, 199 (2011). B. Willman and J. Strader, AJ, 144, 76 (2012).
J. D. Simon et al., Ap. J. 733, 46 (2011) and references therein.
J. Wolf  et al., MNRAS, 406, 1220 (2010) and references therein.
G. D. Martinez et al., Ap J, 738, 55 (2011). 

\item[6] P. Col\'{\i}n, O. Valenzuela, V.  Avila-Reese, Ap J, 542, 622 (2000).
J. Sommer-Larsen, A. Dolgov, Ap J, 551, 608 (2001).
L. Gao and T. Theuns,  Science, 317, 1527 (2007).
A. Schneider et al., MNRAS, 424, 684 (2012).
M. R. Lovell et al., MNRAS, 420, 2318 (2012).
A. V. Tikhonov et al., MNRAS, 399, 1611 (2009).	

\item[7] E. Papastergis et al., Ap J, 739, 38 (2011),
J. Zavala et al., Ap J,	700, 1779 (2009).


\item[8] Dama/Libra: R. Bernabei et al., arXiv:1301.6243,
{\em Eur.Phys.J.} {\bf C67} (2010) 39.
CDMS-II: R. Agnese et al., arXiv:1304.4279, 
Z.~Ahmed et~al., {\em Science} {\bf 327} (2010) 1619, PRL {\bf 106}
  (2011) 131302.
AMS: M. Aguilar et al.,   PRL, 110, 141102 (2013).
CoGeNT: C.~Aalseth et~al., 
  Phys.Rev.Lett. {\bf 107} (2011) 141301.
CRESST: G.~Angloher et~al., {\em Eur.Phys.J.} {\bf C72} (2012) 1971.
XENON10: J.~Angle et~al., PRL {\bf 107} (2011) 051301,
XENON100: E.~Aprile et~al., PRL {\bf 109} (2012)   181301.

\item[9] P. L. Biermann et al. PRL (2009), P. Blasi, P. D. Serpico PRL (2009).

\end{description}

\newpage

\subsection{Elena FERRI, on behalf of the MARE collaboration}

\vskip -0.3cm

\begin{center}

Dip. di Fisica ``G. Occhialini '', Universit\`a di Milano-Bicocca and INFN Sezione di Milano-Bicocca, Milano, Italy\\

\bigskip

{\bf The MARE experiment and its capabilities to measure the mass of light (active) and heavy (sterile)
neutrinos} 

\end{center}

\medskip

In the last decades, the observation of flavour oscillations of solar and atmospheric neutrinos, as well as of reactor and accelerator neutrinos, have provided the evidence of a non-vanishing neutrino mass leading to a new physics beyond the Standard Model (SM). The difficulty in detecting and studying them explains why the neutrinos are the object of many experiments dedicated to determine their mass and nature. As the oscillation experiments are sensitive only to the differences in squared neutrino mass $\Delta m^2_{ij}$, they are not able to determine the absolute neutrino mass scale. The experiments dedicated to effective electron-neutrino mass determination are the ones based on kinematic analysis of electrons emitted in single $\beta$-decay. Relying only on energy-momentum conservation in $\beta$-decay, they are the only model-independent method to measure the neutrino mass scale with a sub-eV sensitivity. These experiments look for a tiny deformation of the beta spectr
 um closed to the end-point energy due to a non-zero neutrino mass.

\bigskip

To date, the most stringent results come from electrostatic spectrometers on tritium decay. The Mainz collaboration has reached m$_\nu\le$ 2.3 eV/c$^2$ [1]. With the Troitsk experiment, an upper limit on neutrino mass of 2.5 eV/c$^2$ has been obtained [2]. The next generation experiment KATRIN is designed to reach a sensitivity of 0.2 eV/c$^2$ in five years [3-4]. Spectrometers allow focusing on a very narrow energy range below the end-point, which can be investigated with a very sharp energy resolution. However a severe limitation of these experiments resides in their configuration since the $^3$H is external to the spectrometer. For that reason the measured electron energy has to be corrected for the energy lost in excited atomic and molecular states, in crossing the source, in scattering through the spectrometer and more. Because of the large weight of systematics, the confidence in the results can be obtained only through confirmation by independent experiments.

\bigskip

An alternative approach to spectrometry is calorimetry where the $\beta$-source is embedded in the detector so that all the energy emitted in the decay is measured, except for the one taken away by the neutrino. In this way the measurement is completely free from systematics induced by any possible energy loss in the source and due to decays into excited final states. The remaining systematics may be due to energy lost in metastable states living longer than the detector response time. However, there is an important inconvenience which represents a serious limitation for this approach. In fact, in contrast to the spectrometric approach, the full beta spectrum is acquired. Therefore, the source activity has to be limited to avoid pile-up which would deform the shape of beta spectrum. As a consequence the statistics near the end-point is limited as well. Since the fraction of decays in a given energy interval $\Delta E$ below the end-point $Q$ is only $\propto (\Delta E/Q)^3$, 
 this limitation may be then partially balanced by using $\beta$-emitting isotopes with an end-point energy as low as possible. The best choice is to use $^{187}$Re as beta source, the
beta-active nuclide with the second lowest known transition energy ($Q \sim  2.5$ keV).  Up to now, only two $\beta$ decay experiments have been carried out with thermal detectors and $^{187}$Re: MANU [5-6] and MIBETA [7-8] experiments. MANU used metallic Rhenium single crystal as absorber, while MIBETA used AgReO$_4$ crystals. Collecting a statistic of about 10$^7$ events, they achieved an upper limit on neutrino mass of about 26 eV/c$^2$ at 95\% CL and 15 eV/c$^2$ at 90\% CL, respectively. In these experiments, the systematic uncertainties are still small compared to the statistical errors. The main sources of systematics are the background, the pile-up, the theoretical shape of the $^{187}$Re $\beta$ spectrum, the detector response function and the Beta Environmental Fine Structure (BEFS) [9-11]. 

\bigskip

\begin{figure}[h!tbp]
\centering
  \includegraphics[scale=0.18]{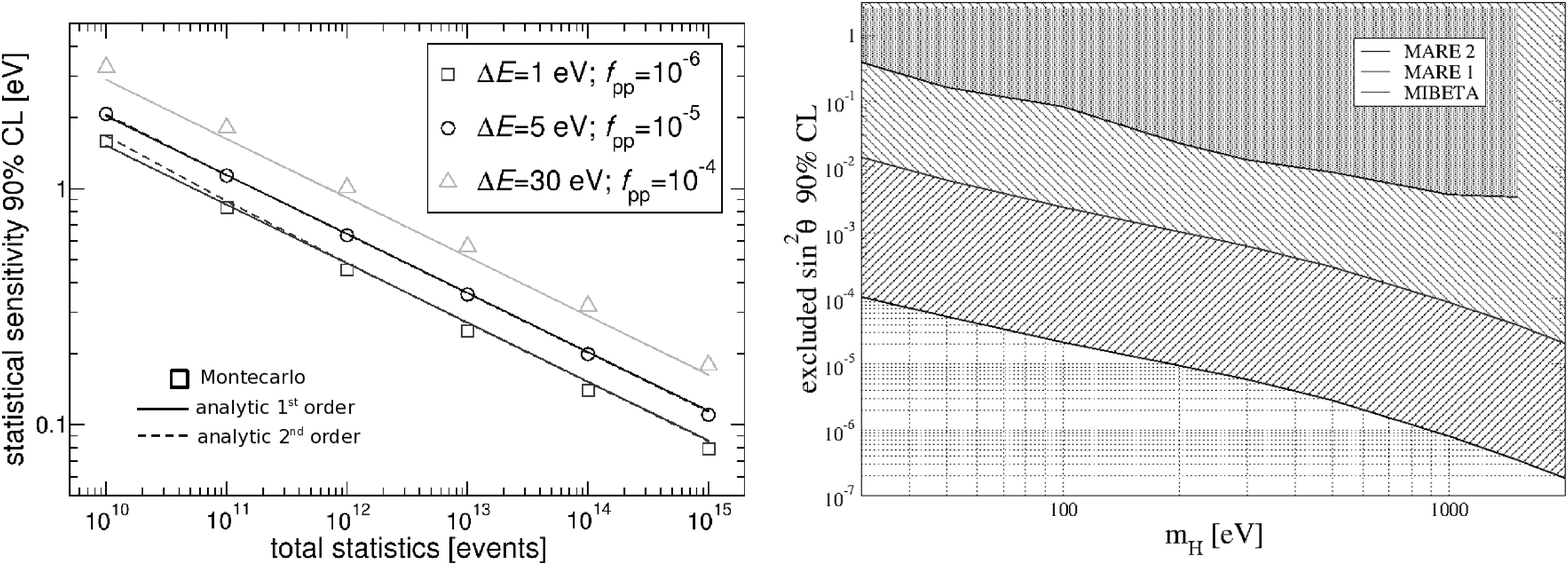}
  \caption{Left Panel: statistical sensitivity of a $^{187}$Re neutrino mass calorimetric experiment estimated with both Montecarlo (points) and analytic (lines) approaches. Right Panel: statistical sensitivity to the emission of heavy neutrinos with mass $m_H$. From lower to upper curve: Montecarlo simulation with $N_{ev}$ = 10$^{14}$, $f_{pp} = 10^{−5}$ and $\Delta E = 1$ eV (MARE 2); Montecarlo symulation with $N_{ev}$ = 10$^{10}$, $f_{pp} = 10^{−4}$ and $\Delta E$ = 30 eV (MARE 1); MIBETA experiment. }
\label{fig:3}
\end{figure}

In this scenario an international collaboration has grown around the project of Microcalorimeter Arrays for a Rhenium Experiment (MARE) for a direct calorimetric measurement of the neutrino mass with sub-electronvolt sensitivity. The experimental requirements of a such experiment are the total statistics  $N_{ev}$, the energy resolution $\Delta E$ and the pile-up level $f_{pp}$ – i.e. the fraction of pile-up events given by the product between the beta activity $A_\beta$ and the rise time of the detector $\tau_R$. As shown in the left panel of figure \ref{fig:3}, the MARE experiment must collect more than 10$^{13}$ events in order to achieve a sub-eV sensitivity on neutrino mass.    Furthermore, the results show that the 90\% confidence level sensitivity is proportional to $N_{ev}^{-1/4}$, this dependence may be exploited to scale the Montecarlo results.
 
\bigskip

Although the baseline of the MARE project consists in a large array of rhenium based thermal detectors, a different option for the isotope is also being considered. In fact, the MARE collaboration is studying the possibility to use $^{163}$Ho. While $^{187}$Re  decays beta, $^{163}$Ho decays via electron capture EC and its $Q$ value is around 2.5 keV. $^{163}$Ho decays to $^{163}$Dy with
a half life of about 4570 years and the capture is only allowed from the M shell or higher. The EC may be only detected through the mostly non radiative atom de-excitation of the Dy atom and from the Inner Bremsstrahlung (IB) radiation. One of the methods to estimate the neutrino mass from the $^{163}$Ho EC  was proposed by De Rujula and Lusignoli in 1982 [12] and it consists in studying the end-point of the total absorption spectrum.  Also in $^{163}$Ho experiments the sensitivity on $m_\nu$ depends on the fraction of events at the end-point which increases for decreasing $Q_{EC}$. Unfortunately,  the $Q_{EC}$ is not well known and the various $Q_{EC}$ determinations span from 2.2 to 2.8 keV [13], with a recommended value of 2.555 keV [14]. $^{163}$Ho represents a very interesting alternative to $^{187}$Re and thanks to its relatively short half life, detectors may be realized by introducing only few $^{163}$Ho nuclei in low temperature microcalorimeters.

\bigskip

The MARE project has a staged approach. The first phase of the project (MARE-1) is a collection of activities with the aim of sorting out both the best isotope and the most suited detector technology to be used for the final experiment [15]. The goal of the last phase (MARE-2) is to achieve a sub-eV sensitivity on the neutrino mass. It will deploy several arrays of thermal microcalorimeters.

\bigskip

One of the MARE-1 activity is starting in Milano using two 6x6 silicon implanted arrays equipped with AgReO$_4$ absorbers. So far, the established energy and time resolutions are about 25 eV and 250 $\mu$s respectively. The experiment, which is presently being assembled, is designed to host up to 8 XRS2 arrays. With 288 detectors and such performances, a sensitivity of 4 eV at 90 $\%$ CL on the neutrino mass can be reached within 3 years. This corresponds to a statistics of about 10$^{10}$ decays. The purposes of this experiment are to reach a sensitivity on neutrino mass of few eV and to investigate the systematics of $^{187}$Re neutrino mass measurements, focusing on those caused by the Beta Environmental Fine Structure (BEFS) and the beta spectrum theoretical shape. Furthermore, superconducting microwave microresonator arrays for calorimetric measurement of the energy spectra of $^{163}$Ho EC decay are currently subject of an R\&D work in Milano.

Finally, the calorimetric spectra measured by MARE with either isotopes are also suitable to investigate the emission of heavy (sterile) neutrinos with a mixing angle $\theta$. In heavy neutrino investigation it is useful to assume that the electron neutrino $\nu_e$ is predominately a linear combination of two mass eigenstates $\nu_L$ and $\nu_H$ of masses $m_L$ and $m_H$ ($m_L \ll m_H$), where $L$ and $H$ stand for light and heavy respectively. So that $\nu_e= \nu_L \cos(\theta) + \nu_H \sin(\theta)$ and the measured energy spectrum is $N(E,m_L, m_H, \theta) = N(E,m_L)\cos^2(\theta) + N(E,m_H) \sin^2(\theta)$. Therefore, the emission of heavy neutrino would manifest as a kink in the spectrum at energy $Q-m_H$ for heavy neutrinos with masses between 0 and $Q-E_{th}$, where $E_{th}$ is the experimental energy threshold. The potential statistical sensitivity to heavy neutrinos for the full scale experiment (MARE-2), for the intermediate scale experiment (MARE-1) and for the MIB
 ETA experiment are displyed in the right panel of figure \ref{fig:3}. 

\bigskip

{\bf References}

\begin{description}

\item[1] Ch. Kraus. Eur. Phys. J. C 40 (2005) 447-468.

\item[2] V. M. Lobashev. Nucl. Phys. B (Proc Suppl.)  91 (2001) 280-286.

\item[3] Osipowic A et~al, {\it Letter of intent} hep-ex/0109033 (2001).

\item[4] Angrik J. et~al, {\it Design Report} \textbf{7090}  http://bibliothek.fzk.de/zb/berichte/FZKA7090.pdf (2004)

\item[5] F. Gatti et~al, {\it Nucl. Phys. B}  \textbf{91} (2001) 293.

\item[6] M. Galeazzi et~al, {\it Phys. Rev. C}  \textbf{63} (2001).

\item[7] M. Sisti  et~al, {\it NIM A}  \textbf{520} (2004) 125.

\item[8] C. Arnaboldi et~al, {\it Phys. Rev. Lett.}  \textbf{91} (2003).

\item[9] Kooning. S.E., Nature No. 354 (1991) 468-470.

\item[10] F. Gatti et~al. Nature 397 (1999) 137.

\item[11] C. Arnaboldi et~al. Phys. Rev. Lett.  96 (2006) 042503.

\item[12] A. De Rujula and M. Lusignoli, {\it Phys. Lett. B} \textbf{118}  (1982) 72.

\item[13 ]C.W. Reich et~al, {\it Nucl. Data Sh.} \textbf{111} (2010).

\item[14] G. Audi and A. H. Wapstra, {\it Nucl. Phys. A} \textbf{595} (1995) 409.

\item[15] A. Nucciotti, proceeding of Neutrino 2010, Athens, Greece, June 14-19, 2010, arXiv:1012.2290v1

\end{description}

\newpage

\newcommand{\apjl}[0]{The Astrophysical Journal}
\newcommand{\aj}[0]{The Astronomical Journal}
\newcommand{\aap}[0]{Astronomy {\&} Astrophysics}
\newcommand{\aaps}[0]{Astronomy {\&} Astrophysics Supplement Series}
\newcommand{\apss}[0]{Astrophysics and Space Science}
\newcommand{\apjs}[0]{The Astrophysical Journal Supplement Series}
\newcommand{\mnras}[0]{Monthly Notices of the Royal Astronomical Society}
\newcommand{\pasp}[0]{Publications of the Astronomical Society of the Pacific}
\newcommand{\araa}[0]{Annual Review of Astronomy and Astrophysics}

\subsection{Ayuki KAMADA}

\vskip -0.3cm

\begin{center}

Institute for the Physics and Mathematics of the Universe, TODIAS,\\
The University of Tokyo, 5-1-5 Kashiwanoha, Kashiwa, Chiba 277-8583, Japan

\bigskip

{\bf Light sterile neutrino as warm dark matter and the structure of galactic dark halos} 

\end{center}

\medskip

{\bf Abstract:} We study the formation of nonlinear structure in $\Lambda$ Warm Dark Matter (WDM) 
cosmology using large cosmological N-body simulations.
We assume that dark matter consists of sterile neutrinos that are generated 
through nonthermal decay of singlet Higgs bosons 
near the Electro-Weak energy scale.
Unlike conventional thermal relics, the nonthermal WDM has a peculiar velocity 
distribution, which results in a characteristic shape of the matter
power spectrum.
We perform large cosmological N-body simulations for the nonthermal 
WDM model. We compare the radial distribution of subhalos 
in a Milky Way size halo 
with those in a conventional thermal WDM model.
The nonthermal WDM with mass of 1 keV predicts
the radial distribution of the subhalos that is
remarkably similar to the observed distribution
of Milky Way satellites.

\medskip

{\bf Introduction}

\medskip

Alternative models to the standard $\Lambda$ Cold Dark Matter model have been suggested as a solution 
of the so-called 'Small Scale Crisis'.
One of them is $\Lambda$ Warm Dark Matter cosmology, in which Dark Matter particles had non-zero velocity dispersions. 
The non-zero velocities smooth out primordial density perturbations below its free-streaming length of sub-galactic sizes. 
The formation of subgalactic structure is then suppressed. Moreover, The large phase space density may prevent dark matter 
from concentrating into galactic center. Particle physics models provide promising WDM candidates such as gravitinos, 
sterile neutrinos and so on. 
Gravitinos with mass of $\sim$keV in generic Supergravity theory are produced in thermal bath immediately 
after reheating and are then decoupled kinematically
from thermal bath similarly to the Standard Model (SM) neutrinos 
because gravitinos interact with SM particles only through gravity. 
The thermal relics have a Fermi-Dirac (FD) momentum distribution. 
Sterile neutrinos are initially proposed in the see-saw mechanism to explain the masses of the SM neutrinos. 

\bigskip

If the mass of sterile neutrinos is in a range of $\sim$keV 
and their Yukawa coupling is of order $\sim 10^{-10}$, then 
it cannot be in equilibrium with SM particles 
throughout the thermal history of the universe. 
There are several peculiar production mechanisms for sterile neutrinos,
such as Dodelson-Widrow (DW) mechanism~[1] and EW scale Boson Decay (BD)~[2]. 
In DW mechanism, sterile neutrinos 
are produced through oscillations of active neutrinos, and its velocity distribution has a Fermi-Dirac form 
just like gravitinos.
In the BD case, sterile neutrinos are produced via decay 
of singlet Higgs bosons and they have generally a nonthermal velocity distribution. 
The nonthermal velocity distribution imprints particular features in the transfer function of the density fluctuation 
power spectrum~[3]. The transfer function has a cut-off at the corresponding free-streaming length,
but it decreases somewhat slowly than thermal WDM models.
In this article, we study the formation of nonlinear structure for a cosmological model with nonthermal sterile neutrino WDM.
We perform large cosmological N-body simulations. 
There are several models of nonthermal WDM. For example, gravitinos can be produced via decay of inflaton~[4] 
or long-lived Next Lightest Supersymmetric Particle (NLSP)~[5]. 
Our result can be generally applied to the formation of nonlinear structure in these models as well.

\medskip

{\bf Nonthermal sterile neutrino}

\medskip

The clustering properties of the above BD sterile neutrino model is investigated by Boyanovsky~[3],
who solved the linearized Boltzmann-Vlasov equation in the matter dominant era when WDM has already 
become non-relativistic. 
Firstly, solving the Boltzmann equation for sterile neutrinos produced by the singlet boson decay process, 
we find the most of contribution to the present number of sterile neutrinos comes when the temperature 
decreases to the EW scale. We then get the velocity distribution of sterile neutrinos. 

The BD distribution has a distinguishable feature from usual Fermi-Dirac distributions at small velocities, $y=P/T(t)$:
\begin{align}
f_{\rm BD}(y)\propto \frac{1}{y^{\frac{1}{2}}}, \ \ \ \ \ f_{\rm FD}(y)\propto 1. \notag
\end{align}
The velocity distribution is imprinted in the power spectrum, which decreases slowly across the free-streaming length scale. 
Following Boyanovsky~[3], we solve the linearlized Boltzmann-Vlasov equation.
 We define the comoving free-streaming wavenumber as
$$ 
 k_{\rm fs}={\Big[}\frac{3H_{0}^2\Omega_{M}}{2\langle\vec{V}^2\rangle(t_{\rm eq})}{\Big]}^{\frac{1}{2}} \notag
$$

akin to the Jeans scale at the matter-radiation equality.
In the BD case, the present energy density of dark matter is determined
by the Yukawa coupling.

\medskip

Assuming the relativistic degree of freedom $\bar{g}$ is the usual SM value 
$\bar{g}\simeq 100$ for the $m=1$ keV sterile neutrino dark matter, we find
\begin{align}
k_{\rm fs}^{\rm BD}=18 \ h \ \mathrm{Mpc^{-1}}, \notag
\end{align}
while for the $m=1$ keV gravitino dark matter, which has the FD distribution, 
we should take $\bar{g}\simeq1000$ to get the present energy density of dark matter.
Then, for $m=1$ keV gravitinos, 
\begin{align}
k_{\rm fs}^{\rm FD}=32 \ h \ \mathrm{Mpc^{-1}}. \notag
\end{align}

\medskip

\begin{figure}[htbp]
\begin{center}
\includegraphics[height=45mm]{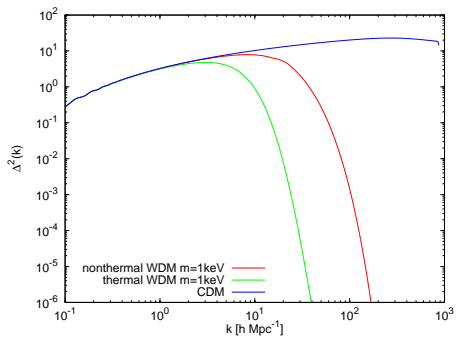}
\end{center}
\caption{The linear power spectra for the nonthermal WDM (red line), thermal WDM (green line) and CDM (blue line). 
The dark matter mass of the both WDM model is $m=1$ keV.}
\label{ak1}
\end{figure}

Although these two models have almost the same free-streaming length, 
their linear power spectra show appreciable differences, as seen in 
Fig. \ref{ak1}. There, we adopt the numerically fit transfer function 
in Bode et al.~[6] for the linear power spectrum of the gravitino 
dark matter.
 
The enhancement of the velocity distribution in the low velocity region leads 
to the slower decrease of the linear power spectrum. This implies that we should care not only free-streaming 
scale or velocity dispersion, but also the shape of the velocity distribution in studying the formation 
of the nonlinear object below the cut-off (free-streaming) scale.

Using these linear power spectrum, we have performed direct numerical simulations to study observational 
signatures of the imprinted velocity distribution in the subgalactic structures. 
We start our simulations from a redshift of $z=9$.

We use $N = 256^{3}$ particles in a comoving volume of $10 \ h^{-1}$ Mpc on a side. The mass of a dark 
matter particle is $4.53\times10^{6} \ h^{-1} \ \mathrm{M_{\odot}}$ and the gravitational softening
length is $2 \ h^{-1}$ kpc.

\medskip

{\bf Simulation Results}

\medskip

\begin{figure}[htbp]
\begin{minipage}{0.32\hsize}
\begin{center}
\includegraphics[width=35mm,height=35mm]{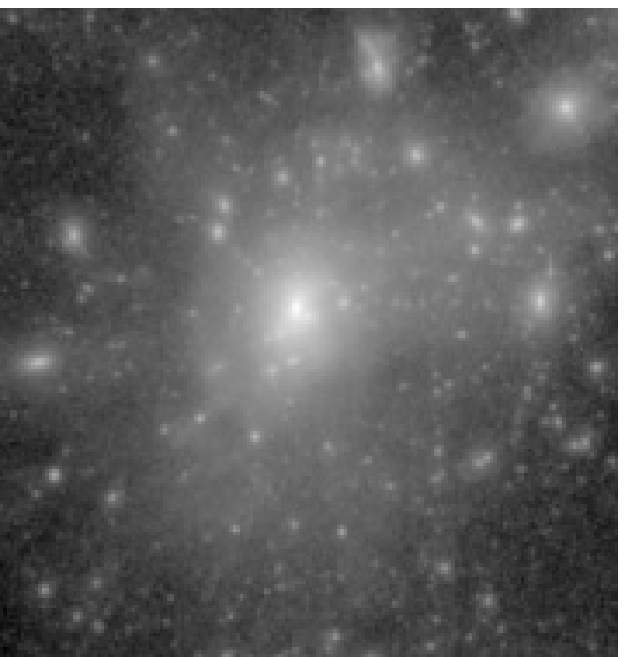}
\end{center}
\end{minipage}
\begin{minipage}{0.32\hsize}
\begin{center}
\includegraphics[width=35mm,height=35mm]{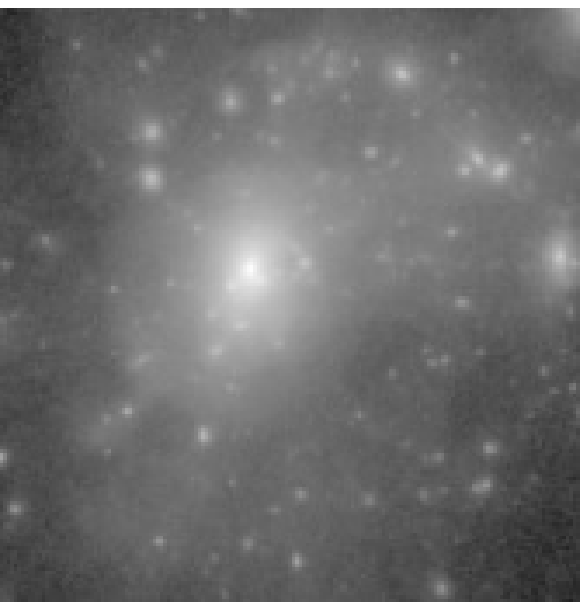}
\end{center}
\end{minipage}
\begin{minipage}{0.32\hsize}
\begin{center}
\includegraphics[width=35mm,height=35mm]{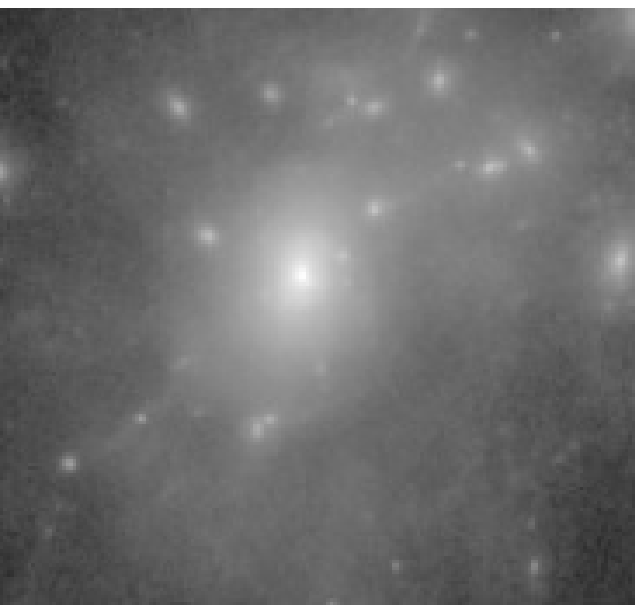}
\end{center}
\end{minipage}
\caption{The projected distributions of the substructures in our 'Milky Way' halo at $z=0$. 
The sidelength of the shown region is $2 \ h^{-1}$ Mpc. 
The results are for three dark matter models, CDM,  nonthermal WDM with 1 keV mass and thermal WDM with 1keV mass, respectively, 
from left to right.}
\label{ak2}
\end{figure}

Fig. \ref{ak2} shows the projected distribution of dark matter in and
 round a Milky Way size halo at $z=0$. 
The plotted region has a side of $2 \ h^{-1}$ Mpc. 
Dense regions appear bright. 
Clearly, there are less subgalactic structures for the WDM models. 
The 'colder' property of the nonthermal WDM shown in the linear 
power spectrum (see Fig. \ref{ak1}) can also be seen in the abundance 
of subhalos.

\begin{figure}[htbp]
\begin{center}
\includegraphics[height=50mm]{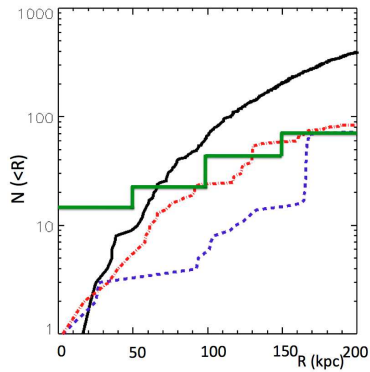}
\end{center}
\caption{The radial distribution of the subhalos in our 'Milky Way' halo at $z=0$. 
For simulation results for CDM (black), 1keV nonthermal WDM (red), and 
1keV thermal WDM (blue). Green bars show the distribution of the observed Milky Way satellites.}
\label{ak3}
\end{figure}

In Fig. \ref{ak3}, we compare the cumulative radial distribution of the subhalos in our 'Milky Way' halo at $z=0$ 
with the distribution of the observed Milky Way satellites~[7]. 
Interestingly, the nonthermal WDM model reproduces the radial distribution of the observed Milky Way satellites 
in the range above $\sim 40$ kpc. 
Contrastingly, the CDM model overpredicts the number of subhalos by a factor of $5$ than the 
observed Milky Way satellites. This is another manifestation of the so-called 
'Missing satellites problem'. The thermal WDM model 
surpresses subgalactic structures perhaps too much, by a factor of $2-4$ than the observation.

\medskip

{\bf Summary}

\medskip

We have studied the formation of the nonlinear structures in a $\Lambda$WDM cosmology using 
large cosmological N-body simulations. We adopt the sterile neutrino dark matter produced via the decay of 
singlet Higgs bosons with a mass of EW scale. The sterile neutrinos have a nonthermal velocity distribution, 
unlike the usual Fermi-Dirac distribution. The distribution is a little skewed to low velocities. 
The corresponding linear matter power spectrum decreases
slowly across the cut-off scale compared to the thermal WDM, such as gravitino dark matter.
Both of the two models have the same mass of $1$ keV and an approximately the same cut-off (free-streaming) scale 
of a few 10 $h \ \mathrm{Mpc^{-1}}$.  We have shown that this 'colder' property of the nonthermal WDM can be seen 
in richer subgalactic structures. The nonthermal WDM model with mass of $1$ keV appears to reproduce the radial 
distribution of the observed Milky Way satellites.

\medskip

{\bf References}

\begin{description}

\item[1] S.~Dodelson and L.~M.~Widrow, Phys.\ Rev.\ Lett.\  {\bf 72}, 17 (1994)
  [arXiv:hep-ph/9303287].

\item[2] K.~Petraki and A.~Kusenko,
Phys.\ Rev.\  D {\bf 77}, 065014 (2008)
  [arXiv:0711.4646 [hep-ph]].

\item[3] D.~Boyanovsky, Phys.\ Rev.\  D {\bf 78}, 103505 (2008)
  [arXiv:0807.0646 [astro-ph]].

\item[4] F.~Takahashi, Phys.\ Lett.\  B {\bf 660}, 100 (2008)
  [arXiv:0705.0579 [hep-ph]].

\item[5] K.~Sigurdson and M.~Kamionkowski, Phys.\ Rev.\ Lett.\  {\bf 92}, 171302 (2004)
  [arXiv:astro-ph/0311486].

\item[6] P.~Bode, J.~P.~Ostriker and N.~Turok, Astrophys.\ J.\  {\bf 556}, 93 (2001)
  [arXiv:astro-ph/0010389].

\item[7] E.~Polisensky and M.~Ricotti, Phys.\ Rev.\  D {\bf 83}, 043506 (2011)
  [arXiv:1004.1459 [astro-ph.CO]].

\end{description}

\newpage

\subsection{I. D. KARACHENTSEV}

\vskip -0.3cm

\begin{center}

Special Astrophysical Observatory of the Russian Academy of Sciences,
 Nizhnij Arkhyz, KChR, 369167, Russia 

\bigskip

{\bf Cosmography of the Local Universe}

\end{center}

\medskip

The Local volume sample of 800 galaxies with distances $D < 10$ Mpc [1] and
a wider sample of 11 000 galaxies situated within a sphere of radius
$\sim 50$~Mpc [2] are used to study the distribution of luminous and dark
matter around us, based on the local velocity field. The average density
of matter in this volume, $\Omega_{m,{\rm loc}}=0.08\pm0.02$, turns
out to be much lower than the global cosmic density $\Omega_{m,{\rm glob}}=
0.28\pm0.03$ [3]. We discuss three possible explanations of this paradox:

\medskip

 1) galaxy groups and clusters are surrounded by extended dark halos, most
mass of them located outside their virial radii;

\medskip

 2) the considered local volume of the Universe is not representative,
being situated inside a giant void;

\medskip

 3) the bulk of matter in the Universe is not related to clusters and
groups, but is rather distributed between them in the form of massive
dark clumps.

  Some arguments in favor of the latter assumption are presented [4].

\medskip
  
  From the dynamic point of view, it seems reasonable to divide the
elements of the large-scale structure into three categories:

\medskip
 
 a) virialized zones of groups and clusters, where the balance
between the kinetic and potential energy was established, and the
galaxies have ``forgotten'' initial conditions of their formation;

\medskip
 
 b) collapsing regions around the  virialized zones, limited by
zero velocity spheres with the $R_0$ radius;

\medskip
 
 c) the remaining, infinitely expanding regions of the ``general
field'' which comprise the population of voids and diffuse filaments.

\medskip
  
  Basic parameters of three dynamic components of the large-scale
structure are presented in the table [5]:

\begin{table}[hbt]
\begin{tabular}{l|c|c|c} \hline
\multicolumn{1}{c|} {Basic parameters} & Virialized  &Collapsing &Remaining \\
 & zones           &regions        &expanding \\
& & &background\\
 \hline
Relative number of galaxies & 54\%   & $\sim20$\% & $\sim26$\% \\
Relative amount of stellar mass  & 82\%   & $\sim8$\%  &   $\sim10$\% \\

Relative fraction of occupied volume   & 0.1\%  & 5\%     &         95\% \\
Contribution to $\Omega_m$  & 0.06   & 0.02       &         0.20 \\
Mean ratio of dark to stellar mass, $M_{\rm DM}/M_*$  & $\sim26$ &$\sim87$&    $\sim690$\\
\hline
\end{tabular}
\end{table}

 We notice that besides the two well-known inconsistencies of the
$\Lambda$CDM model with observational data: the problem of missing
satellites of normal galaxies and the problem of missing baryons,
there arises another one---the issue of missing dark matter.

\bigskip

{\bf References}

\bigskip

[1] I.~D.~Karachentsev, V.~E.~Karachentseva, W.~K.~Huchtmeier and
       D.~I.~Makarov, Astron. J. {\bf 127}, 2031  (2004).\\

\medskip
 
 [2] D.~I.~Makarov and I.~D.~ Karachentsev, MNRAS {\bf 412}, 2498 (2011).\\

\medskip

[3] D.~N.~Spergel  et al., Astrophys. J. Suppl. {\bf 170}, 377 (2007).\\

\medskip

[4] I.~D.~Karachentsev, O.~G.~Nasonova and  H.~M.~Courtois, Astrophys. J {\bf
743}, 123 (2011).\\

\medskip

[5] I.~D.~Karachentsev, Astrophys. Bulletin, 2012, 67, 123\\
\medskip

\newpage

\subsection{Wei LIA0}

\vskip -0.3cm

\begin{center}

Institute of Modern Physics, 
East China University of Science and Technology,\\
130 Meilong Road, Shanghai 200237, P. R. China

\bigskip

{\bf On the detection of keV scale sterile neutrino warm dark matter
in beta decay experiment}

\end{center}

\medskip

keV scale sterile neutrino, $\nu_s$, is a well motivated 
Warm Dark Matter(WDM) candidate.
Its large free-streaming length naturally suppresses 
the small scale structure in the universe and predicts a
small scale power spectrum which agrees well with astronomical observation.
Its keV scale small mass naturally predicts a lifetime 
longer than the age of the universe
despite the fact that it can decay via its mixing to active neutrinos.
So modeling of this Dark Matter(DM) candidate does not requires us to 
include a symmetry by hand to guarantee its long lifetime. This
is a nice feature of $\nu_s$ DM which is in sharp contrast to most 
models of cold DM. $\nu_s$ WDM has received increasing attention
in recent years after the observation that keV scale $\nu_s$
can be realized in seesaw type models and it can be
identified as one of the three right-handed neutrinos in the
seesaw model [1].

\medskip

One of the major concerns to keV scale $\nu_s$ DM is
how to detect it in laboratory. After all the energy scale of
this DM candidate is very low and its basic interaction
is the weak interaction induced by its mixing to active neutrinos
 $\theta_{ls}(l=e,\mu,\tau)$ if no other interaction is assumed.
Since the mixing has to be small to satisfy
constraint from satellite X-ray observation of $\nu_s \to \nu+\gamma$ decay
the interaction of $\nu_s$ is further suppressed by its small mixing
with active neutrinos.
For example, a direct estimate shows that for $m_{\nu_s}=2$ keV and
$|\theta_{es}|^2=10^{-6}$ the cross section of $\nu_s$ scattering with
electron is $\sigma v/c \sim 10^{-55}$ cm$^2$. Apparently the
extremely weak interaction of keV scale $\nu_s$ DM makes it
very hard to be detected in laboratory. 
This difficult topic has been pursued by a number of works.

\medskip

It turns out that there are two possible schemes to detect
keV scale $\nu_s$ in laboratory. The first one is to measure the
distortion of $\beta$ decay spectrum induced by admixture of
$\nu_s$ in $\beta$ decay. The second one is to use radioactive
nuclei to capture $\nu_s$ DM in the universe
and to detect final mono-energetic electrons beyond the $\beta$ decay
spectrum. These two ways to detect $\nu_s$
all use radioactive nuclei and can be performed in $\beta$ decay
experiment, e.g. in upgrade of direct neutrino mass measurement
experiment.

\medskip

Consider a $\beta$ decay process
\begin{eqnarray}
N_1 \to N_2 +e^- +{\bar \nu_e}, \label{betadecay1}
\end{eqnarray}
where $N_1$ and $N_2$ are two types of nuclei. 
At the end point of the $\beta$ spectrum the kinetic energy of electron
is
\begin{eqnarray}
E=Q_\beta. \label{decayenergy1}
\end{eqnarray}
$Q_\beta$ is the decay energy of this $\beta$ decay process.
In (\ref{decayenergy1}) we have taken $m_{\nu_e}=0$ since
masses of active neutrinos are smaller than eV scale and effectively 
we can take them zero in later discussion.

\medskip

The first scheme makes use of the observation that
due to ${\bar \nu}_s-{\bar \nu}_e$ mixing 
the following process is possible

\begin{eqnarray}
N_1 \to N_2 +e^- +{\bar \nu_s}. \label{betadecay2}
\end{eqnarray}

Since $\nu_s$ has keV scale mass the electron produced
in process (\ref{betadecay2}) can not have energy larger than
$Q_\beta - m_{\nu_s}$ and at the end point of the spectrum
of this decay process the electron has energy

\begin{eqnarray}
E=Q_\beta - m_{\nu_s}. \label{decayenergy2}
\end{eqnarray}

The event rate $N(p)$ of $\beta$ spectrum composes two
parts which are caused by decay processes (\ref{betadecay1}) and
(\ref{betadecay2}) respectively:

\begin{eqnarray}
N(p)=\cos^2\theta_{es} ~N_e(p,m_{\nu_e}=0)
+\sin^2\theta_{es} ~N_e(p,m_{\nu_s}). \label{eventrate1}
\end{eqnarray}

\begin{figure}
\begin{center}
\includegraphics[height=3.cm,width=7cm]{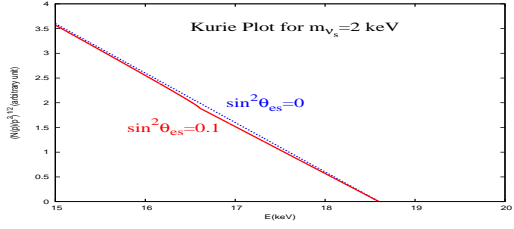}
\end{center}

\caption{Kurie plot for $m_{\nu_s}=2 $ keV in Tritium $\beta$ decay.}
\label{Kurieplot}
\end{figure}

In the energy range $Q_\beta -m_{\nu_s} < E < Q_\beta$,
$N_e(p,m_{\nu_s})=0$ and $N(p)=\cos^2 \theta_{es} ~N_e(p,0)$.
In this energy range the count rate is smaller than 
that expected from the usual $\beta$ decay.
In Kurie plot the signature is that the slope in the
energy range $[Q_\beta -m_{\nu_s}, Q_\beta]$ is smaller than
that in the energy range $[0,Q_\beta-m_{\nu_s}]$, as shown
Fig. \ref{Kurieplot}
This type of measurement has been done before ~[2,3]
and most recent measurement gives [2]
: $\sin^2\theta_{es} < 0.01$, for keV $m_{\nu_s}$.
This type of measurement of keV scale $\nu_s$ can be
done in high precision experiments such as MARE or 
KATRIN. 

\medskip

The above scheme is a direct way to detect $\nu_s$ in laboratory. 
However it is not yet a direct way to detect $\nu_s$ DM in the universe.
The second scheme I proposed is such a new way to do direct laboratory
search of $\nu_s$ DM in the universe [4].
It is based on the observation
that due to the $\nu_s-\nu_e$ mixing $\nu_s$ DM
can be captured by radioactive nuclei:

\begin{eqnarray}
\nu_s + N_1 \to N_2 +e^-. \label{nuscapture}
\end{eqnarray}

This reaction is of no threshold.
Electron produced in this process is mono-energetic:

\begin{eqnarray}
E=Q_\beta+ m_{\nu_s}. \label{captureenergy}
\end{eqnarray}

\begin{figure}
\begin{center}
\includegraphics[height=3.cm,width=7cm]{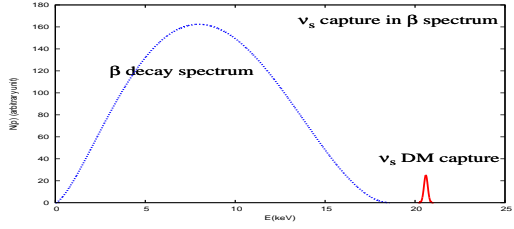}
\end{center}
\caption{Events of $\nu_s$ DM capture are well separated from 
$\beta$ decay events.}
\label{nuscaptureevent}
\end{figure}

Event of this process is keV scale beyond the
end point of the $\beta$ decay spectrum and is
well separated from $\beta$ decay events, 
as shown in Fig. \ref{nuscaptureevent}.

I found that the rate of $\nu_s$ DM capture by
Tritium and $^{106}$Ru are [4]
\begin{eqnarray}
N \approx 0.7 ~\textrm{year}^{-1} \times \frac{n_{\nu_s}}{10^{5} 
~\textrm{cm}^{-3}}
 \frac{|\theta_{es}|^2}{10^{-6}} \frac{^3 \textrm{H}}{10 ~\textrm{kg}}, ~~
N \approx 16 ~\textrm{year}^{-1} \times \frac{n_{\nu_s}}{10^{5} 
~\textrm{cm}^{-3}}
 \frac{|\theta_{es}|^2}{10^{-6}} \frac{^{106} \textrm{Ru}}{10
 ~\textrm{Ton}}. \label{capturerate}
\end{eqnarray}

In ~(\ref{capturerate}) the reference number
$n_{\nu_s} \sim 10^5$ cm$^{-3}$ is obtained using the energy density
of DM at the position of solar system. $n_{\nu_s}$ 
is of this order if keV scale $\nu_s$ DM accounts for all or a significant
part of the DM energy density in the universe.
A detailed analysis of this detection scheme can be found in [5].
One can see that the event rate of $\nu_s$
capture is quite large. Implementing this type of measurement
using $^3$H and $^{106}$Ru is probably challenging.
But it is possible to be done in future experiment.

\medskip

Other proposals to detect keV scale $\nu_s$ include: 1) using electron-capture
decaying matter $^{163}$Ho to capture $\nu_s$ DM [6]; 2) measuring
complete kinetic information of $\beta$ decay [7]; 3) measuring the recoil
of $\nu_s$ scattering with electrons [8]; 4) measuring spin flip of nuclei
induced by interaction with $\nu_s$ DM [8]. These proposals have
difficulties on achieving sufficient event rate or on achieving
enough statistics for high precision. They are hard to be done in near
future.

\medskip

In summary I have shown that there are two possible schemes to detect
keV scale $\nu_s$ in laboratory. One is the old idea that $\nu_s$
can be detected in the effect of its admixture to $\beta$ decay spectrum.
However this is still not a direct way to detect $\nu_s$ DM in the universe.
The second scheme I suggested is a new way to directly 
detect DM in the universe. It detects keV scale $\nu_s$ DM 
by capturing it in target of radioactive nuclei.
Up to now the second scheme is the best way to do direct laboratory search
of keV sterile neutrino DM in the universe. These two ways to
detect $\nu_s$ DM can be done in one $\beta$ decay experiment, or in different
phases of one experiment.

\bigskip

{\bf References}

\begin{description}

\item[1]
T. Asaka, S. Blanchet, M. Shaposhnikov, 
Phys.Lett. B{\bf 631}, 151(2005),
arXiv: hep-ph/0503065

\item[2]
M. Galeazzi, et. al., Phys. Rev. Lett. {\bf 86}, 1978(2001)

\item[3]
References in [2]

\item[4]
W. Liao, Phys. Rev. D{\bf 82}, 073001(2010), arXiv: 1005.3351

\item[5]
Y. F. Li and Z. Z. Xing, Phys. Lett. B{\bf 695}, 205(2011),
arXiv:1009.5870

\item[6]
Y. F. Li and Z. Z. Xing, JCAP {\bf 1108} (2011) 006,
arXiv:1104.4000

\item[7]
F. Bezrukov and M. Shaposhnikov, Phys. Rev. D{\bf 75}, 053005 (2007),
arXiv:hep-ph/0611352

\item[8]
S. Ando and A. Kusenko,  Phys. Rev. {\bf D81}, 113006(2010)
arXiv:1001.5273

\end{description}

\newpage

\subsection{Mark LOVELL}

\vskip -0.3cm

\begin{center}

Institute for Computational Cosmology, Durham University, UK 

\bigskip

{\bf Numerical Simulations of WDM haloes} 

\end{center}

\medskip

The development of the neutrino minimal standard model ($\nu$MSM) of
particle physics, coupled with issues facing supersymmetry, has
increased interest in the possible existence of sterile neutrinos. These as
yet undetected particles could have masses around the keV scale
and would act as warm dark matter (WDM)[1]. The free streaming of WDM
particles in the early Universe would erase small scale density perturbations,
and thus introduce a cutoff in the high redshift matter power spectrum:
cosmologically interesting consequences of this include the suppression of
small scale structures and the delay of structure formation relative
to cold dark matter (CDM). We therefore resimulated the
Aquarius Aq-A halo (Milky Way-mass halo, WMAP1 cosmological
parameters, simulation particle mass $m_{\mathrm{p}}=1.37\times
10^{4}\mathrm{M}_{\odot}$, [2]) with a WDM power spectrum to examine the effects
of the power spectrum cutoff.

\bigskip

The suppression of small scale power in WDM relative to CDM should
prevent small ($<10^{9}\mathrm{M}_{\odot}$, depending on the position of
the power spectrum cutoff) dark matter haloes from forming; however
$\sim 90$ per cent of all haloes in the simulation fall in this mass
range. These low mass haloes have been shown to result from the
spurious fragmentation of filaments, and not from any physical process
[3]. Therefore, to accurately determine the number of genuine haloes in our
simulation we require a process that removes spurious haloes from our
halo catalogues. Our prefered method of doing this is to take each
halo's particles, find their positions in the simulation initial
conditions, and determine the shapes of the resulting particle
distributions. Genuine haloes form from regions that are triaxial,
whereas spurious haloes originate in thin slices of filaments. These disc-like structures
are typically much flatter than the triaxial objects, so we can
apply a cut to our halo catalogue based on measurements of these initial conditions shapes.

\bigskip

We now apply the cleaned halo catalogue to a particular problem of
galaxy formation. It has recently been argued that the largest
dark matter haloes found to orbit Milky Way-like dark matter haloes in
CDM simulations contain more mass in their centres than is observed to be
the case in dwarf spheroidal galaxies [4, 5]. We perform the same
analysis as [4] using our WDM simulation, and find that using WDM
ameliorates the problem [6]. WDM structures collapse later than is the
case for CDM, so
WDM haloes form at later times when the Universe is less dense. This
translates into a lower central density for WDM haloes compared to
CDM. Other effects such as small number statistics, alternative
particle theories, and even lower estimates of the Milky Way halo mass
have been suggested to solve this problem; however the WDM hypothesis
remains competitive and compelling.

\bigskip

\begin{figure}[t]
\includegraphics[scale=0.7]{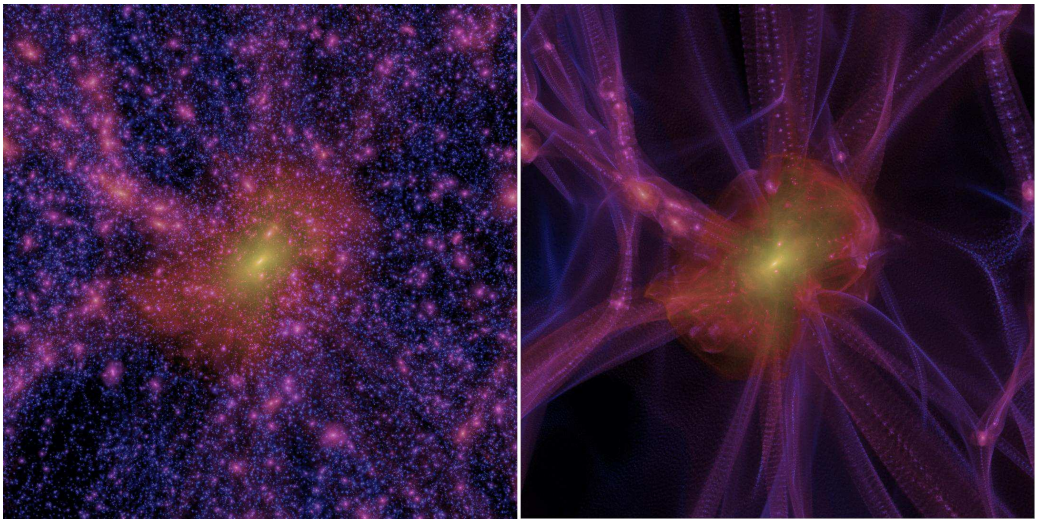}
  \caption{Images of a Milky Way-like dark matter halo at redshift 3
  simulated with CDM (\emph{left}) and WDM (\emph{right}). Each panel is
  2Mpc (comoving) on each side.}
  \label{MLovellFig1}
\end{figure}

Whilst this model was able to solve the problem of satellite central
densities, it did not predict enough dark matter haloes to host the
expected number of satellite galaxies. Our next step was therefore to
resimulate the same halo with
a variety of WDM models, in order to establish what power spectrum
cutoff was required to produce the right number of satellite galaxies. A pair of
images from these simulations are shown in
Figure~\ref{MLovellFig1}. This work is ongoing, and preliminary results
suggest that the effective WDM matter particle mass (the mass of a
thermal relic for which the high redshift matter power spectrum approximately
matches the sterile neutrno spectrum) needs to be greater than 2keV in
order to produce enough dark matter haloes to host the Milky Way dwarf
spheroidals, in agreement with previous work.  

\bigskip

{\bf References}

\begin{description}

\item[1] Boyarsky et al.,  2009, Phys. Rev. Lett., 102,{ \bf 201304}

\item[2] Springel et al.,  2008, MNRAS, 391,{ \bf 1685}

\item[3] Wang \& White, 2007, MNRAS, 380,{ \bf 93}

\item[4] Boylan-Kolchin et al., 2011, MNRAS, 415,{ \bf L40}

\item[5] Boylan-Kolchin et al., 2012, MNRAS, 422,{ \bf 1203}

\item[6] Lovell et al., 2012, MNRAS, 420,{ \bf 2318}

\end{description}

\newpage

\subsection{Emmanouil PAPASTERGIS}

\vskip -0.3cm

\begin{center}
 
Center for Radiophysics \& Space Research, Cornell University, Ithaca NY, USA

\bigskip

{\bf The galaxy-halo connection: insights from ALFALFA} 

\end{center}

\medskip

The Arecibo Legacy Fast ALFA (ALFALFA) survey is a blind, extragalactic HI-line survey performed at the Arecibo Observatory. So far, data over $\approx$ 3000 deg$^2$ of sky (40\% of the final survey area) have been fully reduced, producing the largest HI-selected galaxy sample to date (`$\alpha$.40' catalog,[1]). For each ALFALFA detection, we can extract the width of its 21cm-line emission from its global spectral profile. Since HI is usually the most extended baryonic component in late-type galaxies, it traces the rotation curve at large galactocentric radii. As a result, the galactic velocity width, $w$, is (up to a projection on the line-of-sight) a good indicator of the total dynamical mass within the extent of the HI disk.

\medskip

We use 10744 sources in the $\alpha$.40 catalog with measured velocity widths to estimate the velocity width function (WF) of galaxies, i.e. the space density of HI-bearing galaxies as a function of $w$. We compare the ALFALFA measurement with the predicted distribution in a CDM universe. In particular, we consider semianalyic galaxy catalogs [2,3,6] populating the halos of three CDM N-body simulations [4-6]. Only one of the theoretical works includes explicit modelling of the HI-component [2], but is restricted to relatively massive galaxies ($w_{lim} \approx 200$ kms). The other two [3,6] have the advantage of modelling lower mass galaxies ($w_{lim} \approx 100$ kms \ and $w_{lim} \approx 50$ kms \ for [3] and [6], respectively).

\begin{figure}[htbp]

\includegraphics[scale=0.5]{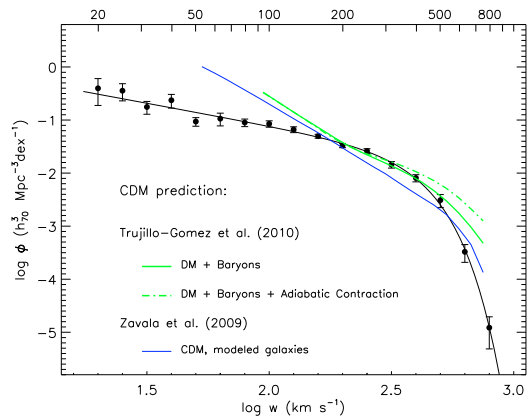}
\caption{\textit{The CDM overabundance problem}: datapoints with errorbars and black solid line represent the measured ALFALFA WF. The green lines represent the WF of a sample of synthetic galaxies modeled by [3], which populate the halos in the Bolshoi CDM simulation [5]. The blue solid line represents the WF of a modeled galaxy population corresponding to the higher resolution CDM simulation of [6]. }
\label{fig:aa_vs_cdm}
\end{figure}

\begin{figure}[htbp]

\includegraphics[scale=0.4]{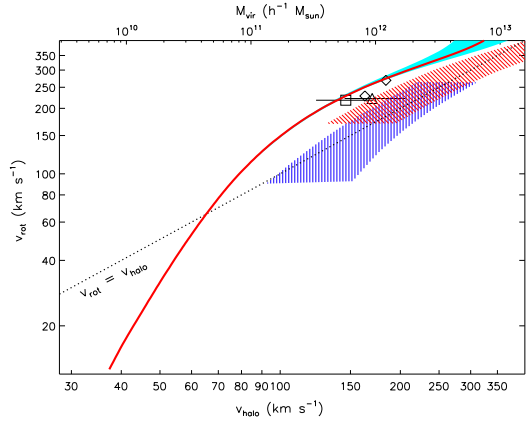}
\caption{$v_{rot}$ - $v_{halo}$ \textit{relation in a CDM universe}: the red solid line corresponds to the relation between the rotational velocity of galaxies measured observationally ($v_{rot}$) and the maximum rotational velocity of the corresponding CDM halo ($v_{halo}$). The relation was obtained by the abundance matching of the velocity distribution of halos in the Bolshoi CDM simulation [5] with the velocity distribution of galaxies inferred from ALFALFA. }
\label{fig:vrot_vh}
\end{figure}

\begin{figure}[htbp]

\includegraphics[scale=0.5]{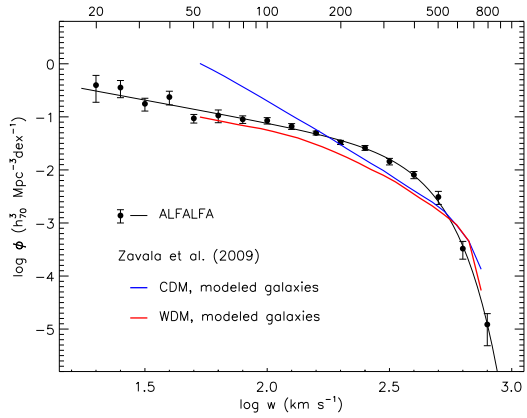}
\caption{Data points with error bars and black solid line represent the measured ALFALFA WF. The blue solid line represents the WF of a modeled galaxy population based on the high resolution CDM simulation of [6], while the red solid line represents the WF corresponding to a second run of the  simulation assuming a 1 keV WDM particle (both simulations employ the same scheme to populate halos with synthetic galaxies). }
\label{fig:aa_vs_wdm}
\end{figure}

Figure \ref{fig:aa_vs_cdm} puts in evidence the significant disagreement at low velocity widths between the CDM expectations and the ALFALFA measurement of the WF. In particular, the difference in abundance is a factor of $\sim 8$ at $w = 50$ kms, which extrapolated to the ALFALFA limit of $w \approx 20$ kms \ would be a factor of $\sim 100$. This problem is just one additional aspect of the general \textit{CDM overabundance problem}, which stems from the fact that CDM structure formation predicts large numbers of low-mass halos, seemingly in contradiction with the relative paucity of visible low-mass galaxies.

\bigskip

One possible solution to the disagreement between the theoretical and observational WFs could arise from the fact that often times the rotation curves of dwarf galaxies are not extended enough to probe the full amplitude of the rotation curve, but are rising to the last measured point. In such cases, the measured HI rotational velocity corrected for inclination ($v_{rot}$) may underestimate the true maximum circular velocity of the host dark matter halo ($v_{halo}$). 

\medskip

Qualitatively, this solution would explain the paucity of low-width ALFALFA galaxies by asserting that such galaxies are hosted by much larger (and hence much rarer) halos than their HI widths imply. Quantitatively, we can infer the average $v_{rot} - v_{halo}$ relation that is necessary to reproduce the ALFALFA WF starting from a CDM halo velocity function. This relation is shown by the red thick line in Fig.~\ref{fig:vrot_vh}. It shows that the measured HI rotational velocity of low-mass galaxies must greatly underestimate the maximum circular velocity of the host halo, e.g. by a factor of $\sim 2$ in a galaxy with $v_{rot} = 20$ kms. Another result of the $v_{rot} - v_{halo}$ relation is that \textit{all} ALFALFA galaxies are expected to be hosted by halos with $v_{halo} > 30$ kms. However, as pointed out by [7], many dwarf galaxies with spatially resolved HI rotation curves seem to be incompatible with this requirement.                

\bigskip

A second solution could involve the modification of the properties properties of dark matter. For example, a dark matter particle with mass in the keV range (warm dark matter; WDM) would suppress  the primordial small scale power, and would result in a lower number of low-mass halos in the present-day universe. As shown in Fig.~\ref{fig:aa_vs_wdm}, the WF of galaxies in a WDM universe is much closer to the ALFALFA result, and displays a much `shallower' low-width slope compared to the CDM case.

\bigskip

{\bf References}

\begin{description}

\item[1] M.P. Haynes, R. Giovanelli, A.M. Martin, K. Hess, and 20 coauthors, AJ, 142:5 (2011)

\item[2] D. Obreschkow, D. Croton, G. De Lucia, S. Khochfar, S. Rawlings, ApJ, 698, 2, 1467-1484 (2009)

\item[3] S.Trujillo-Gomez, A.A. Klypin, J. Primack, A.J. Romanowsky, ApJ, 742, 1, 16 (2011)

\item[4] V. Springel, S.D. White, A. Jenkins, C.S. Frenk, N. Yoshida, and 13 coauthors, Nature, 435, 7042, 629-636 (2005)

\item[5] A.A. Klypin, S. Trujillo-Gomez, J. Primack, ApJ, 740, 2, 102 (2011)

\item[6] J. Zavala, Y.P. Jing, A. Faltenbacher, G. Yepes, Y. Hoffman, S. Gottl\"ober, B. Catinella, ApJ, 700, 2, 1779-1793 (2009)

\item[7] I. Ferrero, M.G. Abadi,  J.F. Navarro, L.V. Sales, S. Gurovich, arXiv: 1111.6609

\end{description}

\newpage

\subsection{Norma G. SANCHEZ and Hector J. DE VEGA}

\vskip -0.3cm

\begin{center}

HJdV: LPTHE, CNRS/Universit\'e Paris VI-P. \& M. Curie \& Observatoire de Paris.\\
NGS: Observatoire de Paris, LERMA \& CNRS

\bigskip

{\bf  Quantum WDM fermions and gravitation determine the observed galaxy structures II}

\end{center}

Quantum  fermionic WDM gives the correct galaxy properties and galaxy profiles.

We treat in ref. [1] the self-gravitating fermionic DM in the Thomas-Fermi approximation.
{\bf Approach}:In this approach, the central quantity to derive is the DM chemical potential $ \mu(r) $,
the chemical potential being the free energy per particle.
We consider a single DM halo in the late stages of structure formation when DM
particles composing it are non--relativistic and their phase--space distribution
function $ f(t, \br,\bp) $ is relaxing to a time--independent form, at least for
$\br$ not too far from the halo center. In the Thomas--Fermi approach such a
time--independent form is taken to be a energy distribution function $f(E)$ of
the conserved single--particle energy $E = p^2/(2m) - \mu $, where $m$ is the
mass of the DM particle and $\mu$ is the chemical potential
$   \mu(\br) =  \mu_0 - m \, \phi(\br) $
with $\phi(\br)$ the gravitational potential and  $ \mu_0 $ some constant.
We consider the spherical symmetric case. 

Here, the Poisson equation for $ \phi(r) $ is a nonlinear and selfconsistent equation
\be \label{pois}
  \frac{d^2 \mu}{dr^2} + \frac2{r} \; \frac{d \mu}{dr} = - 4\pi \, G \, m \, \rho(r)\; , 
\ee
where the mass density $ \rho(r) $ is a function of $ \mu(r) $ and
$ G $ is Newton's constant. $ \rho(r) $ is expressed here as a function of $ \mu(r) $ through the
standard integral of the DM phase--space distribution function over the momentum
for Dirac fermions as
\be \label{den}
  \rho(r) = \frac{m}{\pi^2 \, \hbar^3} \int_0^{\infty} dp\;p^2 
  \; f\left(\displaystyle \frac{p^2}{2m}-\mu(r)\right)\; , 
\ee
Another integral of the DM phase--space distribution function is the pressure
$$
  P(r) = \frac{1}{3\pi^2 \,m\,\hbar^3} \int_0^{\infty} dp\;p^4 
  \,f\left(\displaystyle \frac{p^2}{2m}-\mu(r)\right) \; .
$$
From $ \rho(r) $ and $ P(r) $ other quantities of interest, such as the velocity dispersion
$ \sigma(r) $ and the phase--space density $ Q(r) $ can be determined as
$$
  \sigma^2(r) = \frac{P(r)}{\rho(r)} \quad,\qquad Q(r) = \frac{\rho(r)}{\sigma^3(r)}  \; .
$$
Eqs.(\ref{pois}) and (\ref{den}) provide an ordinary nonlinear
differential equation that determines selfconsistently the chemical potential $ \mu(r) $ and
constitutes the Thomas--Fermi semi-classical approach. We obtain a family of solutions 
parametrized by the value of  $ \mu_0 \equiv \mu(0) $ [1].

\medskip

We integrate the Thomas-Fermi nonlinear differential
equations (\ref{pois})-(\ref{den}) from $ r = 0 $ till the 
boundary $ r = R = R _{200} \sim R_{vir} $ defined as the radius where the 
mass density equals $ 200 $ times the mean DM density [1].

\medskip

We define the core size $ r_h $ of the halo by analogy with the Burkert density profile as
$   \frac{\rho(r_h)}{\rho_0} = \frac14 \quad , \quad  r_h = l_0 \; \xi_h \; .$
where $ \rho_0 \equiv \rho(0) $ and $ l_0 $ is the characteristic length that emerges from 
the dynamical equations (\ref{pois})-(\ref{den}):
\be\label{varsd2}
l_0 \equiv  \frac{\hbar}{\sqrt{8\,G}} \left(\frac{9\pi}{m^8\,\rho_0}\right)^{\! \! \frac16} 
  = R_0 \; \left(\frac{{\rm keV}}{m}\right)^{\! \! \frac43}  \; 
  \left(\rho_0 \; \frac{{\rm pc}^3}{M_\odot}\right)^{\! \! -\frac16} 
  \;,\qquad R_0 = 18.71 \; \rm pc  \; .
\ee
As an example of distribution function $ \Psi(E/E_0) $, we consider the Fermi--Dirac distribution 
$  \Psi_{\rm FD}(E/E_0) = \frac1{e^{E/E_0} + 1} \; $.

\medskip

We define the dimensionless chemical potential $ \nu(r) $ as  
$ \nu(r) \equiv \mu(r)/E_0 \quad {\rm and} \quad   \nu_0 \equiv \mu(0)/E_0 \quad . $
Large positive values of the chemical potential at the origin $ \nu_0 \gg 1 $ 
correspond to the degenerate 
fermions limit which is the extreme quantum case, and oppositely, $ \nu_0 \ll -1 $ gives 
the diluted regime which is the classical limit. In this classical regime the Thomas-Fermi equations
(\ref{pois})-(\ref{den}) become exactly the equations for a self-gravitating Boltzmann gas.

\medskip

\begin{figure}[h]
\begin{center}
\includegraphics[width=14.cm]{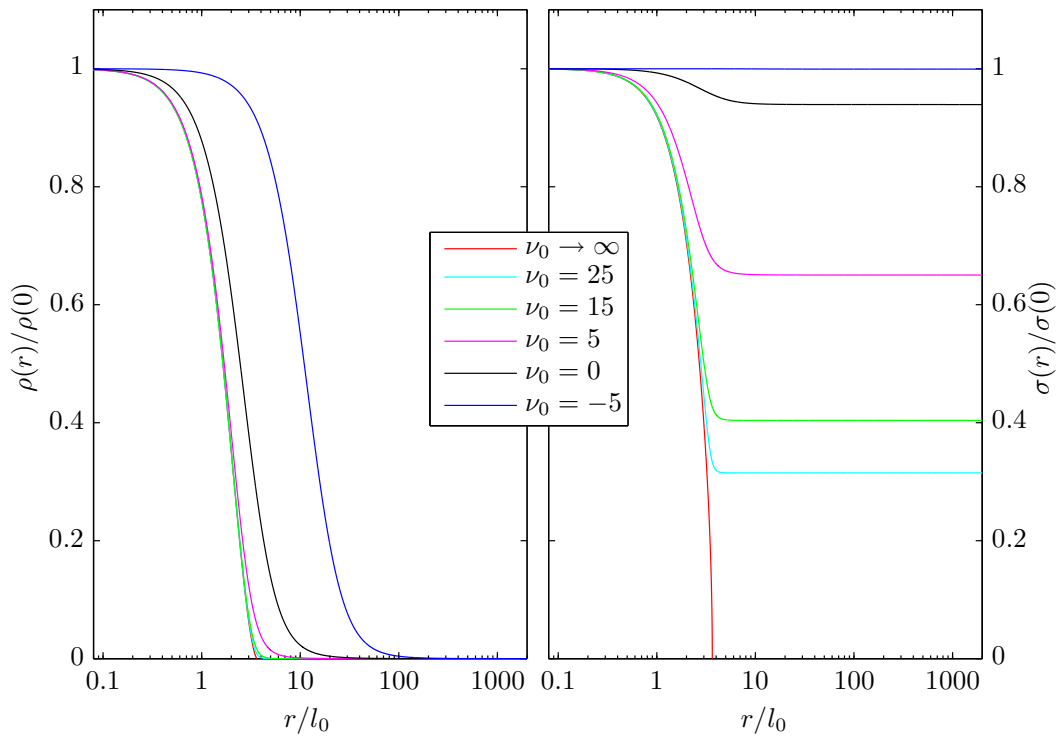}
\caption{Density and velocity profiles, $ \rho(r)/\rho_0 $ and $ \sigma(r)/\sigma(0) $, 
as functions of $ r/l_0 $ for different values of the chemical potential
at the origin $ \nu_0 $ [1]. Large positive values of $ \nu_0 $ correspond
to compact galaxies, negative values of $ \nu_0 $ correspond to the classical regime
describing spiral and elliptical galaxies.
All density profiles are cored. The sizes of the cores $ r_h $ 
are in agreement with the observations, from the compact galaxies where $ r_h \sim 35 $ pc till
the spiral and elliptical galaxies where $ r_h \sim .2 - 60 $ kpc. The larger and positive is 
$ \nu_0 $, the smaller is the core. The minimal one arises in
the degenerate case  $ \nu_0 \to +\infty $ (compact dwarf galaxies).}
\label{deg}
\end{center}
\end{figure}

{\bf Results} We display in fig. \ref{deg} the density and velocity profiles [1]. Namely,
we plot $ \rho(r)/\rho_0 $ and $ \sigma(r)/\sigma(0) $ as functions of 
$ r/R $ for $ \nu_0 \equiv \nu(0) = -5, \; 0, \; 5, \; 15, \;  25 $ and 
for the degenerate fermion limit $ \nu_0 \to +\infty $.
The obtained fermion profiles are always cored. 
The sizes of the cores $ r_h $ 
are in agreement with the observations, from the compact galaxies where $ r_h \sim 35 $ pc till
the spiral and elliptical galaxies where $ r_h \sim 0.2 - 60 $ kpc. The larger and positive is 
$ \nu_0 $, the smaller is the core. The minimal core size arises in
the degenerate case  $ \nu_0 \to +\infty $ (compact dwarf galaxies).

\begin{figure}[h]
\begin{center}
\includegraphics[width=14.cm]{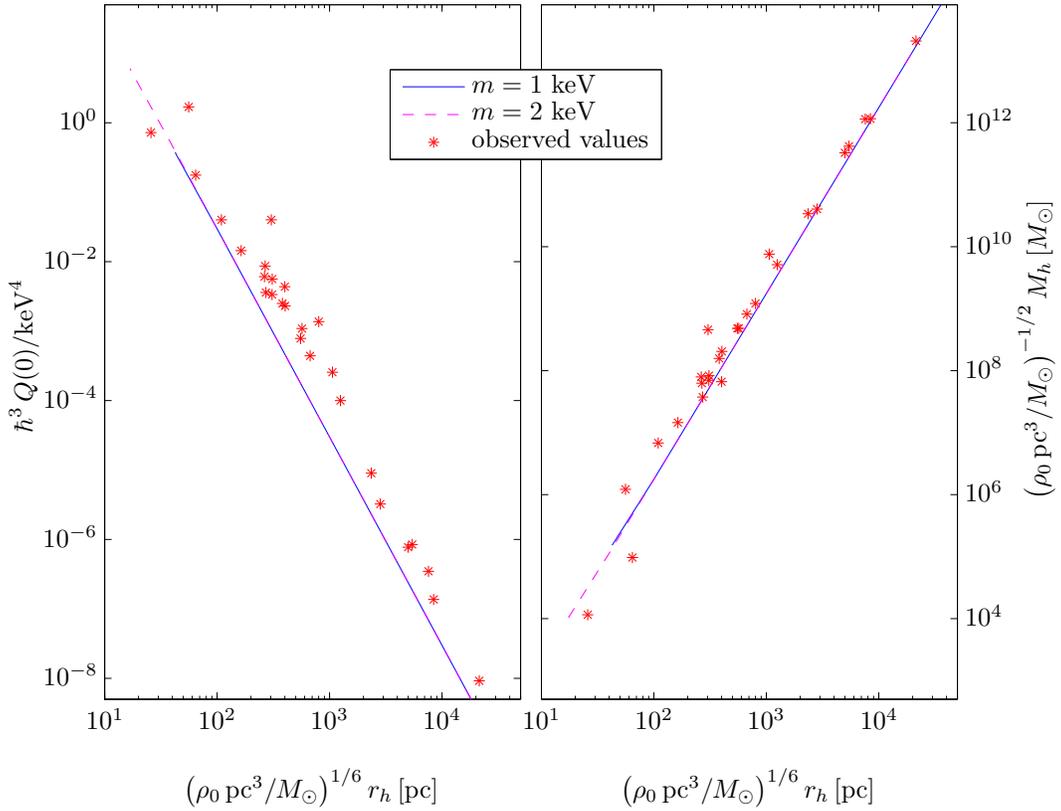}
\caption{In the left panel we display the galaxy phase-space density 
$ \hbar^3 \; Q(0)/({\rm keV})^4 $
obtained from the numerical resolution of the Thomas-Fermi eqs. (\ref{pois})-(\ref{den})
for WDM fermions of mass $ m = 1 $ and $ 2 $ keV 
versus the ordinary logarithm of the product $ \log_{10}\{r_h \; 
[{\rm pc}^3 \; \rho_0/ M_\odot]^{\frac16} \} $ in parsecs [1].
The red stars $ * $ are the observed values of $ \hbar^3 \; Q(0)/({\rm keV})^4 $ from Table 1, p.14.
Notice that the observed values $ Q_h $ from the stars' velocity 
dispersion are in fact upper bounds for the DM $ Q_h $ and therefore the theoretical curve is slightly below them.
In the right panel we display the galaxy mass 
$ (M / M_\odot) \sqrt{M_\odot / [\rho_0 \; {\rm pc}^3]} $ obtained from the numerical resolution of the 
Thomas-Fermi eqs.(\ref{pois})-(\ref{den}) for WDM fermions of mass $ m = 1 $ and $ 2 $ keV 
versus the product $ r_h \; [{\rm pc}^3 \; \rho_0/ M_\odot]^{\frac16} $
in parsecs [1].  The red stars $ * $ are the observed values of 
$ (M / M_\odot) \sqrt{M_\odot / [\rho_0 \; {\rm pc}^3]} $ 
from Table  1, p.14. Notice that the error bars 
of the observational data are not reported here but they are at least about $ 10-20 \%$.}
\label{halo}
\end{center}
\end{figure}

In the left panel of fig. \ref{halo} we plot the dimensionless quantity  [1]
\be\label{cute1}
\frac{\hbar^3}{({\rm keV})^4} \; Q(0) \; . 
\ee
In the right panel of fig. \ref{halo}, we plot instead the dimensionless product
\be\label{cute2}
\frac{M_h}{M_\odot} \sqrt{\frac1{\rho_0} \; \frac{M_\odot}{{\rm pc}^3}} \; ,
\ee
where $ M_h $ is the halo mass, namely the galaxy mass inside the core radius $ r_h
$. In both cases we consider the two values $ m = 1 $ and $ 2 $ keV and we put in
the abscissa the product $
r_h  \; \left(\frac{\rm pc^3}{M_\odot} \; \rho_0\right)^{\! \! \frac16} $
in ~ parsecs, where $ r_h $ is the core radius.

\medskip

 The phase-space density $ Q(0) $ and the galaxy 
mass $ M_h $ are obtained by solving the Thomas-Fermi eqs.(\ref{pois})-(\ref{den}).
We have also superimposed the
observed values $ \hbar^3\, Q_h/({\rm keV})^4 $ and $ M_h \sqrt{M_\odot / [\rho_0
  \; {\rm pc}^3]} \; \; \left(m/{\rm keV}\right)^4 $ from Table I in page 14.
Notice that the observed values $ Q_h $ from the stars' velocity dispersion are
in fact upper bounds for the DM $ Q_h $. This may explain why the theoretical
Thomas-Fermi curves in the left panel of fig. \ref{halo} appear slightly below
the observational data. Notice also that the error bars of the observational
data are not reported here but they are at least about $ 10-20 \%$.

\medskip

The phase space density decreases from its maximum value for the
compact dwarf galaxies corresponding to the limit of degenerate fermions till
its smallest value for large galaxies, spirals and ellipticals, corresponding to
the classical dilute regime. On the contrary, the halo radius $ r_h $ and the halo mass $ M_h $
monotonically increase from the quantum (small and compact galaxies) to the classical regime
(large and dilute galaxies).

Thus, the whole range of values of the chemical potential at the origin $ \nu_0
$ from the extreme quantum (degenerate) limit $ \nu_0 \gg 1 $ to the classical
(Boltzmann) dilute regime $ \nu_0 \ll -1 $ yield all masses, sizes, phase space
densities and velocities of galaxies from the ultra compact dwarfs till the
larger spirals and elliptical in agreement with the observations (see Table 1, p.14).

\medskip

{\bf Results on the WDM particle mass}: From figs. \ref{halo} we can extract important information on the fermion particle WDM mass.

We see from the left panel fig. \ref{halo} that decreasing the DM particle mass $ m $ moves the theoretical curves
$ \hbar^3 \; Q(0) / ({\rm keV})^4 $ towards smaller $ Q(0) $ values and larger galaxy sizes, one over each other.
In the right panel of fig. \ref{halo} we see that decreasing the DM particle mass $ m $ 
displaces the theoretical curves $ (M_h / M_\odot) \sqrt{M_\odot / [\rho_0 \; {\rm pc}^3]} $ 
towards larger galaxy masses and sizes, one over each other.

\medskip

The small galaxy endpoint of the curves in figs. \ref{halo}  corresponds to
the degenerate fermion limit $ \nu_0 \to +\infty $ and its value depends on the
WDM particle value $ m $. For increasing $ m $, the small galaxy endpoint moves towards
smaller sizes while for decreasing $ m $, it moves towards larger sizes.

We see from figs. \ref{halo} that decreasing the particle mass beyond a given value,
namely for particle masses $ m \lesssim 1 $ 
keV,  the theoretical curves do not reach the more compact galaxy data.
Therefore, $ m \lesssim 1 $ keV is ruled out as WDM particle mass.

\medskip

For growing $ m \gtrsim $ keV the left part of the theoretical curves corresponding to the lower galaxy 
masses and sizes, will not have observed galaxy counterpart. Namely, increasing  
$ m \gg $ keV would show an overabundance of small galaxies (small scale structures) 
which do not have counterpart in the data.
This is a further indication that the WDM particle mass is approximately around 2 keV
in agreement with earlier estimations [2,3].

\medskip

The overabundance of small scale structures that appears here for $ m \gg $ keV
is a consequence {\bf only} of the DM particle mass value.
This result is independent of the cosmic evolution, namely the primordial 
power spectrum and the structure formation dynamics. 

In addition, the galaxy velocity dispersions turn to be fully consistent 
with the galaxy observations in Table 1, p.14.

\medskip

{\bf Baryons}: Adding baryons to CDM simulations have been often invoked to solve the
serious CDM problems at small scales. It must be noticed however that the
excess of substructures in CDM happens in DM dominated
halos where baryons are especially subdominat and hence the effects of
baryons cannot drastically modify the overabundance of substructures of the
pure CDM results.

The influence of baryon feedback into CDM cusps of density profiles
depends on the strength of the feedback. For normal values of the
feedback, baryons produce adiabatic contraction and the cusps in the density
profiles become even more cuspy.

Using the baryon feedback as a free parameter, it is possible
to exagerate the feedback such to destroy the CDM cusps
but then, the star formation ratio disagrees with the
available and precise astronomical observations.
Moreover, "semi-analytic (CDM + baryon)  models" have been introduced
which are just empirical fits and prescriptions to some galaxy observations.

In addition, there are serious evolution problems in CDM galaxies:
for instance pure-disk galaxies (bulgeless) are observed whose formation
through CDM is unexplained.

In summary, adding baryons to CDM simulations bring even more serious
discrepancies with the set of astronomical observations.

\medskip

{\bf Symmetry}: In ref. [1], spherical symmetry is considered for simplicity to determine 
the essential physical galaxy properties as the classical or 
quantum nature of galaxies, compact or dilute galaxies, 
the phase space density values, the
cored nature of the mass density profiles, the galaxy masses and 
sizes.  It is clear that DM halos are not perfectly 
spherical but describing them as spherically symmetric is a first
approximation to which other effects can be added.
In ref. [1] we estimated the angular momentum 
effect and this yields small corrections. The quantum or classical 
galaxy nature, the cusped or cored nature of the density profiles in the 
central halo regions can be well captured in the spherically symmetric treatment.

\medskip

Our spherically symmetric treatment captures the essential features
of the gravitational dynamics and agree with the observations.
Notice that we are treating the DM particles quantum mechanically through
the Thomas-Fermi approach, so that expectation values are independent
of the angles (spherical symmetry) but the particles move and fluctuate
in all directions. Namely, this is more than treating purely classical orbits
for particles in which only radial motion is present. 
The Thomas-Fermi approach can be generalized to
describe non-spherically symmetric and non-isotropic situations,
by considering  distribution functions which include other 
particle parameters like the angular momentum.

\medskip

{\bf Conclusion}: To conclude, the galaxy magnitudes: halo radius, galaxy masses and velocity dispersion
obtained from the Thomas-Fermi quantum treatment for WDM fermion masses in the keV scale are
fully consistent with all the observations for all types of galaxies (see Table I, pag 14). 
Namely, fermionic WDM treated quantum mechanically, as it must be, is able to reproduce
the observed DM cores and their sizes in galaxies [1].

It is highly remarkably that in the context of fermionic WDM, the simple stationary
quantum description provided by the Thomas-Fermi approach is able to reproduce such broad variety of galaxies.

Baryons have not yet included in the present study. This is fully justified for dwarf compact 
galaxies which are composed today 99.99\% of DM. In large galaxies the baryon fraction can
reach values up to  1 - 3 \%. Fermionic WDM by itself produces galaxies and structures in 
agreement with observations for all types of galaxies, masses and sizes. Therefore, the effect of including 
baryons is expected to be a correction to these pure WDM results, consistent with the fact that dark matter 
is in average six times more abundant than baryons.

\bigskip

{\bf References}

\begin{description}

\item[1] C. Destri, H. J. de Vega, N. G. Sanchez, arXiv:1204.3090,
New Astronomy {\bf 22}, 39 (2013) and arXiv:1301.1864, Astroparticle Physics, {\bf 46}, 14 (2013).

\item[2] H. J. de Vega,  P. Salucci, N. G. Sanchez, 
New Astronomy {\bf 17}, 653 (2012)

\item[3] H. J. de Vega, N. G. S\'anchez, MNRAS {\bf 404}, 885 (2010) and
Int. J. Mod. Phys. {\bf A 26}, 1057 (2011).

\end{description}

\newpage

\subsection{Patrick VALAGEAS}

\vskip -0.3cm

\begin{center}

Institut de Physique Th\'eorique,\\
CEA, IPhT, F-91191 Gif-sur-Yvette, C\'edex, France\\

\bigskip

{\bf Impact of a Warm Dark Matter late-time velocity dispersion on large-scale structures} 

\end{center}

\medskip

{\bf Abstract:} {\it We investigate whether the late-time (at $z\leq 100$) velocity dispersion expected in
Warm Dark Matter scenarios could have some effect on the cosmic web (i.e., outside
of virialized halos).
We consider effective hydrodynamical equations, with a pressure-like term that agrees
at the linear level with the analysis of the Vlasov equation. Then, using analytical 
methods, based on perturbative expansions and the spherical dynamics,
we investigate the impact of this term for a $1$keV dark matter particle.
We find that the late-time velocity dispersion has a negligible effect
on the power spectrum on perturbative scales and on the halo mass function.
However, it has a significant impact on the probability distribution function of the density
contrast at $z \sim 3$ on scales smaller than $0.1 h^{-1}$Mpc, which correspond to
Lyman-$\alpha$ clouds. Finally, we note that numerical simulations should start
at $z_i\geq 100$ rather than $z_i \leq 50$ to avoid underestimating gravitational clustering
at low redshifts.}

\medskip

{\bf Introduction:}

\medskip

At early times and on large scales, the formation of large-scale structures within
WDM scenarios is studied through the linearized Vlasov equation [1-4].
At low redshift and on small scales, the nonlinear regime of gravitational clustering is
investigated through numerical simulations [5-6]
and halo models based on such simulations.
In practice, one often uses the same N-body codes as for CDM
scenarios and the only
difference comes from the density power spectrum that is set at the initial redshift $z_i$
of the simulations. This means that one takes into account the high-$k$ cutoff due to
free-streaming during the relativistic era but neglects the small velocity dispersion at
low redshifts, $z\leq z_i$.
In some cases, one adds to the ``macro-particles'' used in the N-body simulations
an additional initial random velocity drawn from the thermal distribution
of the WDM particles [7-9], even though 
these ``macro-particles'' have a much larger mass than the
WDM particles and a clump of so many elementary dark matter particles would
have a much smaller (almost zero) mean velocity.
In fact, setting such initial conditions at $z_i=100$ leads to spurious power at high $k$ in
the power spectrum measured at later times (starting at $z_i=40$ appears to avoid this
problem because $v_{\rm rms}$ is smaller) [7].
This can be understood from the fact that adding these random velocities is equivalent
to model a CDM scenario, but with the damped WDM power spectrum, to which is
added a $k^2$ high-$k$ tail associated with the random velocity component.
Indeed, this adds a white-noise component to the initial velocity power spectrum,
which corresponds to a $k^2$ tail for the density power spectrum (when we decompose
over growing and decaying modes).

This shows that it is not easy to include the WDM velocity dispersion in
numerical simulations in a realistic fashion.
Therefore, it is interesting to have a quantitative estimate of the impact of such a 
late-time velocity dispersion.

\medskip

{\bf Set-up:}

\medskip

First, we note that in the matter dominated era one can derive from the linearized Vlasov 
equation an evolution equation for the matter density contrast of the form [1,3]
\begin{equation}
\frac{\partial^2 \tilde{\delta}}{\partial\tau^2} + {\cal H} \frac{\partial\tilde{\delta}}{\partial\tau}
- \left( \frac{3}{2} \Omega_{\rm m} {\cal H}^2 - k^2 c_s^2 \right) \tilde{\delta} = S[\tilde{\delta},\tau] ,
\end{equation}
which becomes identical to the CDM case if we set $c_s=0$ and $S=0$.
The new term in the left hand side is similar to a pressure term
(at the linear level) and $c_s$ would be the comoving sound speed. The comoving Jeans
wavenumber $k_J$ would thus correspond to the comoving free-streaming wavenumber
$k_{\rm fs}$, with $k_{\rm fs}^2 = 3\Omega_{\rm m}{\cal H}^2/(2 c_s^2)$.
However, this is only a formal analogy because there is no true pressure
as we consider a collisionless fluid. This term arises from the nonzero velocity dispersion
and its evolution with time is set by the comoving free-streaming wavenumber 
rather than by a thermodynamical equation of state. 
At this linear level, $k_{\rm fs}(\tau)$ is obtained from the analysis of the linearized Vlasov equation [1,3].
On the other hand, the new source term in the right hand side of Eq.(1) is subdominant
on large scales and at late times. 
Therefore, we consider the approximate equations of motion [10]
\begin{equation}
\frac{\partial \delta}{\partial\tau} + \nabla\cdot [(1+\delta) {\bf v}] = 0 ,
\hspace{2cm} 
\frac{\partial{\bf v}}{\partial\tau} + {\cal H} {\bf v} 
+ ({\bf v}\cdot\nabla){\bf v} = - \nabla \Phi 
- \frac{c_s^2 \nabla\rho}{\bar{\rho}}  ,
\end{equation}
where $\Phi$ is the gravitational potential.
The second Eq.(2) is a modified Euler equation, which only differs from the CDM case by the
last term. 
This approximation is the simplest closure of the hierarchy of fluid equations that agrees with the linear theory (when we neglect the source term in Eq.(1)).

\medskip

{\bf Results:}

\medskip

Using analytical methods, based on perturbative expansions and the spherical dynamics,
we compare the predictions for various statistics of large-scale structures obtained
when we either include or neglect the ``pressure-like'' term of the second Eq.(2).
This allows us to estimate the impact of the late-time WDM velocity dispersion.

\begin{figure*}
\begin{center}
\epsfxsize=5.5 cm \epsfysize=5. cm {\epsfbox{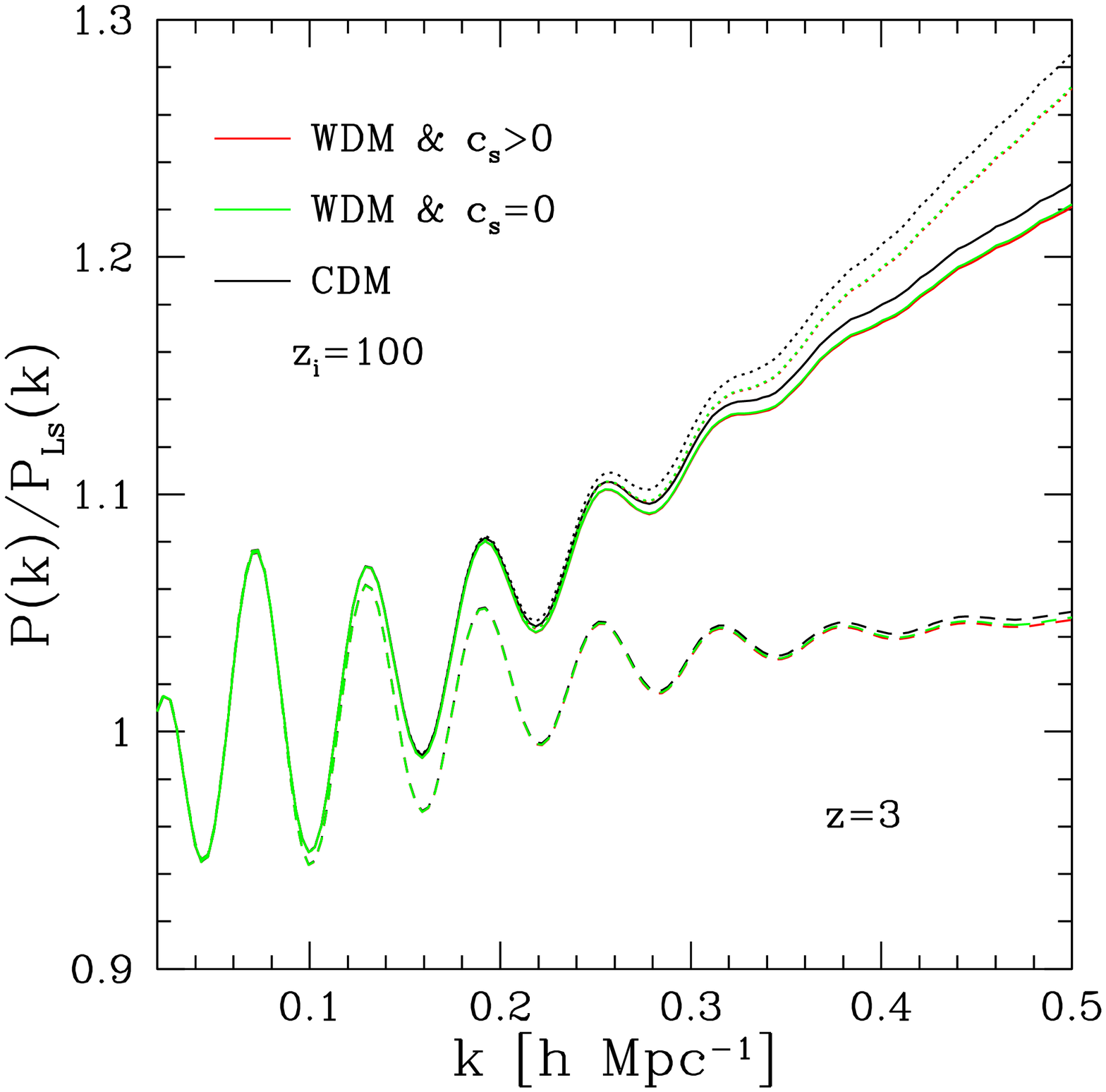}}
\epsfxsize=5.5 cm \epsfysize=5. cm {\epsfbox{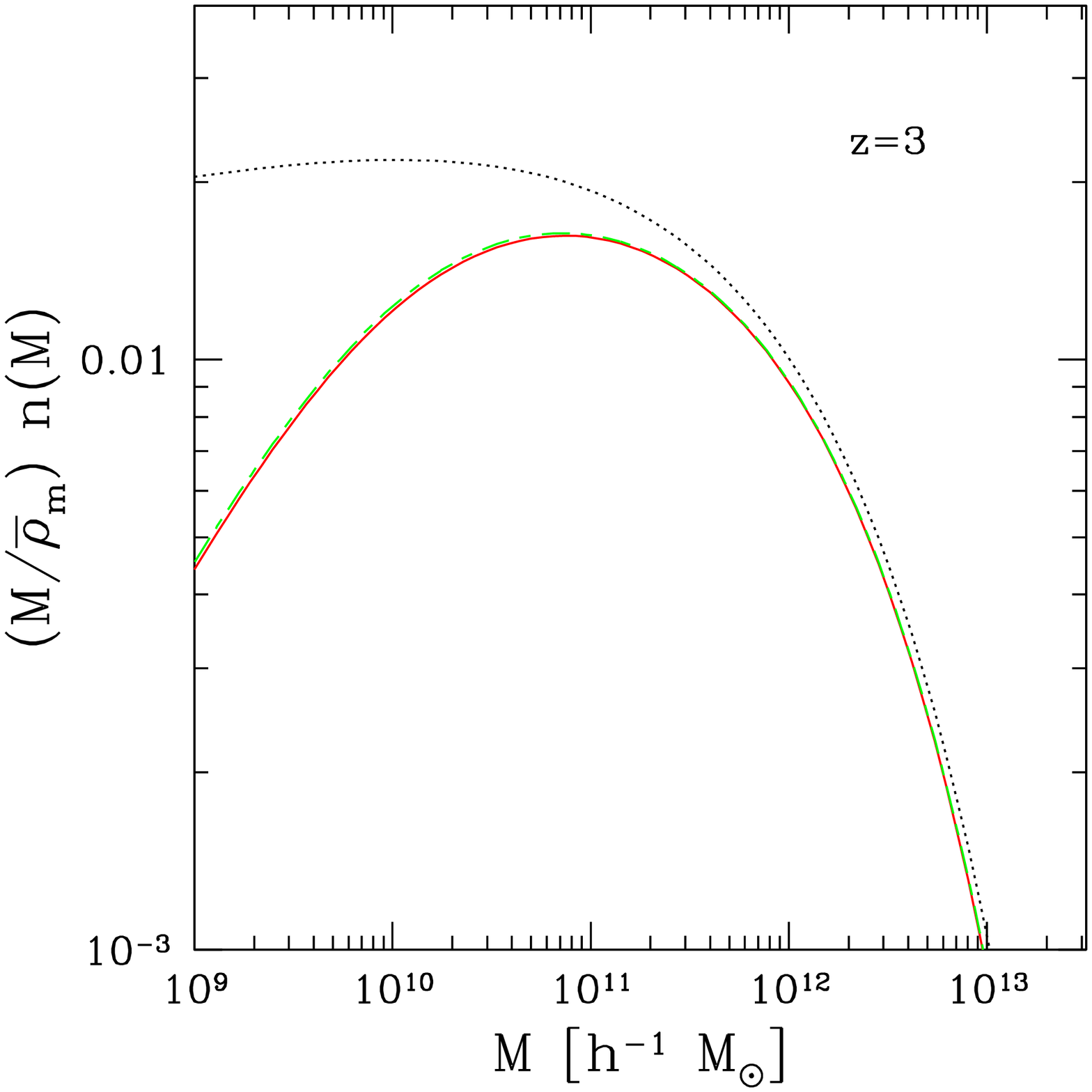}}
\epsfxsize=5.5 cm \epsfysize=5. cm {\epsfbox{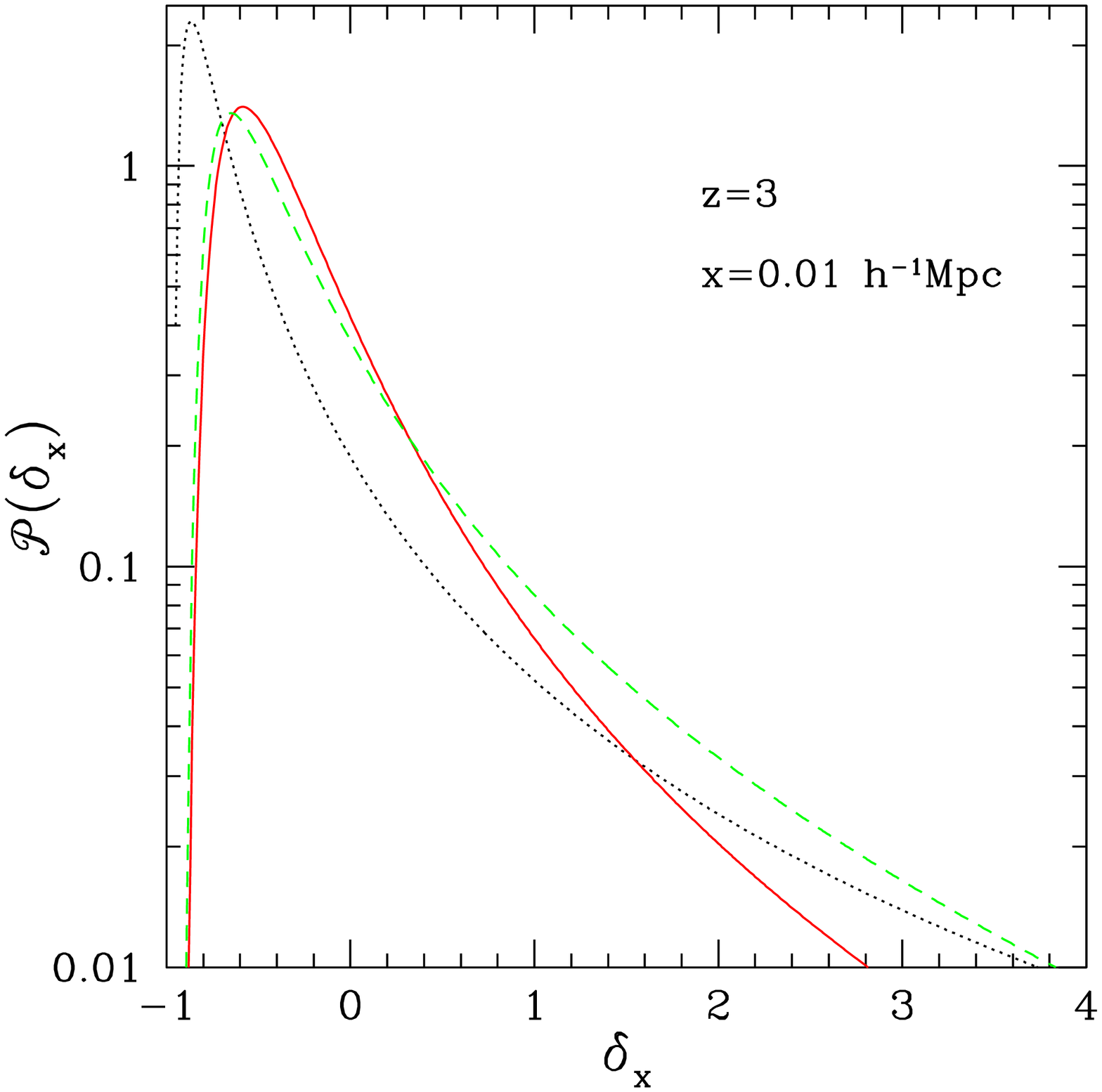}}
\end{center}
\caption{{\it Left panel:} ratio of the power spectrum $P(k)$ to a smooth
$\Lambda$CDM linear power spectrum $P_{Ls}(k)$ without baryonic oscillations.
We show our results for the reference CDM scenario
(black lines), the WDM scenario with $c_s=0$ (green lines) and with $c_s\neq 0$
(red lines). In each case, we plot the linear power (lower dashed lines), the nonlinear
one-loop ``steepest descent'' resummation (solid lines), and the ``standard'' 1-loop
result (upper dotted lines).
{\it Middle panel:} halo mass functions at $z=3$.
{\it Right panel:} Probability distribution ${\cal P}(\delta_x)$ of the matter density 
contrast within spheres of radius $x=0.01 h^{-1}$Mpc at $z=3$.}
\end{figure*}

\medskip

We show our results for a $1$keV dark matter particle in Fig.1 (with an ``initial'' redshift
$z_i=100$ where we set the initial conditions, as in numerical simulations).
The left panel shows that the deviation of the density power spectrum from CDM
is very small on large perturbative scales and below the accuracy of 
standard perturbation theory, as shown by the comparison between the standard
1-loop perturbative result (upper dotted lines) and an improved resummation scheme
(solid lines).
Moreover, the deviation between the cases $c_s=0$ and $c_s>0$ is even smaller.
This implies that on these scales, the effects of the late-time WDM velocity dispersion
can be neglected.
The middle panel shows that the same conclusion holds for the halo mass function

\medskip

The right panel, which displays the probability distribution function of the density contrast
on smaller scales, shows as expected a larger deviation from CDM.
The lack of small scale power in the WDM scenario leads to a
less advanced stage of the nonlinear evolution: the peak shifts closer to the mean
$\langle\delta_x\rangle=0$ and the tails are sharper.
The late-time velocity dispersion even further impedes the nonlinear evolution and
makes the large density tail sharper.
This suggests that accurate measures of the probability distribution of the flux decrement
of distant quasars, due to Lyman-$\alpha$ absorption lines, which is closely related to the 
probability distribution of the matter density on these scales [11,12],
could be sensitive to this late-time velocity dispersion.
Thus, numerical simulations that do not include this effect are likely to underestimate
somewhat the difference between the CDM and WDM scenarios with respect to
Lyman-$\alpha$ absorption lines.

\medskip

Using the same tools, we can estimate the impact of the ``initial'' redshift $z_i$ of the
simulations and we find that a choice $z_i \leq 50$ leads to a small underestimate of 
gravitational clustering at low $z$.

\bigskip

{\bf References}

\begin{description}

\item[1] D. Boyanovsky, H. J. de Vega, N. G. Sanchez, Phys. Rev. D, 78, 063546 (2008)

\item[2] D. Boyanovsky, Phys. Rev. D, 83, 103504 (2011)

\item[3] D. Boyanovsky, J. Wu, Phys. Rev. D, 83, 043524 (2011)

\item[4] H. J. de Vega, N. G. Sanchez, Phys. Rev. D, 85, 043516 (2012)

\item[5] P. Bode, J. P. Ostriker, N. Turok, ApJ, 556, 93B (2001)

\item[6] A. Schneider, R. E. Smith, A. V. Maccio, B. Moore, arXiv:1112.0330 (2011)

\item[7] P. Col{\'{\i}}n, O. Valenzuela, V. Avila-Reese, ApJ, 673, 203 (2008)

\item[8] A. Boyarsky, J. Lesgourgues, O. Ruchayskiy, M. Viel, JCAP, 05, 012 (2009)

\item[9] M. Viel, K. Markovic, M. Baldi, J. Weller, MNRAS, 421, 50 (2012)

\item[10] P. Valageas, arXiv:1206.0554 (2012)

\item[11] P. Valageas, R. Schaeffer, J. Silk, Astron. \& Astrophys., 345, 691 (1999)

\item[12] M. Viel, S. Matarrese, H. J. Mo, T. Theuns, M. G. Haehnelt, MNRAS, 336, 685 (2002)

\end{description}

\newpage

\subsection{Casey R. WATSON}

\vskip -0.3cm

\begin{center}

Department of Physics and Astronomy,\\

Millikin University, Decatur, Illinois 62522,USA \\

\bigskip

{\bf Using X-ray Observations to Constrain Sterile Neutrino Warm Dark Matter} 

\end{center}

\medskip

To detect or place the most restrictive constraints on radiatively decaying dark matter, 
it is critical to (1) choose a target that maximizes
$\Sigma^{\rm FOV}_{\rm DM} = M^{\rm FOV}_{\rm DM}D^{-2}$, 
i.e., maximum dark matter mass within the detector field of view (FOV) at minimum distance, $D$
and (2) minimize the X-ray background from baryonic sources,

such as diffuse hot gas and X-ray binaries. To accomplish (1) and (2) and place the most restrictive constraints 
on the sterile neutrino mass and mixing to date, Watson et al. analyzed
{\it Chandra} {\rm~ACIS-I} observations of a $12'-28'$ annulus surrounding the center
of the Andromeda galaxy (M31),(Watson,2012).

\bigskip

In the context of the Dodelson-Widrow (DW) Model (Dodelson 1994; AFT; Abazajian) and the model presented
in Asaka,2006, the unresolved X-ray spectrum from this FOV requires
$m^{\rm M}_{\rm s} < 2.2$ keV (Majorana) and $m^{\rm D}_{\rm s} < 2.4$ keV (Dirac) (95\% C.L.)
to avoid more than doubling the amplitude of the measured
spectrum at $E_{\gamma, \rm s} = m_{\rm s, lim}/2$ = 1.1 keV and 1.2 keV, respectively
Watson:2011dw.  Lifting the DW-model restriction, this spectrum also yields improved constraints
in the $m_{s~} -$~sin$^{2}2\theta$ (mass-mixing) plane (Fig. 4 in Watson, 2012).

\bigskip

Most of the currently available sterile neutrino mass-mixing 
parameter space remains open only for scenarios in which $\Omega_{\rm s} \sim 0.3$ can be 
generated at very small mixing, (e.g. Shi,1998; AFP; Laine, 2008; Abazajian, Petraki, 2007).
The DW scenario now remains
viable only between the Tremaine-Gunn (T-G) bound (T-G, 1979) and the limit from Ref.(Watson,2011dw):
0.4 keV $ \lesssim m_{s~} \lesssim$ 2.2 keV, which interestingly falls within the range of dark matter particle
masses that best explains the core of the Fornax Dwarf Spheroidal galaxy (Strigari,2006).
This result underscores the need to continue to carefully and independently pursue all possible constraints
on sterile neutrino properties. However, because of the decreasing sterile neutrino signal
at lower $m_{\rm s}$ values and the large number of atomic emission features at energies $\lesssim 1$ keV,
it will be difficult to improve upon the limits in Ref.(Watson, 2012).

\bigskip

One of the most promising routes toward better radiative constraints
includes the use of much higher effective area and
spectral resolution instruments than those currently available, as discussed, e.g.,
Abazajian, 2009.  In the meantime, the environments between nearby merging galaxies, like the M81/M82 system - where significant dark matter mass is likely to be spatially separated from hot baryons - may provide opportunities for near-term improvements.

\bigskip

Adopting the initial conditions of Sofue, 1998, Chris Purcell ran a preliminary 500k particle simulation of the M81/M82 system over the last $\sim 1.1$ Gyrs (private communication).  He found that after only one near pass with M81, the mass of M82 fell from its initial value of $10^{11}M_\odot$ to its present value of $10^{10}M_\odot$ (Greco,2012). This result alone suggests the presence of a significant dark matter distribution between the two galaxies. Perigalacton occurred $\sim 0.25$ Gyrs from the end of the simulation, which matches the time of closest approach suggested by the starburst activity in M82, (deMello,2007), and the final separation distance of the two galaxies in the simulation is about 80\% of the observed separation.  Although there is clearly room for improvement in terms of mass resolution and reproducing observed properties of the system more exactly, the results of the simulation are broadly consistent with observations and should therefore provide a reasonable estimate of the dark matter distribution between the two galaxies.

\bigskip

The dark matter mass found within a typical $15' \times 15'$ ACIS-I FOV ($\simeq 15 \rm{kpc} \times 15 \rm{kpc}$ at $D_{\rm M81} = 3.6 \pm 0.2$ Mpc (Gerke,2011) in the region between the two galaxies is
$M^{\rm FOV}_{\rm DM} \simeq 2.5\times10^{10}M_\odot$.
The corresponding value of $\Sigma^{\rm FOV}_{\rm DM}$ in this region is:
$\Sigma^{\rm FOV}_{\rm DM,M81-M82} \simeq 0.019\times 10^{11}M_\odot \rm Mpc^{-2}$.
Adopting the low-mass model of the Milky Way from Battaglia,2005, and integrating the density along the line of sight towards M81-M82 yields $\Sigma^{\rm FOV}_{\rm DM,MW} \simeq 0.009\times 10^{11}M_\odot \rm Mpc^{-2}$.
Therefore, a fairly conservative estimate for the total dark matter column density within a typical
ACIS-I FOV in the region between M81 and M82 is 
$\Sigma^{\rm FOV}_{\rm DM,Tot} \simeq 0.028\times 10^{11}M_\odot \rm Mpc^{-2}$.

\begin{figure}
\includegraphics[height = .35\textheight, width = .5\textwidth]{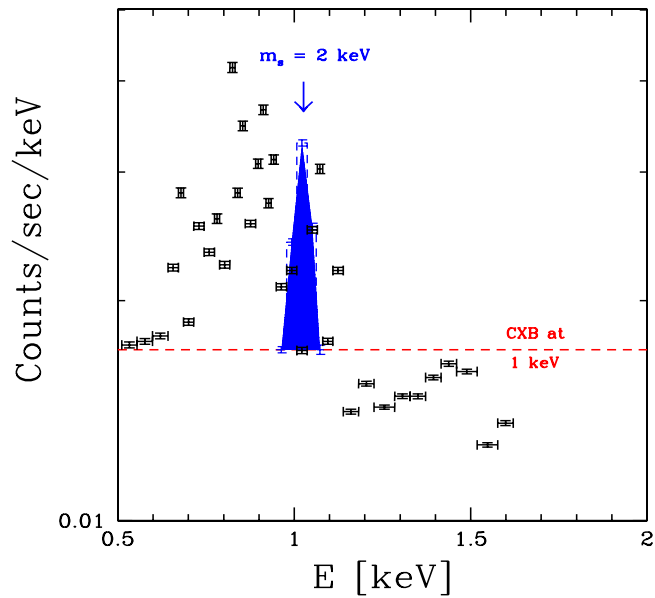}
\caption {Here we estimate the Cosmic X-ray Background spectrum from a typical $15'\times 15'$ ACIS-I FOV
between M81 and M82 in the absence of sterile neutrinos (black data points) and show the
first statistically significant DW Majorana $\nu_s$ decay
peak (blue, shaded), which would double the projected level of the CXB data at 1 keV (red, dashed), thereby excluding
$m_{\rm s} \geq 2$ keV (95\% C.L.) if no such feature is observed.
\label{CXB_nus_sig}}
\end{figure}

\bigskip

Provided that the baryons between M81 and M82 are too cold to produce a significant amount of X-rays, the prospective sterile neutrino decay signatures from this region would only have to compete against the Cosmic X-ray Background (CXB). Fig.~\ref{CXB_nus_sig} shows the estimated CXB spectrum from a typical $15'\times 15'$ ACIS-I FOV and the prospective decay signature (Eqn. 4 of Ref.Watson,2012) from 2 keV DW Majorana sterile neutrinos with a surface density of $\Sigma^{\rm FOV}_{\rm DM} \simeq 0.028\times 10^{11}M_\odot \rm Mpc^{-2}$ within the same region.
This prospective decay signature doubles the CXB data at 1 keV, so the M81/M82 system should allow for detection or exclusion of DW sterile neutrinos down to $m_{\rm s} = 2$ keV.

\bigskip

Although this improvement is incremental relative to the $m_{\rm s} \geq 2.2$ keV limit of Watson, 2012, the steadily growing amount of astronomical data that favors keV-scale dark matter suggests that any opportunity for improvement is worthwhile, as it may lead to a detection. In addition to M81/M82, other systems with the potential for large amounts of spatially separated dark matter, such as the M31/M32/NGC205 environment, will be examined more thoroughly in future work.  Next-generation detectors, e.g., (Abazajian:2009) will, however, be necessary to achieve dramatically better keV dark matter constraints and to significantly increase the chances of finding this more and more observationally favored candidate, particularly for very low mixing production scenarios, (e.g., Shi:1998; AFP; Petraki,2007; Laine,2008; Abazajian).

\bigskip

{\bf Acknowledgments}

\medskip

I would like to thank Zhiyuan Li, Nick Polley, and Chris Purcell for their collaboration on the work I presented, the conference organizers, Hector J. de Vega and Norma Sanchez, for hosting an interesting and productive meeting, and all the participants, particularly Peter Biermann and Jorge Penarubbia, for helpful discussions.

\bigskip

{\bf References}

\begin{description}

\item C.~R.~Watson, Z.~Li and N.~K.~Polley, JCAP {\bf 1203}, 018 (2012)
[arXiv:1111.4217 [astro-ph.CO]].
                                                                                                
\item 
S.~Dodelson and L.~M.~Widrow,
Phys.\ Rev.\ Lett.\  {\bf 72}, 17 (1994)
[arXiv:hep-ph/9303287].

\item K.~Abazajian, G.~M.~Fuller and W.~H.~Tucker,
Astrophys.\ J.\  {\bf 562}, 593 (2001)
[arXiv:astro-ph/0106002].
  
\item Phys.\ Rev.\  D {\bf 73}, 063506 (2006)
[arXiv:astro-ph/0511630].

\item K.~Abazajian, Phys.\ Rev.\  D {\bf 73}, 063513 (2006)
[arXiv:astro-ph/0512631].

\item  T.~Asaka, M.~Laine and M.~Shaposhnikov,
JHEP {\bf 0701}, 091 (2007)
[arXiv:hep-ph/0612182].

\item  X.~D.~Shi and G.~M.~Fuller, Phys.\ Rev.\ Lett.\  {\bf 82}, 2832 (1999)
[arXiv:astro-ph/9810076].

\item  K.~Abazajian, G.~M.~Fuller and M.~Patel, Phys.\ Rev.\  D {\bf 64}, 023501 (2001)
  [arXiv:astro-ph/0101524].

\item  M.~Laine and M.~Shaposhnikov, JCAP {\bf 0806}, 031 (2008)
  [arXiv:0804.4543 [hep-ph]].
  
\item  K.~Abazajian and S.~M.~Koushiappas, Phys.\ Rev.\  D {\bf 74}, 023527 (2006)
  [arXiv:astro-ph/0605271].

\item  K.~Petraki and A.~Kusenko, Phys.\ Rev.\  D {\bf 77}, 065014 (2008)
  [arXiv:0711.4646 [hep-ph]].

\item S. Tremaine and J. E. Gunn, Phys.\ Rev.\ Lett.\ {\bf 42}, 407 (1979).

\item L.~E.~Strigari, J.~S.~Bullock, M.~Kaplinghat, A.~V.~Kravtsov, O.~Y.~Gnedin, K.~Abazajian and A.~A.~Klypin, Astrophys.\ J.\  {\bf 652}, 306 (2006) [arXiv:astro-ph/0603775].

\item K.~N.~Abazajian, arXiv:0903.2040 [astro-ph.CO].

\item  Y.~Sofue, astro-ph/9803012.

\item J.~P.~Greco, P.~Martini and T.~A.~Thompson, arXiv:1202.0824 [astro-ph.CO].

\item D.~F.~de Mello, L.~J.~Smith, E.~Sabbi, J.~S.~Gallagher, M.~Mountain and D.~R.~Harbeck,
arXiv:0711.2685 [astro-ph].

\item J.~R.~Gerke, C.~S.~Kochanek, J.~L.~Prieto, K.~Z.~Stanek and L.~M.~Macri,
Astrophys.\ J.\  {\bf 743}, 176 (2011) [arXiv:1103.0549 [astro-ph.CO]].

\item  G.~Battaglia, A.~Helmi, H.~Morrison, P.~Harding, E.~W.~Olszewski, M.~Mateo, K.~C.~Freeman and J.~Norris {\it et al.}, Mon.\ Not.\ Roy.\ Astron.\ Soc.\  {\bf 364}, 433 (2005)
  [Erratum-ibid.\  {\bf 370}, 1055 (2006)]
  [astro-ph/0506102].

\end{description}

\newpage

\subsection{Jes\'us ZAVALA}

\vskip -0.3cm

\begin{center}

CITA National Fellow\\
Department of Physics and Astronomy, University of Waterloo, Waterloo, Ontario, N2L 3G1, Canada\\
Perimeter Institute for Theoretical Physics, 31 Caroline St. N., Waterloo, ON, N2L 2Y5, Canada

\bigskip

{\bf The velocity function of galaxies in the local environment from
Cold and Warm Dark Matter simulations} 

\end{center}

\medskip

Despite the remarkable success of the Cold Dark Matter (CDM) paradigm in describing the large scale structure of the Universe, there are important challenges that remain at smaller scales. One of these is the 
striking CDM prediction of a vast number of dark matter substructures residing within our own Milky Way (MW)
halo, which, at first sight, seems to be at odds with actual observed abundance of MW satellites (e.g. [1]).
Although this ``missing satellite'' problem can be solved by invoking astrophysical processes that are expected
to suppress galaxy formation in the smallest subhalos (e.g. [2]), it is also interesting to consider the 
possibility of adopting a different dark matter particle to alleviate this problem. Warm Dark Matter (WDM) 
particles with masses of $\mathcal{O}(1$~keV$)$ can accomplish this since they have free streaming masses of 
$\mathcal{O}(10^{10}$~M$_{\odot})$ (e.g. [3]).

\bigskip

We show how this overabundance of small scale structure is also observed for isolated dwarf galaxies, which
are in principle free of the complex gas removal processes that satellite galaxies are subjected to. This is shown
by analyzing the early data release of the HI ALFALFA survey [4], which preferentially detects galaxies 
with high gas fractions and gives a robust measurement of the maximum rotational velocity of a galaxy,
$V_{\rm max}$, using the 21cm line width. We make 
a detailed comparison of the predictions of the CDM and WDM cosmogonies for the abundance of low-mass 
dark matter halos in our local environment using constrained N-body simulations (e.g. [5]). These simulations 
are constructed to reproduce the gross features of the nearby Universe, such as the Local Supercluster and 
the Virgo cluster, and we found that they are able to match the observed abundance of HI galaxies at the
intermediate and high mass end of the velocity function. However, at the low-mass end 
($V_{\rm max}<60$~kms$^{-1}$, corresponding to $M_{\rm halo}<4\times10^{10}$~M$_\odot$), the CDM model clearly
predicts and overabundance of low-mass galaxies that reaches a factor of $\sim10$ at $V_{\rm max}~35$~kms$^{-1}$, whereas
the WDM simulation (corresponding to a thermal relic with a mass of $1$~keV) is consistent with the data at all 
masses [6].  

\bigskip

Another important challenge to the CDM paradigm is the apparent presence of central density cores in the
inferred mass profiles of low surface brightness galaxies (e.g [7]) and MW dwarf spheroidals (dSphs, e.g. [8]).
We have recently shown that a velocity-dependent self-interacting dark matter model (motivated by a Yukawa-like 
new gauge boson interaction[9]), can create sizable cores in the dark matter subhalos that host the MW dSphs, 
without violating any current constraints on the scattering cross section [10]. 

\bigskip

{\bf References}

\begin{description}

\item[1] Klypin, A., Kravtsov, A. V., Valenzuela, O., \& Prada, F. 1999, ApJ, 522, 82

\item[2] Guo Q. et al., 2011, MNRAS, 413, 101

\item[3] Col\'{i}n, P., Avila-Reese, V., \& Valenzuela, O. 2000, ApJ, 542, 622

\item[4] Giovanelli, R. et al., 2005, AJ, 130, 2613

\item[5] Klypin, A., Hoffman, Y., Kravtsov, A. V., \& Gottl\"ober, S. 2003,
ApJ, 596, 19

\item[6] Zavala, J. et al., 2009, ApJ, 700, 1779

\item[7] de Blok W. J. G. \& McGaugh S. S., 1997, MNRAS, 290, 533

\item[8] Walker M. G. \& Pe\~narrubia J., 2011, ApJ, 742, 20

\item[9] Loeb A. \& Weiner N., 2011, Physical Review Letters, 106, 171302

\item[10] Vogelsberger, M., Zavala, J. \& Loeb, A., MNRAS, 2012, arXiv:1201.5892

\end{description}

\newpage

\subsection{He ZHANG}

\vskip -0.3cm

\begin{center}

Max--Planck--Institut f\"ur Kernphysik,\\
Saupfercheckweg 1, 69117 Heidelberg, Germany

\bigskip

{\bf Sterile neutrinos for Warm Dark Matter in flavor symmetry
models}

\end{center}

\medskip

The presence of Dark Matter could be viewed as an indirect proof of
the physics beyond the Standard Model (SM). Another solid evidence
of new physics beyond the SM is the neutrino mass, which is not
allowed in the simplest SM framework. It has been shown that the
sterile neutrino hypothesis could solve the above two puzzles
simultaneously (see a recently review in Ref. [1]). However, from
the theoretical point of view, the sterile neutrino mass scale needs
to be explained in a natural way. Possible ways to accommodate light
sterile neutrinos include the extra dimension scenario [2], the
Froggatt-Nielsen (FN) mechanism [3-4], flavor symmetries [5-7],
mirror world models, Axino models, and the extended seesaw model
[8]. 

\bigskip

We will show that the neutrino mass generation and the sterile
neutrino Warm Dark Matter (WDM) can be connected in the framework of
flavor symmetries, e.g. the tetrahedral group $A_4$. Namely, a
sterile neutrino with mass at the keV scale and with small mixing to
the active neutrinos is a WDM candidate, which is compatible with
observation and in fact could solve some of the problems of the Cold
Dark Matter paradigm. In addition, the other two sterile neutrinos
lead to visible contributions to active neutrino masses and they can
be located at very different energy scales, from the eV scale to
superheavy scales.

\bigskip

The particle contents and assignments of the $A_4$ seesaw model are
shown in Table I [9].
\begin{table}[h]
\label{table:afssmodel_a} \vspace{8pt}
\begin{tabular}{c|ccccc|ccccccc|ccc}
\hline \hline  Field & $L$ & $e^c$ & $\mu^c$ & $\tau^c$ & $h_{u,d}$ & $\varphi$ & $\varphi'$ & $\varphi''$ & $\xi$ & $\xi'$ & $\xi''$ & $\Theta$ & $\nu^c_{1}$ & $\nu^c_{2}$ & $\nu^c_{3}$ \\
\hline $SU(2)_L$ & $2$ & $1$ & $1$ & $1$ & $2$ & $1$ & $1$ & $1$ & $1$ & $1$ & $1$ & $1$ & $1$ & $1$ & $1$ \\
$A_4$ & $\underline{3}$ & $\underline{1}$ & $\underline{1}''$ & $\underline{1}'$ & $\underline{1}$ & $\underline{3}$ & $\underline{3}$ & $\underline{3}$ & $\underline{1}$ & $\underline{1}'$ & $\underline{1}$ & $\underline{1}$ & $\underline{1}$ & $\underline{1}'$ & $\underline{1}$  \\
$Z_3$ & $\omega$ & $\omega^2$ & $\omega^2$ & $\omega^2$ & $1$ & $1$ & $\omega$ & $\omega^2$ & $\omega^2$ & $\omega$ & $1$ & $1$ &  $\omega^2$ & $\omega$ & $1$  \\
$U(1)_{\rm FN}$ & - & $3$ & $1$ & $0$ & - & - & - & - & - & - & - & $-1$ & $F_1$ & $F_2$ & $F_3$ \\[1mm] \hline \hline
\end{tabular}
\centering \caption{Particle assignments of the $A_4$ type I seesaw
model, with three right-handed sterile neutrinos. The additional
$Z_3$ symmetry decouples the charged lepton and neutrino sectors;
the $U(1)_{\rm FN}$ charge generates the hierarchy of charged lepton
masses and regulates the mass scales of the sterile states.}
\end{table}
Apart from the SM fermions, three right-handed neutrinos $\nu^c_i$
($i=1,2,3$) are introduced transforming as singlets under $A_4$. In
addition, flavon fields $\varphi$, $\varphi'$ and $\varphi''$
transforming as $A_4$ triplets are needed to construct the charged
lepton mass matrix and the Dirac neutrino mass term. Furthermore,
singlet flavons $\xi$, $\xi'$ and $\xi''$ are introduced in order to
give masses to the right-handed neutrinos and keep the right-handed
neutrino mass matrix $M_R$ diagonal at leading order. The Lagrangian
invariant under the SM gauge group and the additional $A_4 \otimes
Z_3 \otimes U(1)_{\rm FN}$ symmetry is

\begin{align}
\nonumber -{\cal L}_{\rm Y} &=  \frac{y_e}{\Lambda}\lambda^3
\left(\varphi L h_d\right) e^c + \frac{y_\mu}{\Lambda}
\lambda\left(\varphi L h_d\right)' \mu^c + \frac{y_\mu}{\Lambda}
\left(\varphi L h_d\right)''
\tau^c  \\
&+  \frac{y_1}{\Lambda}\lambda^{F_1}(\varphi L h_u) \nu^c_1 +
\frac{y_2}{\Lambda}\lambda^{F_2} (\varphi' L h_u)'' \nu^c_2 +
\frac{y_3}{\Lambda}\lambda^{F_3} (\varphi''L h_u) \nu^c_3
\label{eq:seesaw_lag} \\
\nonumber
 &+ \frac{1}{2} \left[w_1 \lambda^{2F_1}\xi \nu^c_1 \nu^c_1 + w_2\lambda^{2F_2} \xi' \nu^c_2 \nu^c_2 + w_3 \lambda^{2F_3}\xi'' \nu^c_3 \nu^c_3
 \right] + {\rm h.c.},
\end{align}

at leading order, where the notation $(ab)'$ refers to the product
of $A_4$ triplets transforming as $\underline{1}'$, etc., and
$y_\alpha$, $y_i$ and $w_i$ are coupling constants. $\lambda \equiv
\langle \Theta\rangle/\Lambda < 1$ is the FN suppression parameter,
and for simplicity we assume $\Lambda$ to be the cutoff scale of
both the $A_4$ symmetry and the FN mechanism. We choose the vacuum
alignment $\langle \varphi \rangle = (v,0,0)$, which ensures that
the charged lepton mass matrix is diagonal at leading order, i.e.,
$M_\ell=\frac{v_d v}{\Lambda}{\rm diag}(y_e \lambda^3,y_\mu
\lambda,y_\tau)$, with $v_d=\langle h_d\rangle$. Furthermore,
$\nu^c_1$ is assumed to be the WDM candidate in our model with a
mass given by $M_1=\omega_1 u \lambda^{2F_1}$, where $u=\langle \xi
\rangle$. The vacuum alignment of $\varphi$ means that at leading
order the first column of the Dirac neutrino mass matrix is $(y_1 v
v_u \lambda^{F_1}/\Lambda,0,0)^T$, indicating that the sterile
neutrino $\nu^c_1$ only mixes with the electron neutrino. The
active-sterile mixing is given by

\begin{equation}
\theta_{e1} \simeq \frac{y_1 v v_u}{w_1 u \Lambda}  \lambda^{-F_1}
\; ,
\end{equation}

so the active-sterile mixing is actually controlled by both the
flavon VEVs and the FN charge $F_1$. In what regards the keV sterile
neutrino,  we give a rough numerical estimate by choosing the mass
scales $ v \simeq 10^{11}\ {\rm GeV}$, $u \simeq 10^{12}\ {\rm
GeV}$, $\Lambda \simeq 10^{13}\ {\rm GeV}$, $v_u = \langle h_u
\rangle \simeq  174~{\rm GeV}$ together with $\lambda \simeq 0.1$.
For the choice $F_1 = 9$, one obtains a sterile neutrino of mass
$M_1 \simeq 1$~keV with the desired mixing angle $\theta_1^2 \simeq
10^{-8}$, with $y_1,w_1 \leq {\cal O}(1)$. Such a mixing angle is
favored in the Dodelson-Widrow scenario, in which the sterile
neutrino WDM are produced through neutrino oscillations. Note that
$\nu^c_1$ is essentially decoupled in the seesaw mechanism due to
the tiny mixing angle.

\bigskip

In the model, the other two sterile neutrinos $\nu^c_2$ and
$\nu^c_3$ generate active neutrino masses via the seesaw mechanism.
For the normal mass ordering case, we need the vacuum alignments
$\langle \varphi' \rangle = (v',v',v')$ and $\langle \varphi''
\rangle = (0,v'',-v'') $ which results in the tri-bimaximal neutrino
mixing pattern (TBM) at leading order. In contrast, we assume
$\langle \varphi' \rangle = (v',v',v')$ and $\langle \varphi''
\rangle = (2v'',- v'',- v'') $ for the inverted mass ordering case
in order to produce the TBM. The FN changes of $\nu^c_2$ and
$\nu^c_3$ remain undetermined, which allows us to alter the spectrum
of sterile neutrinos making the model versatile. We summarize in
Table II the consequent neutrino phenomenology for different choices
of $F_2$ and $F_3$.

\begin{table}[t]
\vspace{8pt}
\begin{tabular*}{0.98\textwidth}{@{\extracolsep{\fill}}c|cccccc|c}
\hline \hline  \multirow{2}{*}{} & \multirow{2}{*}{$F_1$, $F_2$,
$F_3$} & \multirow{2}{*}{Mass spectrum} &
\multirow{2}{*}{$|U_{\alpha 4}|$} & \multirow{2}{*}{$|U_{\alpha
5}|$} & \multicolumn{2}{c|}{$\langle m \rangle_{ee}$} &
\multirow{2}{*}{Phenomenology}
\\ & & & & & NO & IO & \\ \hline  I & $9$, $10$, $10$ & $M_{2,3} =
{\cal O}({\rm eV})$, & ${\cal O}(0.1)$ & ${\cal O}(0.1)$ & 0 & 0 &
$3+2$ mixing
\\[1mm] \hline
 \multirow{2}{*}{IIA} &  \multirow{2}{*}{$9$, $10$, $0$} & $M_{2} =
{\cal O}({\rm eV}) $ &  \multirow{2}{*}{${\cal O}(0.1)$} &
\multirow{2}{*}{${\cal O}(10^{-11})$} & \multirow{2}{*}{0} &
\multirow{2}{*}{$\dfrac{2\sqrt{\Delta m^2_A}}{3}$ 
}  & \multirow{5}{*}{$3+1$
 mixing}\\[1mm] && $M_{3} = {\cal O}(10^{11}\,{\rm
GeV})$ &&&& \\[1mm] \cline{1-7}  \multirow{2}{*}{IIB} &
\multirow{2}{*}{$9$, $0$, $10$} & $M_{2} = {\cal O}(10^{11}\,{\rm
GeV})$ & \multirow{2}{*}{${\cal O}(10^{-11})$} &
\multirow{2}{*}{${\cal O}(0.1)$} & \multirow{2}{*}{$\dfrac{\sqrt{\Delta m^2_S}}{3}$
} &
\multirow{2}{*}{$\dfrac{\sqrt{\Delta m^2_A}}{3}$
} &  \\[1mm] && $M_{3} = {\cal O}({\rm eV})$ &&&&  \\[1mm] \hline  III & $9$, $5$, $5$ & $M_{2,3} = {\cal O}(10\,{\rm GeV})$ & ${\cal O}(10^{-6})$ & ${\cal O}(10^{-6})$ & $\dfrac{\sqrt{\Delta m^2_S}}{3}$
& $\sqrt{\Delta m^2_A}$
& Leptogenesis\\[3mm]
\hline \hline
\end{tabular*}
\centering \caption{Summary of the different scenarios discussed in
the $A_4$ seesaw model. In each case the WDM sterile neutrino has a
mass $M_{1} = {\cal O}({\rm keV})$, and the corresponding active
neutrino is approximately massless. }
\end{table}

\medskip

In scenario I, we assign the FN charges $F_{2,3}=10$, so that the
right-handed neutrino masses are lowered down to the eV scale, and
there are no heavy neutrinos right-handed neutrinos that could be
used to explain the matter-antimatter asymmetry of the Universe via
leptogenesis. Neutrinoless double beta decay is also vanishing since
the contributions from active and sterile neutrinos exactly cancel
each other, unless there are other new physics contributions. 

\medskip

In scenario II, either $F_2$ or $F_3$ is 0, indicating that one
right-handed neutrino is located at a superheavy scale, e.g.
$10^{11}~{\rm GeV}$, and the right-handed neutrino mass spectrum is
splitted. The neutrinoless double beta decay is allowed in this
scenario while the reactor neutrino anomaly could also be explain
since there exists one very light sterile neutrino. 

\medskip

Finally, in scenario III, we assume $F_2=F_3=5$, which reproduces the $\nu$MSM
paradigm. Baryogenesis then proceeds via oscillations between
$\nu_2^c$ and $\nu_3^c$, which need to be sufficiently degenerate
$\left(|M_2-M_3|/M_2 \simeq 10^{-6}\right)$ to give the correct
baryon asymmetry. Moreover, similar to the ordinary type-I seesaw,
neutrinoless double beta decay is allowed and the right-handed
neutrinos make no contribution to this process.

\bigskip

The addition of sterile right-handed neutrinos to the SM is a
natural way to explain the active neutrino masses via the seesaw
mechanism, while a keV sterile neutrino turns out to be a very
attractive WDM candidates. Starting from a flavor symmetry model
based on the tetrahedral group $A_4$, we described different ways to
introduce sterile neutrinos. In both cases the FN mechanism is used
to suppress the masses of the right-handed neutrinos, and a keV
sterile neutrino playing the role of WDM can be accommodated.
Further experimental data in the years to come will put the presence
of sterile neutrinos at the eV and/or keV scale/s to the test, thus
determining whether it is indeed a useful enterprise to further
pursue this avenue of research.

\bigskip

{\bf References}

\begin{description}

\item[1] K.N. Abazajian, {\it et al.}, arXiv:1204.5379.

\item[2] A. Kusenko, F. Takahashi, T.T. Yanagida, arXiv:1006.1731,
Phys. Lett. B 693, 144 (2010).

\item[3] J. Barry, W. Rodejohann, H. Zhang, arXiv:1105.3911, JHEP
1107, 091 (2011).

\item[4] A. Merle, V. Niro, arXiv:1105.5136, JCAP 1107, 023 (2011).

\item[5] M. Shaposhnikov, hep-ph/0605047, Nucl. Phys. B 763, 49 (2007).

\item[6] M. Lindner, A. Merle, V. Niro, arXiv:1011.4950, JCAP 1101, 034 (2011).

\item[7] R. Mohapatra, S. Nasri, H.B. Yu, hep-ph/0505021, Phys. Rev.
D 72, 033007 (2005).

\item[8] H. Zhang, arXiv:1110.6838.

\item[9] J. Barry, W. Rodejohann, H. Zhang, arXiv:1110.6382, JCAP
1201, 052 (2012).

\end{description}

\newpage

\section{Summary and Conclusions of the Workshop by H. J. de Vega and N. G. Sanchez}

\subsection{GENERAL VIEW AND CLARIFYING REMARKS}

Participants came from Europe, North and South America, Canada, Russia, Japan, Australia, China, Korea.
Discussions and lectures were outstanding.
WDM research evolves fastly in both astronomical, numerical, theoretical, particle and experimental research. 
The Workshop allowed to make visible the work on WDM made by different groups over the world and cristalize 
WDM as the viable component of the standard cosmological model in agreement with CMB + Large Scale
Structure (LSS) + Small Scale Structure (SSS) observations, $\Lambda$WDM, in contrast to 
 $\Lambda$CDM which only agree with CMB+LSS observations and is plagued with SSS problems.  

\medskip

The participants and the programme represented the different communities doing
research on dark matter:

\begin{itemize}
\item{Observational astronomers}
\item{Computer simulators}
\item{Theoretical astrophysicists not doing simulations}
\item{Physical theorists}
\item{Particle experimentalists}
\end{itemize}

\medskip

WDM refers to keV scale DM particles. This is not Hot DM (HDM). (HDM refers to eV scale DM particles, 
which are already ruled out). CDM refers to heavy DM particles (so called wimps of GeV scale or any scale 
larger than keV). 

\medskip 

It should be recalled that the connection between 
small scale structure features and the mass of the DM particle 
follows mainly from the value of the free-streaming
length $ l_{fs} $. Structures 
smaller than $ l_{fs} $ are erased by free-streaming.
WDM particles with mass in the keV scale 
produce $ l_{fs} \sim 100 $ kpc while 100 GeV CDM particles produce an
extremely small $ l_{fs} \sim  0.1 $ pc. While the keV WDM $ l_{fs} \sim 100 $ kpc
is in nice agreement with the astronomical observations, the GeV CDM $ l_{fs} $ 
is a million times smaller and produces the existence of too many
small scale structures till distances of the size of the Oort's cloud
in the solar system. No structures of such type have ever been observed.

\medskip
 
Also, the name CDM  precisely refers to simulations with heavy DM particles in the GeV scale.
Most of the literature on CDM simulations do not make explicit
the relevant ingredient which is the mass of the DM particle (GeV scale wimps in the CDM case). 

\medskip
 
The mass of the DM particle with the free-streaming length naturally enters in the initial power spectrum 
used in the N-body simulations and in the initial velocity. The power spectrum for large scales 
beyond 100 kpc is identical for WDM and CDM particles,
while the WDM spectrum is naturally cut off at scales below 100 kpc, 
corresponding to the keV particle mass free-streaming length. In contrast, the CDM spectrum 
smoothly continues for smaller and smaller scales till $\sim$ 0.1 pc, which gives rise
to the overabundance of predicted CDM structures at such scales.

\medskip
 
CDM particles are always non-relativistic, the initial velocities
are taken zero in CDM simulations, (and phase space density is unrealistically infinity 
in CDM simulations), while all this is not so for WDM.

\medskip

Since keV scale DM particles are non relativistic for $ z < 10^6 $
they could also deserve the name of cold dark matter, although for historical reasons the name WDM
is used. Overall, seen in perspective today, the reasons why CDM does not work are simple: the heavy wimps 
are excessively non-relativistic (too heavy, too cold, too slow), and thus frozen, which preclude them to 
erase the structures below the kpc scale, while 
the eV particles (HDM) are excessively relativistic, too light and fast, (its free streaming length is too large), 
which erase all structures below the Mpc scale; in between, WDM keV particles produce the right answer. 

\medskip

\subsection{CONCLUSIONS}

Some conclusions are:

\begin{itemize}

\item{Sterile neutrinos with mass in the keV scale (1 to 4 keV with 2 kev clearly favored) 
emerge as leading candidates for the dark matter (DM)
particle from theory combined with astronomical observations.

DM particles in the keV scale (warm dark matter, WDM)
naturally reproduce (i) the observed
galaxy structures at small scales (less than 50 kpc), (ii) the observed
value of the galaxy surface density and phase space density
(iii) the cored profiles of galaxy density profiles seen in
astronomical observations.

Heavier DM particles (as wimps in the GeV mass scale) do not
reproduce the above important galaxy observations and run into
growing and growing serious problems (they produce satellites problem,
voids problem, galaxy size problem, unobserved density cusps and other
problems).}

\item {Sterile neutrinos are serious WDM candidates:
Minimal extensions of the Standard Model of particle physics 
include keV sterile neutrinos which are very weakly coupled to the standard model particles
and are produced via the oscillation of the light (eV) active neutrinos, with their mixing
angle governing the amount of generated WDM. The mixing angle theta between active and sterile neutrinos
should be in the $ 10^{-4} $ scale to reproduce the average DM density in the
Universe.\\  
Sterile neutrinos are usually produced out of thermal equilibrium. 
The production can be non-resonant (in the absence of lepton asymmetries) 
or resonantly ennhanced (if lepton asymmetries are present). The usual X ray bound together
with the Lyman alpha bound forbids the non-resonant mechanism in the $\nu$MSM model. }

\item{ Sterile neutrinos can decay into an active-like neutrino and an X-ray photon. Abundance and phase 
space density of dwarf spheroidal galaxies constrain the mass to be in the 
$ \sim $ keV range.  
Small scale aspects of sterile neutrinos and different mechanisms of their production 
were presented: The transfer function and power spectra obtained by solving the 
collisionless Boltzmann equation during the radiation and matter dominated eras feature new WDM acoustic
oscillations on mass scales $ \sim 10^8-10^{9} \, M_{\odot} $.}

\item{Lyman-alpha constraints
have been often misinterpreted or superficially invoked in the past
to wrongly suggest a tension with WDM, but those constraints have been by now clarified
and relaxed, and such a tension does not exist: keV sterile neutrino dark matter (WDM) 
is consistent with Lyman-alpha constraints within a 
{\it wide range} of the sterile neutrino model parameters. {\bf Only} for sterile
neutrinos {\bf assuming} a {\bf non-resonant} (Dodelson-Widrow model) 
production mechanism, Lyman-alpha constraints provide a lower bound for the 
mass of about 4 keV . For thermal WDM relics (WDM particles decoupling at
thermal equilibrium) the Lyman-alpha lower particle mass bounds are 
smaller than for non-thermal WDM relics (WDM particles decoupling out of
thermal equilibrium). The number of Milky-Way satellites indicates lower bounds 
between 1 and 13 keV for different models of sterile neutrinos.}

\item {WDM keV sterile neutrinos can be copiously produced in the supernovae cores. Supernovae stringently 
constraint the neutrino mixing angle squared to be $ \lesssim 10^{-9}$ for sterile neutrino masses
$ m > 100$ keV (in order to avoid excessive energy lost) but for smaller sterile neutrino masses the 
SN bound is not so direct. Within the models 
worked out till now, mixing angles are essentially unconstrained by SN in the favoured WDM mass range, namely 
$ 1 < m  < 10 $ keV. Mixing between electron and keV sterile neutrinos could help SN explosions, case 
which deserve investigation}

\item{Signatures for a right-handed few keV sterile neutrino should be Lyman alpha emission and absorption 
at around a few 
microns; corresponding emission and absorption lines might be visible from molecular Hydrogen H$_2$  
and H$_3$  and their ions in the far infrared and sub-mm wavelength range.  The detection at very 
high redshift of massive star formation, stellar evolution and the formation 
of the first super-massive black holes would constitute the most striking and testable prediction of 
WDM sterile neutrinos.}

\item{The effect of keV WDM can be also observable in the statistical 
properties of cosmological Large Scale Structure. Cosmic shear 
(weak gravitational lensing) does not strongly depend on baryonic 
physics and is a promisingig probe. First results in a simple thermal 
relic scenario indicate that future weak lensing surveys could see a WDM 
signal for $m_{WDM} \sim 2$ keV or smaller. The predicted limit beyond 
which these surveys will not see a WDM signal is $m_{WDM} \sim 2.5$ keV 
(thermal relic) for combined 
Euclid + Planck. More realistic models deserve investigation and are 
expected to relaxe such minimal bound. With the real data, the 
non-linear WDM model should be taken into account}

\item {The possibility of laboratory detection of warm dark matter
is extremely interesting. Only a direct detection of the DM particle can give a clear-cut answer
to the nature of DM. At present,  only the {\bf Katrin and Mare experiments}
have the possibility to do that for sterile neutrinos.
{\bf Mare} bounds on sterile neutrinos have been reported in this Workshop.
Namely, bounds from the beta decay of Re187 and EC decay of Ho163.
{\bf Mare} keeps collecting data in both.}

\item {The possibility that {\bf Katrin} experiment can look to sterile neutrinos in the tritium
decay did appeared in the discussions.
Katrin experiment have the potentiality to detect warm dark matter if its set-up 
would be adapted to look to keV scale sterile neutrinos.
Katrin experiment concentrates its attention right now
on the electron spectrum near its end-point
since its goal is to measure the active neutrino mass.
Sterile neutrinos in the tritium decay will affect the electron
kinematics at an energy about $m$ below the end-point of the
spectrum ($m$ = sterile neutrinos mass). Katrin in the future
could perhaps adapt its set-up to look to keV scale sterile neutrinos.
It will be a a fantastic discovery to detect dark matter
in a beta decay}

\item{Astronomical observations strongly indicate that
{\bf dark matter halos are cored till scales below 1 kpc}. 
More precisely, the measured cores {\bf are not} hidden cusps.
CDM Numerical simulations -with wimps (particles heavier than $ 1 $ GeV)-
without {\bf and} with baryons yield cusped dark matter halos.
Adding baryons do not alleviate the problems of wimps (CDM) simulations,
on the contrary adiabatic contraction increases the central density of cups
worsening the discrepancies with astronomical observations. 
In order to transform the CDM cusps into cores, the baryon+CDM simulations
need to introduce strong baryon and supernovae feedback which produces a
large star formation rate contradicting the observations.
None of the predictions of CDM simulations at small scales
(cusps, substructures, ...) have been observed. The discrepancies of CDM 
simulations with the astronomical observations at small scales $ \lesssim 100 $ kpc {\bf is staggering}: 
satellite problem (for example, only 1/3 of satellites predicted by CDM
simulations around our galaxy are observed), the surface density problem (the value 
obtained in CDM simulations is 1000 times larger than the
observed galaxy surface density value),  the voids problem, size problem 
(CDM simulations produce too small galaxies).}

\item{The use of keV scale WDM particles in the simulations instead of the GeV CDM wimps, alleviate all the
above problems. For the core-cusp problem, setting
the velocity dispersion of keV scale DM particles seems beyond
the present resolution of computer simulations. 
However, the velocity dispersion of WDM particles is negligible
for $ z < 20 $ where the non-linear regime and structure formation starts.
Analytic work in the
linear approximation produces cored profiles for keV scale DM particles
and cusped profiles for CDM. Model-independent analysis of DM from phase-space density
and surface density observational data plus theoretical analysis
points to a DM particle mass in the keV scale.
The dark matter particle candidates with high mass (100 GeV, `wimps') 
are strongly disfavored, while cored (non cusped) dark matter halos and warm (keV scale mass) 
dark matter are strongly favoured from theory and  astrophysical observations.}

\item{An `Universal Rotation Curve' (URC) of spiral galaxies emerged from  
3200 individual observed Rotation Curves (RCs) and 
reproduces remarkably well out to the virial radius the Rotation Curve of 
any spiral galaxy. The URC is the observational counterpart of the  circular  
velocity profile from cosmological  simulations.
$\Lambda$CDM simulations give the well known NFW cuspy halo profile. A careful analysis 
from about 100 observed high quality rotation curves has now {\bf ruled out} the disk + NFW halo 
mass model, in favor of {\bf cored profiles}. 
The observed galaxy surface density (surface gravity acceleration) appears to be universal within 
$ \sim 10 \% $ with values around $ 120 \; M_{\odot}/{\rm pc}^2 $
, irrespective of galaxy morphology, luminosity and Hubble types, spanning over 14 magnitudes in 
luminosity and mass profiles determined by several independent methods.}

\item{Interestingly enough, a constant surface density  (in this case column density) with value around 
$ 120 \; M_{\odot}/{\rm pc}^2 $ similar to that found for galaxy systems  is found too for the interstellar 
molecular clouds, irrespective of size and compositions over six order of magnitude; this universal surface 
density in molecular clouds
is a consequence of the Larson scaling laws. This suggests the role of gravity on matter (whatever DM or 
baryonic) as a dominant underlying mechanism to produce such universal surface density in galaxies and molecular 
clouds. Recent re-examination of different and independent (mostly millimeter) molecular cloud data sets show 
that interestellar clouds do follow Larson law $Mass \sim (Size)^{2}$ exquisitely well, and therefore very 
similar projected mass densities at each extinction threshold. Such scaling and universality should play a key 
role in cloud structure formation.}

\item{Visible and dark matter observed distributions (from rotation curves of high resolution 2D velocity fields) 
of galaxies in compact groups, and the comparison with those of galaxies in clusters and field galaxies show 
that: (i) The central halo surface density is constant with respect to the total absolute magnitude similar to 
what is found for the isolated galaxies, suggesting that the halo density is independent of the galaxy type 
and environment. (ii) Core density profiles fit better the rotation curves than cuspy profiles. Dark halo 
density profiles are found almost the same in field galaxies, cluster and compact group galaxies. Core halos 
observed using high-resolution velocity fields in dark matter galaxies are {\bf genuine} and cannot be ascribed 
to systematic errors, halo triaxiality, or non-circular motions. (iii) The halo mass is high (75 to 95 \% of 
the total mass) for both field galaxies and compact groups galaxies, living modest room for a dark mass disk. 
No relation between the disk scale length and the halo central density is seen, 
the halo being independent of galaxy luminosity.}

\item{Besides the DM haloes, $\Lambda$CDM models ubiquituosly predict cold dark matter disks, 
[also believed as the consequence of merging events]: CDM simulations predict that 
a galaxy such as the Milky Way should host a CDM disk 
(with a scale height of $2.1-2.4$ kpc and a local density at the solar position $0.25-1.0$ times that of the 
DM halo; some models reach a height of 4 kpc from the galactic plane), which is thicker than any visible disk 
(the scale height of the galactic old thick stellar disk is $\sim$ 0.9 kpc; young stars and ISM form even 
thinner structures of height 0.3 and 0.1 kpc respectively). However, careful astronomical observations 
performed to see such disk have found no evidence of such CDM disk in the Milky Way.}

\item{A key part of any galaxy formation process and evolution involves dark matter. Cold gas accretion and 
mergers became important ingredients of the CDM models but they have little observational evidence. DM properties 
and its correlation with stellar masses are measured today up to $z =2$; at $z > 2$ observations are much less 
certain. Using kinematics and star formation rates, all types of masses -gaseous, stellar and dark- are measured 
now up to $z =1.4$. The DM density within galaxies declines at higher redshifts.  Star formation is observed 
to be more common in the past than today. More passive galaxies are in more 
massive DM halos, namely most massive DM halos have lowest fraction of stellar mass. CDM predicts high
overabundance of structure today and under-abundance of structure in the past with respect to observations. 
The size-luminosity scaling relation is the tightest of all purely photometric correlations used to 
characterize galaxies; its environmental dependence have been highly debated but recent findings show that the 
size-luminosity relation of nearby elliptical galaxies is well defined by a fundamental line and is environmental 
independent. Observed structural properties of elliptical galaxies appear simple and with no environmental 
dependence, showing that their growth via important mergers -as required by CDM galaxy formation- is not 
plausible. Moreover, observations in brightest cluster galaxies (BCGs) show little changes in the sizes of 
most massive galaxies since $z =1$ and this scale-size evolution appears closer to that of radio galaxies over 
a similar epoch. This lack of size growth evolution, a lack of BCG stellar mass evolution is observed too, 
demonstrates that {\bf major merging is not an important process}. Again, these observations put in serious trouble 
CDM `semianalytical models' of BCG evolution which require about $70\%$ of the final BCG stellar mass to be 
accreted in the evolution and important growth factors in size of massive elliptical massive galaxies.}

\item{Recent $\Lambda$WDM N-body simulations have been performed by different groups. 
High resolution simulations for different types of DM (HDM, WDM or CDM), allow to visualize the effects of 
the mass of the corresponding DM particles: free-streaming length scale, initial velocities and associated 
phase space density properties: for masses in the eV scale (HDM), halo formation occurs top down on all scales 
with the most massive haloes collapsing first; if primordial velocities are large enough, free streaming erases 
all perturbations and  haloes  cannot form (HDM). The concentration-mass halo relation for mass of hundreds eV 
is reversed with respect to that found for
CDM wimps of GeV mass. For realistic keV WDM these simulations deserve investigation: it could be expected from 
these HDM and CDM effects that combined free-streaming and velocity effects in keV WDM simulations could produce 
a bottom-up hierarchical scenario with the right amount of sub-structures (and some scale at which transition 
from top-down to bottom up regime is visualized).}

\item{Moreover, interestingly enough, recent large high resolution $\Lambda$WDM N-body simulations allow to 
discriminate among thermal and non-thermal WDM (sterile neutrinos): Unlike conventional thermal relics, 
non-thermal WDM has a peculiar velocity distribution (a little skewed to low velocities) which translates 
into a characteristic linear matter power spectrum decreasing slowler across the cut-off free-streaming scale 
than the thermal WDM spectrum. As a consequence, the  radial distribution of the subhalos predicted by WDM 
sterile neutrinos remarkably reproduces the observed distribution of Milky Way satellites in the range above 
$\sim 40$ kpc, while the thermal WDM supresses subgalactic structures perhaps too much, by a factor $2-4$ 
than the observation. Both simulations were performed for a mass equal to 1 keV. 
Simulations for a mass larger than 1 keV  (in the range between 2 and 10 keV, say) should still improve these results.}

\item{The predicted $\Lambda$WDM galaxy distribution in the local universe (as performed by
CLUES simulations with a
a mass of $ m_{\rm WDM}=1 $ keV)  agrees well 
with the observed one in the ALFALFA survey. On the
contrary, $\Lambda$CDM predicts a steep rise in the velocity
function towards low velocities and thus forecasts much  more
sources than the ones observed by the ALFALFA survey (both in Virgo-direction as well as in 
anti-Virgo-direction). These results show again the $\Lambda$CDM problems, also shown in the spectrum of 
mini-voids. $\Lambda$WDM 
provides a natural solution to these problems.  WDM physics effectively acts as a truncation of the 
$\Lambda$CDM power spectrum. $\Lambda$WDM CLUES simulations with 
1 keV particles gives much better answer than $\Lambda$CDM when reproducing sizes of local 
minivoids. The velocity function of 1 keV WDM Local Volume-counterpart reproduces the observational 
velocity function remarkably well. 
Overall, keV WDM particles deserve dedicated experimental detection efforts and simulations.}

\item{The features observed in the cosmic-ray spectrum by Auger, Pamela, Fermi, HESS, 
CREAM and others can be all quantitatively well explained with the action of cosmic rays
 accelerated in the magnetic winds of very massive star explosions such as Wolf-Rayet stars, 
without any significant free parameter. All these observations of cosmic ray positrons and 
electrons and the like are due to normal astrophysical sources and processes, and do not require 
an hypothetical decay or annihilation of heavy CDM particles (wimps). 
The models of annihilation or decay of heavy CDM wimps
are highly tailored to explain these normal 
astrophysical processes and their ability to survive observations is more than reduced.}

\item{Theoretical analytic perturbative approachs for large scales in Eulerian and Lagrangian frameworks 
can be combined with phenomenological halo models to build unified schemes for describe all scales. The 
large $k$ Gaussian decay of Eulerian-space propagators (in agreement with simulations) is not a true loss of 
memory of the density field but a "sweepping" effect of the velocity field modes which coherently move density 
structures. Lagrangian-space propagators are not sensitive to such effect. Extending the Zeldovich  
approximation ("adhesion model") show that Eulerian propagators are in fact the velocity probability 
distribution, whereas Lagrangian propagators explicitely relate to the halo mass function and are most 
sensitive probes of the density field than their Eulerian conterparts. Extensions to the Vlasov-Poisson 
system deserve investigation.\\
Theoretical analytic modelling of the halo mass function on small scales gains insight with the use of 
the mathematical excursion set theory in a path integral formulation, which allows direct comparison with 
numerical simulations and observational results. Stochastic modeling of the halo collapse conditions can 
be easily implemented in this formalism.}

\item{The galaxy magnitudes: halo radius, galaxy masses and velocity dispersion
obtained from the Thomas-Fermi quantum treatment for WDM fermion masses in the keV scale are
fully consistent with all the observations for all types of galaxies (see Table I, pag 14). 
Namely, fermionic WDM treated quantum mechanically, as it must be, is able to reproduce
the observed DM cores and their sizes in galaxies.

It is highly remarkably that in the context of fermionic WDM, the simple stationary
quantum description provided by the Thomas-Fermi approach is able to reproduce such broad variety of galaxies.

Baryons have not yet included in the present study. This is fully justified for dwarf compact 
galaxies which are composed today 99.99\% of DM. In large galaxies the baryon fraction can
reach values up to  1 - 3 \%. Fermionic WDM by itself produces galaxies and structures in 
agreement with observations for all types of galaxies, masses and sizes. Therefore, the effect of including 
baryons is expected to be a correction to these pure WDM results, consistent with the fact that dark matter 
is in average six times more abundant than baryons.}

\item{ WDM quantum effects play a fundamental role in the inner galaxy regions. WDM quantum pressure -due to the combined effect of the Pauli and the Heinsenberg principle- is crucial in the formation of the galaxy cores. Moreover, dwarf galaxies turn to be quantum macroscopic objects supported against gravity due to the WDM quantum pressure. A measure of the compactness of galaxies and therefore of its quantum character is the phase space density.  The phase space density decreases from its maximum value for the
compact dwarf galaxies corresponding to the limit of degenerate fermions till
its smallest value for large galaxies, spirals and ellipticals, corresponding to
the classical dilute regime. On the contrary, the halo radius $ r_h $ and the halo mass $ M_h $
monotonically increase from the quantum (small and compact galaxies) to the classical regime
(large and dilute galaxies).

Thus, the whole range of values of the chemical potential at the origin $ \nu_0
$ from the extreme quantum (degenerate) limit $ \nu_0 \gg 1 $ to the classical
(Boltzmann) dilute regime $ \nu_0 \ll -1 $ yield all masses, sizes, phase space
densities and velocities of galaxies from the ultra compact dwarfs till the
larger spirals and elliptical in agreement with the observations}.

\item{All evidences point to a dark matter particle mass around 2 keV.
Baryons, which represent 16\% of DM, are expected to give a correction to pure WDM results.
The detection of the DM particle depends upon the particle physics model.
Sterile neutrinos with keV scale mass (the main WDM candidate) can be detected in 
beta decay for Tritium and Renium and in the electron capture in Holmiun.
The sterile neutrino decay into X rays can be detected observing DM
dominated galaxies and through the distortion of the black-body CMB spectrum.
The effective number of neutrinos, N$_{\rm eff}$ measured by WMAP9 and Planck satellites
is compatible with one or two Majorana sterile neutrinos in the eV mass scale.
The WDM contribution to  N$_{\rm eff}$ is of the order $ \sim 0.01 $ and therefore
too small to be measurable by CMB observations.

So far, {\bf not a single valid} objection arose against WDM.}

\item{{\bf As an overall conclusion}, CDM represents the past and WDM represents the future in the DM research. 
CDM research is more than 20 years old. CDM simulations and their proposed baryonic solutions, 
and the CDM wimp candidates ($ \sim 100$ GeV)
are strongly pointed out by the galaxy observations as the {\it wrong} solution to DM. 
Theoretically, and placed in perspective after more than 20 years, the reason why CDM does not work appears simple and clear 
to understand and directly linked to the excessively heavy and slow CDM wimp, which determines an excessively small 
(for astrophysical structures) free streaming length, and unrealistic overabundance of structures at these scales.  
On the contrary, new keV WDM research, keV WDM simulations, and keV scale mass  
WDM particles are strongly favoured by galaxy observations and theoretical analysis, they naturally {\it work} 
and agree with the astrophysical observations at {\it all} scales, (galactic as well as cosmological scales). 
Theoretically, the reason why WDM works  so well
is clear and simple, directly linked to the keV scale mass and velocities of the WDM particles,
and free-streaming length.  The quantum nature of the DM fermions must be taken into account for WDM. The quantum effects however are negligeable for CDM: the heavy (GeV) wimps behave classically. The quantum pressure of the WDM fermions solves the core size problem and provides the correct observed galaxy masses and sizes covering from the compact dwarfs to the larger and dilute galaxies, spirals, ellipticals. Dwarf galaxies are natural macroscopic quantum objects supported against gravity by the WDM
quantum pressure. The experimental search for serious WDM particle candidates (sterile neutrinos) 
appears urgent and important: it will be a fantastic discovery to detect dark matter in a beta decay. 
There is a formidable WDM work to perform ahead of us, these highlights point some of the directions where 
it is worthwhile to put the effort.}

\subsection{THE PRESENT CONTEXT AND FUTURE IN DM RESEARCH.}

\item{Facts and status of DM research: Astrophysical observations
point to the existence of DM. Despite of that, proposals to
replace DM  by modifing  the laws of physics did appeared, however
notice that modifying gravity spoils the standard model of cosmology
and particle physics not providing an alternative.
After more than twenty active years the subject of DM is mature, (many people is involved in this problem, 
different groups perform N-body cosmological simulations and on the other hand direct experimental particle 
searches are performed by different groups, an important
number of conferences on DM and related subjects is held regularly). DM research  
appears mainly in three sets:
(a) Particle physics DM model building beyond the standard
model of particle physics, dedicated laboratory experiments,
annhilating DM, all concentrated on CDM and CDM wimps.
(b) Astrophysical DM: astronomical observations, astrophysical models.
(c) Numerical CDM simulations. : The results of (a) and (b)
do not agree and (b) and (c) do not agree neither at small scales.
None of the small scale predictions of CDM simulations 
have been observed: cusps and over abundance of substructures
differ by a huge factor with respect to those observed.
In addition, all direct {\it dedicated} searchs of CDM wimps from more than twenty years 
gave {\it null results}. {\it Something is going wrong in the CDM research and the right answer is: 
the nature of DM is not cold (GeV scale) but warm (keV scale)}.}

\item{Many researchers continue to work with heavy CDM candidates
(mass $ \gtrsim 1 $ GeV) despite the {\bf staggering} evidence that these
CDM particles do not reproduce the small scale astronomical observations
($ \lesssim 100 $ kpc). Why? [It is known now that the keV scale DM particles naturally produce the
observed small scale structure]. Such strategic
question is present in many discussions, everyday and off of the record (and on the record) talks in the field. 
The answer deals in large part with the inertia (material, intellectual, social, other, ...) that structured research 
and big-sized science in general do have, which involve huge number of people, huge budgets, longtime planned 
experiments, and the "power" (and the conservation of power) such situation could allow to some of 
the research lines following the trend; as long as budgets will allow to run wimp experimental searches and CDM simulations 
such research lines could not deeply change, although they would progressively decline.}

\item{Notice that in most of the DM litterature or conferences, wimps are still "granted" as "the" DM particle, 
and CDM as "the" DM; is only recently that the differences and clarifications are being clearly recognized 
and acknowledged. While wimps were a testable hypothesis at the beginning of the CDM research, 
one could ask oneself why they continue to be worked out and "searched" experimentally in spite of 
the strong astronomical and astrophysical  evidence against them. \\
Similar situations (although not as extremal as the CDM situation) happened in other branches of physics and cosmology:
Before the CMB anisotropy observations, the issue of structure formation was plugged with several alternative 
proposals which were afterwards ruled out. Also, string theory passed from being considered "the theory of 
everything" to "the theory of nothing" (as a physical theory), as no physical experimental evidence have been 
obtained and its cosmological implementation and predictions desagree with observations. (Despite all 
that, papers on such proposals continue -and probably will continue- to appear. But is clear that  big dedicated 
experiments are not planned or built to test such papers).  In science, what is today `popular' can be discarded 
afterwards; what is today `new' and minoritary can becomes `standard' and majoritarily accepted if verified 
experimentally. }
\end{itemize}

\bigskip

\begin{center}

 {\bf \em  `Examine the objects as they are and you will see their true nature;
look at them from your own ego and you will see only your feelings;
because nature is neutral, while your feelings are only prejudice and obscurity'}

\medskip

[Gerry Gilmore quoting Shao Yong, 1011-1077 in the 14th Paris Cosmology Colloquium Chalonge 2010
http://chalonge.obspm.fr/Programme\_Paris2010.html, arXiv:1009.3494].

\end{center}

\bigskip

\begin{center}

The Lectures of the Workshop can be found at:

\bigskip

{\bf http://chalonge.obspm.fr/Programme\_CIAS2012.html}

\bigskip

The photos of the Workshop can be found at:

\bigskip

{\bf http://chalonge.obspm.fr/albumCIAS2012/index.html}

\end{center}

\section {Live Minutes of the Workshop by Peter Biermann}

Hector de Vega (Paris): Warm DM from theory and galaxy observations: spatial fluctuation in WDM and CDM; always see cored profiles, so density flat in center; 0.05 pc for CDM, 50 kpc cores for keV WDM; phase space density $\rho/\sigma^3$;$ Q(r) = \rho(r)/\sigma(r)^3$; $K \simeq 1$; Pauli principle gives upper limit $K m^4/\hbar^3$; range in Q; CDM scale is essentially zero, WDM observed range, HDM Mpc so too large; the fact that we observe cored profiles disallows bosonic DM. (PLB comment: bosonic structures are unstable). Structures are not formed below about 50 kpc in WDM; quantum limits consistent with dwarf spheroidal galaxies; $10^6 M_{\odot}$ for the very smallest dwarf spheroidal galaxies; these sterile neutrinos fill in gap in standard particle tables; beta decay experiments may detect the effect of sterile neutrinos; HdV + NoSa PRD 2012; HdV et al. 1109.3452; MARE experiment (Nucciotti et al.); keV WDM reproduces density profiles as observed; alleviates satellite problem (Markovic et al. 2011); alleviates void problem; velocity width of galaxies favors WDM over CDM (Papastergis et al. 2011 ApJ;  Zavala et al. 2009 ApJ); HdV et al., New Astronomy 17. 653 (2012); problems with helioseismology (Asplund et al. 2009); could KATRIN join the search for sterile neutrinos; Destri et al. 1204.3090;

\bigskip

Igor Karachentsev (SAO-Special Astrophysical Observatory of the Academy of Science, 
Nizhnii Arkhyz, Russia): Cosmography of the local universe:  Two branches of observational cosmology, 
1)  high redshift objects, and 2)  local cosmography; List of nearby galaxies, less than 10 Mpc, 
Karachentsev et al. AJ 127, 2031, 2004); new updates version about 800 galaxies; Sloan Digital Sky 
Survey has an average distance between galaxies with measured redshift is 8 Mpc; estimates the 
local group mass, gets $2 \; 10^{12} M_{\odot}$, 
so the same number as Kahn \& Woltjer (1959) (PLB comment); 
almost no galaxies detectable at $ < 1 $ Mpc are detectable at larger distances; next go to 50 Mpc; 
calls it the local volume of homogeneity; sample of 10900 galaxies with $V_{LG} < 3500 $ km/s, 
over the entire sky with abs{b}$  > 15 $ deg; special procedure of finding groups and pairs, so 509 pairs, 
168 triplets, 395 groups, 520 especially isolated galaxies; finds that local $\Omega_{m} \simeq 0.08 \pm 0.02$, 
while the global value is $0.28 \pm 0.03$. Several possible explanations: a) All galaxy systems
 extends much farther than virial radius, b)  True homogeneity scale is not 50 Mpc, but more like 300 Mpc, 
c)  The basic fraction of DM is outside somewhere; (a) is excluded, (b) is excluded, suggests (c) is correct. 

\medskip

Find weird looking galaxies without any visible neighbor, so suggests that invisible galaxies exist, so DM galaxies. Find evidence for attractor, ($1/x$)-shaped distribution in V as a fct of distance. Weak lensing finds evidence for invisible galaxies. Three dynamically different components in structure: a) groups and clusters, virial theorem holds, b) collapsing regions around groups and clusters, c)  remaining infinitely expanding field of galaxies; fraction of galaxies in zone (a) 54 \%, (b) 20 \%, (c) 26 \%, fraction of stellar mass (a) 82 \%, (b) 8 \%, (c) 10 \%, fraction of occupied volume (a) 0.1 \%, (b) 5 \%, (c) 95 \%, input to $\Omega_{m}$ (a) 0.06, (b) 0.02, (c) 0.20, mean dark to luminous ratio (a) 26, (b) 87, (c) 690 in solar units; $\Lambda CDM$ problems, missing satellites (1:30), missing baryons (1:10), missing DM (1:3); DM and luminous matter distributed very differently; suggests that dark galaxies exist, or dark halos with baryonic matter inside;

\bigskip

Ayuki Kamada (IPMU Inst. Physics and Mathematics of the Universe, Univ. of Tokyo): Structure formation in WDM models:  small scale CDM crisis now; work by Okamoto\& Frenk 2009; missing dark satellites may not be visible; Ishiyama et al. 2008: there could be a very large variation in subhalo density; Zavala et al. 2009: density of galaxies as fct of velocity width, CDM disagrees, WDM agrees; Thomson scattering can smooth out baryon anisotropy (Silk damping), so small scales cannot be observed; strategy is to go to nonlinear growth and fight with baryon physics; for WDM amplitude of perturbation is damped, less for nonthermal WDM; Landau damping provides the cut at small scales; thermal WDM cuts off small scales by a lot, while nonthermal WDM cuts off less; observed satellites fit to nonthermal WDM simulations, while thermal WDM does not if using 1 keV ; goes through many variants of WDM, thermal and nonthermal; final conclusion that $m_{WDM}$ about or $> 2$ keV;  PLB comment: that is just what Ana told me.

\bigskip

Wei Liao (Inst. Modern Physics, East China Univ. for Science \& Technology, Shanghai): Beta decay experiments:  compares GeV DM with keV DM; mixing active neutrinos with sterile neutrinos key; production proportional to mixing angle squared $(\sin^2 \theta 's)$; Asaka, Blanchet, Shaposhnikov 2005; He, Li \& Liao 2009; Shaposhnikov 2007; splitting of keV and GeV for the other two GeV sterile neutrinos is natural; mixing angle limit from X-ray observation; Lyman $\alpha$ observation suggest $1 - 10$ keV; Liao 2010; reheating can weaken all constraints on sterile neutrinos; says model independent $ m_s > 1$ keV, $\theta^2 < 1.8 10^{5} m_x$ (keV)$^{-5}$; cross section $\sigma \beta \simeq 10^{-55} cm^2$, $\beta velocity$; $\beta-decay$; at the end point of the decay spectrum $Q_{\beta} - m_{\nu_s}$; J.J. Simpson 1981 PRD 24, 2971; Liao PRD 82, 073001 (2010); detect sterile neutrinos by catching with radioactive nuclei; produces mono-energetic electrons beyond end point of $\beta$ decay; finds $0.7 yr^{-1}$ for sterile neutrino density of $10^5 cm^{-3}$ mixing angle squared $10^{-6}$, and $10 kg$ Tritium; analogously 10 tons of $^{106}$Ru gives $16 yr^{-1}$; local density of sterile neutrinos $10^{5}$ per cc for 3 keV (inverse with mass); probably difficult in the near future; Ando \& Kusenko 2010 electron recoil, spin flip of nuclei; event rate does not depend on velocity, since velocity enters into the cross-section estimate, and so cancels in the end.

\bigskip

Subinoy Das (Inst. fuer Theoretische Teilchenphysik und Kosmologie, RWTH Aachen): Cosmological limits on hidden WDM, model independent constraints on warm hidden dark matter; WIMP miracle, that with natural scales of mass and interaction you get $\Omega_{DM} \simeq 0.1$; Feng ARAA 2010; but no experimental hint; so focus on WDM on keV scale with broader perspective; J. Feng \& J. Kumar PRL recently;  using "hidden" sector; Tremaine \& Gunn 1979: Fermi particle phase space density gives lower bound on mass, as does free streaming length scale; Manuel Drees (Bonn) found formula for hidden sector freeze out of WDM; Drees, Kakizaki, Kulkarni 2009 (PRD 80, 043505, 2009); use three constraints, a) density constraint, b) free streaming bound, and c) phase space bound; find lower bound of $1.5$ keV; this is valid also for the "hidden" sector;  S. Das \& Kris Sigurdson PRD 2012; Sigrudson, Kamiokowski 2004; Strigari, Kaplighat, Bullock 2006; Cembranos, Feng, Rajamraman, Takayama 2005; talks about keV particles which are not sterile neutrinos; Fardon, A.E.N., Neal Weiner astro-ph/0309800; Kaplan, A.E.N., Weiner hep-ph/0401099 = PRL; S. Das, Neal Weiner PRD 2011; "late forming DM"; invents bubbles of low mass particles that together act as WDM; assumes neutrino has extra interaction with scalar, fifth force winds over free-streaming and neutrinos get trapped in bubbles, fifth force balanced by Fermi pressure, ..; natural scale about $100$ eV.

\bigskip

Jun 7: Manolis Papastergis (Center for Radiophysics \& Space Research, Cornell Univ.): Velocity width function of galaxies: dark matter implications:  ALFALFA measures redshift, integrated flux in HI, HI mass, velocity width; $10^4$ galaxies, overlap with Sloan; out to $200$ Mpc; ignores hot gas completely; peak efficiency of star formation at peak of distribution of conversion only 30 percent, and lower at higher and lower mass, down to a few percent; small galaxies have trouble turning gas into stars, and also have trouble retaining any gas; total discrepancy almost a factor of $30$; ApJ 739, 38, 2011; modified Schechter function number of galaxies per volume as a fct of the velocity width observed, slope $-3$ in powerlaw range in model; observed slope $-0.85$; a new way to see the "missing small galaxies"; in small galaxies the rotation curve is not flat, but still rising at the last measured point; therefore the measured width could be underestimated; define what is "needed" to match CDM models to observations; this would mean that HI widths must be more than a factor of $2$ off in order to "work"; Tikhonov \& Klypin 2009; Ferrero et al. 2011; Papastergis et al. 2011; all galaxies ought to be $35$ km/s or larger; alternative much larger scale; the Sagittarius galaxies defy this expectation;  so this solution does not work; Boylin-Kolchin et al. 2012; Lovell et al. 2012; so let us see what happens of you change the DM model; Zavala et al. 2009; WDM simulation reproduces the ALFALFA data; no longer at cutoff at $35$ km/s; NorSa comments that you need nonthermal production of WDM; HdV comments that a thermal particle of $1$ keV behaves like a nonthermal particle at several keV; discussion: Obreschkow et al. 2009;  Zwaan et al. 2010;
    At the coffee break Casey Watson (Millikan Univ., Dep Phys. \& Astron., Decatur, IL) mentioned that at FermiLab they are constructing a new experiment to measure the wiggles of time - relevant for the paper with Ben Harms.
    
\bigskip   
    
Nicola Amorisco (Inst. of Astronomy, Univ. of Cambridge, UK): Dark matter cores and cusps, multiple stellar populations in dwarf spheroidals:  Walker 2012; luminous density, gravitational potential, anisotropy structure parameter $\beta$: Jeans equation; King profile, Plummer profile; not all possible combinations give a positive density at the center; Tolstoy et al. 2004; star formation erratic, with red horizontal branch stars different distributions from blue horizon branch stars; Battaglia et al. 2008 - stellar populations different as a fct of metallicity; Walker et al. 2009, Wolf et al. 2010, Amorisco \& Evans 2011; almost all mass profile models give the same mass at a middling radius; Walker \& Penarrubia 2011; he does admit that this is effectively the King approach; for each stellar populations he uses three parameters, $R_h, r_t, \beta$; Amorisco \& Evans 2012; matches the data on both stellar populations quite well; you cannot embed a cored stellar population into a NFW profile of DM; cusped halos ruled out at $99.95$ percent for $r^{-1}$ profile, and similar at 98.6 percent for $r^{-1/2}$ profile;  Agnello \& Evans 2012: same results essentially: any core smaller than 120 pc is not compatible with the Virial Theorem; Gerhard 1993, van der Marel \& Franx 1993; proposes a Bayesian framework; uses Carina dwarf Spheroidal (dSph), data from Walker et al. 2009; then Sextans dSph; Sculptor dSp; Fornax dSph; evidence for radial anisotropy in Sextans, Carina, and Sculptor; phase space modelling excludes an NFW halo in Sculptor.

\bigskip

Elena Ferri (INFN Milano Bicocca): MARE experiment to measure light and sterile neutrinos:  $^{187}$Re vs $^{163}$Ho; deformation of $\beta$ decay spectrum near cutoff; question energy resolution, background, pile-up; effective rate at end-point $F \simeq (\frac{\Delta E}{E})^{3}$; compares spectrometer vs calorimeter; current experiments "electrostatic integrating spectrometers", Mainz, Troitsk, Karlsruhe (KATRIN); calorimeters measure entire spectrum at once; cryogenic detectors as calorimeters; use $AgReO_4$ dielectric Rhenium compound; MIBETA experiment $m_{\nu} < 15 eV $; new experiment MARE, MARE 1 2-4 eV sensitivity, with 100 detectors; MARE $2 0.2$ eV sensitivity, $1000$ detectors; MARE 2 close to KATRIN, but still improvable; Rujula \& Lusignoli PLB 118, 429 (1982); $^{163}$ Ho production from neutron irradiated $^{162}$Er enriched Er; discussion of various errors.

\bigskip

Jorge Penarrubia (Inst. Astrofisica Andalucia, IAA-CSIC, Granada): DM mass profiles of dwarf spheroidal galaxies:  lists all kinds of constraints on DM particles; 63 orders of magnitude in cross section, 51 orders of magnitude in mass - all uncertainty; average distance travelled before it falls into potential well = free streaming length; so no structure below; dwarf spheroidal galaxies most interesting to check on structure formation; Milky Way about about 25 dSph galaxies found, in M31 about 23 dSph; ionization and cooling suggests that above about $10^8 M_{\odot} 
 \rightarrow $ you should see more than 100 such galaxies in our Galaxy in CDM, Koposov et al. 2008, Tollerud et al. 2008; dSph allow to test CDM and WDM predictions; Simon \& Geha 2007.

\medskip

Some galaxies have only $10^6 M_{\odot}$; no gas, no rotation;  Walker et al. 2009; DM density in dSph up to $10^4$ times Solar neighborhood; Strigari et al. 2007, Penarrubia et al. 2008, Walker et al. 2009, Wolf et al. 2009, NFW 1997, Diemand et al. 2007, Springel et al. 2008; DM substructures in dSph have less than $0.01 M_{vir}$ ; using wide binaries, so even small perturbations can disrupt them $ \rightarrow $
 probes clumpiness; used in Galactic halo already; Penarubbia et al. 2010; if dark galaxies exist you would expect a cutoff in the distribution of bound wide binary stars; then inner structure of CDM halos; Bode et al. 2001, Dubinsky \& Carlberg 1991, Moore et al. 1998, Diemand et a. 2005. baryons may alter the inner DM profile; can cusps be erased by baryon effects; results are sofar, that cusps cannot be erased by baryons; so a prediction from CDM is that cusps exist in dSph; unknown $\beta(r)$, unknown $M(r)  \rightarrow $ nasty degeneracy; Joe Wolf et al. 2008; characteristic special radius for all models that give mass; Battaglia et al. 2008; Walker \& Penarubbia 2011; if dSph have two stellar populations then one has two measurements, so get mass profile.

\medskip

They recover published data on 1) proper motions, 2) mean velocity dispersion, 3) average $R_{half}$, 4) mean metallicity; use Carina, Fornax, and Sculptor; two components detected in Sculptor and Fornax, in Carina one component sufficient; Walker \& Penarrubia 2011; cores fit much better, NFW excluded basically; data suggest core DM profile in dSph; uses a simulation to estimate bias, makes case even stronger; 99.98 percent excluded NFW; cored profiles can be more easily disrupted in the tidal field of a galaxy; the disk destroys these galaxies quite efficiently; "tension" with CDM hydrodynamical simulations, Sawala et al. 2010, Parry et al. 2012; all in Walker \& Penarrubia 2011;

\bigskip

In discussion I (PLB) learnt, that the King profiles do not match very deep observations: the observations show a powerlaw tail, with $r^{-5}$, saying that the phase space is filled at very small energies, so no tidal cut visible; (PLB): I wonder whether that is in fact another argument against graininess of dark halo in Galaxy;
In another discussion (PLB) I learnt that John Kormendy is coming in July, and he will argue that galaxies do not merge that often; I wonder whether that might mean that they just merge earlier in their evolution, than what we can observe.

\bigskip

Marc Lovell (Inst. Comput. Cosmology, Univ. Durham): Numerical simulations of WDM halos:  "My adventures in a WDM universe: work with Jenkins, Carlos Frenk, Eke, Gao, Theuns, Wang, Simon White, Boyarsky, ...; Boyarsky et al. 2009; mix of resonant and non-resonant components in the particle physics; he uses WDM; he uses a thermal relic; WDM power spectrum picked to approximate M2L25 model of Boyarsky et al. 2009; resimulate Aquarius Aq-A model of Springel et al. 2008; transfer fct from WDM to CDM given by ($1 + (\alpha k)^{2 \nu})^{-5/\nu}$; use WMAP1 cosmology; Lovell et al. 2012; Wang \& White 2007 show that finite resolution gives spurious features, like funny stripes; shows example, formation of two big objects, with a string of small objects: going to better resolution gives rise to a string of smaller objects, 4 relatively large ones at low resolution, six smaller ones at better resolution; Tollerud et al. 2011;  shows diagram of $R_{max}$ vs $V_{max}$, identifies $R_{max}$ with $V_{infall}$; all using seven simulations in CDM; in WDM simulations you get later formation times of small structures; subhalo structure mass fct: in CDM about $10^{3.5}$ objects at low max velocity, in WDM as low as 30; in  the coolest WDM about 200; and the distributions stop near $10$ km/s, so only have structures above; massive satellite problem alleviated by late formation of WDM halos compared to CDM; Maccio et al. 2012.

\bigskip

Jesus Zavala (Univ. Waterloo, Dep. Phys. \& Astron., Ontario): Velocity function of galaxies in local environment from CDM and WDM simulations:  collaboration with Yi-Peng Jing (SHAO, Shanghai), Andreas Faltenbacher (UWC Cape Town), Gustavo Yepes (UAM, Madrid, Yehuda Hoffman (Jerusalem), Stefan Gottloeber (Potsdam), Barbara Catinella (..); alternative to WDM is self-interacting CDM;  free streaming length $ \sim m_X^{-1} $; velocity dispersion $\sim m_X^{-1/2}$; Boylan-Kolchin et al. 2009; abundance at low mass should be $M^{-2}$; Guo et al. 2010; Papastergis et al. 2011; Kravtsov et al. 2010, Guo et al. 2010, Koposov et al. 2008; Zavala et al. 2009; use Giovanelli et al. 2005 ALFALFA survey trying to reproduce the observations; Mo et al. 1998; Zavala et al. 2008; comparison suggests that WDM does much better than CDM within the range of the distribution which can be trusted; emphasizes that the constrained simulations do much better in Virgo direction - constrained to simulate the local environment; simulate also anti-Virgo direction - confirms; Stewart et al. 2009 (?); Sawala et al. 2012; HdV notes that one should not trust the Lyman-alpha constraints; says that current constraints from Lyman-alpha seem to barely admit a viable solution (I think he means to say that CDM could be saved); self-interacting DM alternative to WDM: shows bullet cluster, Clowe et al. 2006, Randall et al. 2008; Buckley \& Fox 2010; Loeb \& Weiner 2011; try to get what "you want"; two parameters; Vogelsberger, Zavala, \& Loeb 2012; get more spherical halo shapes; finds several allowed versions of cross-section models; the model does NOT help with abundance of sub-halos, still too many; NoSa and HdV argue that such models are "leptophilic" and contain various phantasies.

\bigskip

Jun 8: Marco Lombardi (Univ. of Milano, Dep of Physics, Milan): Star formation rates 
and the nature of the extragalactic scaling relations:  
Star formation  as a fct of z, see Bouwens et al. 2010; Schmidt (1959) law; Schmidt-Kennicutt law 
$\Sigma_{SFR} \sim (\Sigma_{gas})^{1.6}$; Gao-Solomon law SFR $\sim M_{dense} \;  \rho $ Ophiuchi 
cloud and Pipe cloud in GC region; similar total mass in clouds; Pipe cloud has 21 YSO and $8000 M_{\odot}$, while the other has 316 YSO and $14000 \; M_{\odot}$; time scale about 2 million years; other examples, two clouds, same mass, same shape, different star formation rates by factor order 15; uses color dependence of extinction; Alves et al. 2000; Onishi et al. 1999; dust extinction vs CO; number of YSO versus cloud mass scatter diagram; led to assume that the star formation rate scales with material with at least 1 magnitude absorption in the K band; Kainulainen et al. 2009; number of YSO scales linearly with cloud mass with more 0.8 magnitude absorption; numbers can be understood as efficiency 0.1, and star formation time scale of 2 million years; Lombardi et al. 2011; the scaling determined corresponds to a density threshold of $10^4$ per cc; lots of details from some black box code on star formation; SFR scales with FIR, exactly as I worked out in 1976 (PLB); SFR correlates with molecular Hydrogen, not so well with atomic gas HI; (PLB) comment: his typical time scale is exactly the lifetime of very massive stars, and so one might guess that this is in fact the self-regulating process, stars live, make$ > HII$ regions, explode, so destroy their local environment.

\bigskip

He Zhang (MPI fuer Kernphysik, Heidelberg): Sterile neutrinos for WDM in flavor symmetry models:  Lots of particle physics following the lines of Manfred Lindner and Alex Kusenko, using see-saw mechanism; changes neutrino oscillations, as already at distance zero oscillations visible; see-saw with Higgs; Mention, Fechner et al. 2011, active-sterile neutrino mixing, odd feature in reactor neutrinos; proposes active-sterile mixing; MiniBooNE also odd feature, also explainable with sterile neutrinos; Kopp et al. 1103.4570; Hamann et al. 1006.5276; Mangano et al. 1103....; Shaposhnikov et al., Asaka et al., Dodelson \& Widrow 1993; Bezrukov, Hettmannsperger, Lindner 2009; Kusenko et al. 2010 use other dimensions; Lindner, Merle, Niro 2010; Friedberg \& T.D. Lee 2006; He, Li, Liao 2009; He Zhang 1110.6838; goes through various particle physics phantasies to get just keV neutrino without upsetting baryon genesis, baryon asymmetry and leptogenesis; Barry, Rodejohann, He Zhang 2012; active-sterile mixing $10^{-4}$ in this model; discussion about evidence from reactor neutrinos; this problem and WDM cannot be helped with the same neutrino.

\bigskip

Note (PLB) after comment by NoSa:  Jun 8, 2012: If the centers of small galaxies WDM are degenerate, then we come back to the mechanism proposed by Faustin and myself.  That means that the merging of massive stars to make super-massive stars, and then super-massive black holes is in competition.  The decision between these two mechanisms could be another constraint for the mass of the sterile neutrino.

\bigskip

Patrick Valageas (Inst. de Phys. Theorique, Orme de Merisiers, CEA-Saclay, Gif-sur-Yvette): Perturbation approaches and halo models:  
Impact of a WDM late-time velocity dispersion on the cosmic web; P. Valageas 1206.0554; Vlasov-Poison equation 
$ \rightarrow $  non-linear $ \rightarrow $ linearize it; development into moments in momentum - hierarchy; so better work with Vlasov eq directly; start at redshifts like 50 to 100; use macro-particle like $10^5 M_{\odot}$; this approach can lead to spurious effects like a high-end k tail; Colin et al. 2008; Boyanovsky et al. 2008, Boyanovsky \& Wu 2011; no thermodynamical pressure, since no collisions; pressure-like term with expansion trick to slow down the collapse; in WDM linear growing mode depends on scale; WDM leads to less-advanced of non-linear evolution; probably small impact on the Lyman-alpha distribution; using initial redshift of 50 instead of 10 you under-estimate the large mass tail of the halo-mass function; various papers by others with speaker 2008, 2004, 12007a, b, 2011, 2012, etc; complementary to numerical simulations.

\bigskip

Pier Stefano (CNRS LUTH Obs. de Paris, Meudon): Excursion set theory:  formal theoretical approach for virialized objects, like dark matter clumps; "shell-crossing" has already happened, and one looks afterwards; Bond et al. 1991, Maggiore \& Riotto 2010; think of a density perturbation field, and filter it; Press-Schechter (1974 ApJ) formalism; example of problem, in case when a smaller mass is already embedded in a larger mass; in Press-Schechter formalism not seen; excursion set theory takes care of this problem; diffusion problem with boundary condition; refers to Chandrasekhar's review article in RevModPh 1943; only considers spherical collapse; (PLB) comment: does this contradict Zeldovich? - Gunn \& Gott 1973; Doroshkevich 1970; Eisenstein \& Loeb 1995; then he starts talking about ellipsoidal fields;  Sheth, Mo \& Tormen 2001; Sheth \& Tormen 2001; Maggiore \& Riotto 2010b; P.St. \& I. Achitouv 2011a,b; Maggiore \& Riotto 2010a; mentions Chapman-Kolmogorov equation; Tinker et al. 2008; same references again M \& R and P.St. \& I.A.

\bigskip

Casey Watson (Millikan Univ., Dep Phys. \& Astron., Decatur, IL): X-ray observations and WDM:  (PLB) he told me that the Kusenko \& Loewenstein peak in X-rays from DM decay is no longer there; I had invited him to the DM meeting in Berlin at Marcel-Grossmann 2006; "Using X-ray observation to constrain sterile neutrino warm dark matter"; collaborators Zhiyuan Li (UCLA), Nick Polley (Millikan), and Chris Purcell (Pittsburg);  we are coming close to the limits of current detectors.

His paper on Andromeda contains all key references; goes through the standard argument on sterile neutrino decay; the sterile neutrino typically has a "cold" phase space distribution; these telescopes are sensitive to sterile neutrinos between 1 and $10$ keV; Abazajian et al. 2001a, b; three neutrino decay larger than radiative decay, which is given by $1.36\;  10^{-32} s^{-1}  \left( \sin^2\theta /10^{-10} \right) \; \left(m_s/keV\right)^5 $; 
the factor in front is $1.74 10^{-30}$ for three neutrino decay; therefore the X-ray luminosity 
$ 1.2 \; 10^{33} \; (erg/s) (M_{DM}/10^{11} M_{\odot}) \left(\sin^2 \theta/10^{-10} \right)\,
\left(m_s/keV\right)^5 $; flux $ 10^{-17} \; erg \; cm^{-2} s^{-1} $ 
at 1 Mpc; using the assumption that sterile neutrinos explains DM gives a relationship between mixing angle 
and mass, so then we get a dependence on the mass $m_s$; Dodelson \& Widrow model agrees with Asaka et al. 
model between 1 and $10$ keV; runs through all the astrophysical backgrounds, that interest other people; 
Boyarsky et al. 2006, $m_s < 9.3$ keV - using Dodelson \& Widrow model; $m_s < 8.2$ keV (Virgo), 
$m_s < 6.3$ keV (Virgo + Coma) - again using Dodelson \& Widrow; Watson et al. 2006, 2012: 
Andromeda: Klypin et al. 2002, Seigar et al. 2007 for rotation curves; Shirey et al 2001 
subtracted the bright point sources $ \rightarrow $ down by factor of about 3; Andromeda eliminates 
$3.5$ keV, $6.3$ keV, and $8.2$ keV; so new limit $< 3.5$ keV; all using Dodelson \& Widrow.

\medskip

Highlights issue small field of view relative to DM distribution in dwarf galaxies; Watson et al require 8 sigma to rule things out, Boyarsky et al. use 1 sigma; their Andromeda study uses a ring of $12$ to $28$ arc min; all archival data; use lowest reliable estimate of DM mass in field of view; get $M_{DM}^{FOV} \simeq (0.49 \pm 0.05 ) \times 10^{11} M_{\odot}$; Majorana$ m_s < 2.2$ keV, Dirac $< 2.4 keV$; Loewenstein \& Kusenko 2010 strongly excluded; all other X-ray spectra also exclude this; using Tremaine \& Gunn 1979 $m_s > 0.4$ keV; Strigari et al. 2006 give from the cores of dwarf elliptical masses in the same range, $0.4$ to $2.2$ keV; $2.13$ keV Majorana sterile neutrino decay simulates a $1.07$ keV Neon IX line.

\bigskip
 
(PLB) I suggested to try M32 as a highly stripped galaxy, so with a very distributed DM mass, following his idea about M81/M82 system; (PLB) Ana just sent me (email acaramete) a lower limit of 2.08 keV from WMAP analysis; using upper limit from Casey and lower limit from Ana, we obtain range basically 2.1 to 2.2 keV.

\bigskip

Norma Sanchez (CNRS LERMA Obs de Paris): WDM galaxy formation in agreement 
with observations: goes over history of the debate; DM physics must be 
included in all discussion of galaxy formation and cosmology; mentions 
WMAP7 data and further MWBG data; DM outside outside the Standard Model of 
particle physics; CDM problems include "clumpy halo problem", "satellite 
problem", "cusp problem"; free streaming length $57.2$ kpc $ 
keV/m_s \left(100/g_d\right)^{1/3} $, with $g_d$ 
relativistic degrees of freedom; free streaming length depends on the 
redshift as $(1+z)^{-1/2}$ ; mentions Carlos Frenk group results; argues 
once again for quantum treatment of galaxy cores; dwarf galaxies as quantum 
macroscopic WDM objects; mentions that galaxies often contain baryon matter 
disks, that CDM + baryon models cannot explain without a huge number of 
mergers (Kormendy, Bender et al. 2010); CDM naturally produces dark disks, 
in our Galaxy no dark disk (Moni Bidin et al. 2010); many papers find cored 
profiles of galaxies (e.g. van Eymeren et al. 2009 AA, Walker \& Penarrubia 
2012, de Blok 2010, Salucci \& Frigerio Martins 2009); de Vega, Salucci, 
Sanchez 2010;  Gilmore et al. 2007/8; ; central surface density of galaxies 
$120 M_{\odot} \; pc^{-2}$, Donato et al. 2009, Gentile et al. 2009; WDM 
reproduces this observed value while CDM one again strongly desagrees; 
primordial inflationary fluctuations are $\Phi \sim k^{(n_s-1)2} \; , \; 1 + 
n_s/2 = 1.482$, computed density profile with these fluctuations agrees 
with empirical behavior $r^{-1.6 \pm 0.4}$; Destri, de Vega \& Sanchez 
1204.3090; quantum physics with galaxy cores rule out bosonic DM, and need 
WDM; (PLB comment) similar to what Faustin Munyaneza did; NoSa comment: we 
have a complete framework, not only describing degenerate fermions but all 
types of galaxies explicitely related to the WDM state and the inner 
regions(inner quantum or compact, outside semiclassical or dilute,...) 
reproducing galaxy masses, sizes, phase space density,velocity dispersions 
fully in agreement with observations. All these results strongly points to 
a DM particle of 2 keV.

\newpage

\section{List of Participants}

AGRAWAL	 Rashmi,	SSR College, University  of Pune,	Silvassa,	India\\
\medskip

AMORISCO	Nicola C.	,Inst of Astronomy, Univ of Cambridge,	Cambridge,  	UK\\
\medskip

BAUSHEV Anton,	DESY Zeuthen,	Berlin,	Germany\\
\medskip

BIERMANN Peter,	MPIfR Bonn, Germany and Univ. Alabama,Tuscaloosa;	Bonn,	Germany\\
\medskip

BILEK	Michal,	Charles University in Prague,	Prague,	Czech Republic\\
\medskip

BOUCHARD Philippe	Gatineau,	Canada\\
\medskip

BOUDOU Jean Paul,	CNRS Orsay,	France\\
\medskip

CHAUDHURY	Soumini,	Saha Institute of Nuclear Physics, Kolkata,India\\
\medskip

CNUDDE	Sylvain,	 LESIA-Observatoire de Paris,	Meudon,	France\\
\medskip

CORASANITI	Pier Stefano,	CNRS LUTH Observatoire  de Paris, 	Meudon, 	France\\
\medskip

DAS	Pranita,	Gauhati University, Guwahati,  India\\
\medskip

DAS Subinoy, Inst Theor Phys \& Cosmology RWTH-Aachen Univ., 	Aachen,	Germany \\
\medskip

DESTRI Claudio,	Dipt di Fisica G. Occhialini INFN-Univ Milano Bicocca,	Milan,	Italy\\
\medskip

DE VEGA Hector, LPTHE UPMC CNRS, Paris, France\\

\medskip
DUFFY
	Alan,	ICRAR, Perth,	Australia\\
\medskip

DVOEGLAZOV	Valeriy,	Universidad de Zacatecas,	         Zacatecas ,	
Mexico\\
\medskip

ECHAURREN	Juan,	Codelco Chile	, Calama	,Chile\\
\medskip

EVOLI Carmelo,	Institut für Theoretische Physik, Universität 	 Hamburg, 	
Germany\\
\medskip

FERRI	Elena, Universit\`a Milano-Bicocca \& INFN 	  Milano,	Italy\\
\medskip

GOLCHIN	Laya, 	
Cosmology Group-Physics, Dept. Sharif University,	Tehran	Iran \\
\medskip

GUPTA Anirvan,	SSR College, University of Pune,	  Silvassa,	India\\
\medskip

KALLIES Walter,	 JINR Dubna,     	Dubna,	Russia\\
\medskip

KAMADA	Ayuki, IPMU Inst. Physics \& Maths of the Universe-Univ of Tokyo, Tokyo,	Japan\\
\medskip

KARACHENTSEV Igor,	Special Astrophysical Observatory,	N.Arkhyz,	 
Russia\\
\medskip

KEUM	Yong-Yeon,	Seoul National University,	Seoul,	
Republic of Korea\\
\medskip

LACY-MORA Gerardo,	Tuorla Observatory, University of Turku,	Turku,	Finland\\
\medskip

LALOUM	Maurice	CNRS/IN2P3/LPNHE Paris 	
France\\
\medskip

LATTANZI  Massimiliano,	INFN and Dipartimento di Fisica, Università di Milano-Bicocca	Milan,	Italy\\
\medskip

LETOURNEUR	Nicole	LESIA \& CIAS, Observatoire de Paris,	Meudon,	France\\
\medskip

LIAO	Wei,	East China University of Science and Technology,	           Shanghai,	
China\\
\medskip

LOMBARDI	Marco,	Dipt di Fisica, Università degli Studi di Milano,	 Milan,	
Italy\\
\medskip

LOVELL	Mark	
Inst. for Computational     Cosmology Durham, Univ of	Durham ,                	UK\\
\medskip

MALAMUD	
Ernie,	Fermi National Accelerator Laboratory	,       Batavia, IL	
USA\\
\medskip

MENCI	Nicola,	INAF-OSservatorio Astronomico di Roma,	Rome,	
Italy\\
\medskip

MONTESINOS	Matias	Pontificia Universidad Católica de Chile	
Santiago de Chile	Chile\\
\medskip

NIKITIN
	Vladimir,	Joint Institut for Nuclear Research,	Dubna	,Russia\\
\medskip

NISPERUZA	Jorge Luis,	National University  of Colombia,	Bogota    D.C	,
Colombia\\
\medskip

PADUROIU Sinziana,	Geneva Observatory,	Geneva,	         
 Switzerland\\
\medskip			

PAPASTERGIS	Manolis,	
Center Radiophysics \& Space Research, Cornell Univ,
	Ithaca NY,	USA\\
\medskip

PE\~NARRUBIA	Jorge	Inst.Astrofísica Andalucía- IAA-CSIC,	Granada,  	Spain\\
\medskip

PERES	Clovis,	UFRG	 Porto Alegre,         Brasil\\
\medskip

PEREZ-GARCIA	M. Angeles,	
University of Salamanca and IUFFYM,	 Salamanca,	          
Spain\\
\medskip

QUIROGA PELAEZ	Luis Fernando	,Universidad de Antioquia	,Medellin ,	
Colombia\\
\medskip

SCOTT	Robert,	Universite de Bretagne Occidentale,	Brest	,
France\\
\medskip

SANCHEZ	Norma G.	
CNRS LERMA, Observatoire de Paris,	
Paris	France\\
\medskip

SANES	Sergio	
Universidad de Antioquia,	
Medellin 	
Colombia\\
\medskip

SEYOUM
	Daniel,	Tternopil Technical University, Ukraine	Ternopil,	Ukrania,\\
\medskip

SHAPIRO	Ilya,	Universidade Federal de Juiz de Fora,	Juiz de Fora ,	
Brazil\\
\medskip

SINGHAL	Pratik,	Venkateshwar International School,	New Delhi,	
India\\
\medskip

SOUNDER	Selvaraj,	
Sri Vidya Mandir Arts and Science College,	Uthangarai,	
India\\
\medskip

STEINBRINK	Nicholas	Insitut für Kernphysik, WWU Münster	Münster	
Germany\\
\medskip

THOMPSON	Graham,	Queen Mary, University of London,	London,	UK\\
\medskip

URBANOWSKI	Krzysztof	Institute of Physics, University of Zielona Gora	Zielona Gora	
Poland\\
\medskip

URTADO Olivier,	Education Nationale,	Trappes	, France\\
\medskip

VALAGEAS Patrick, Institut de Phys Théorique, Orme de Merisiers, CEA-Saclay	Gif-sur-Yvette,	
France\\
\medskip

WEINER	Richard,	LPT Orsay/University of Marburg	Paris/
Marburg	 ,France/Germany\\
\medskip

WATSON Casey,	Millikin Univ, Dept Physics \& Astron., Decatur, Illinois	Decatur IL,	USA\\
\medskip

WITTEN	Louis,	Department of Phsyics, University of Cincinnati	Cincinnati, OHI	
USA\\
\medskip

ZAVALA  FRANCO,	Jesus	Department of Physics and Astronomy, University of Waterloo	Waterloo	
Canada\\
\medskip

ZHANG	He,	Max Planck Institut für Kernphysik Heidelberg, Germany\\

\medskip
ZHU Ming,	National Astronomical Observatory of China	Beijing	China \\
\medskip

ZIDANI	Djilali,	LERMA Observatoire de Paris-CNRS,  	Paris,	France\\
\medskip

\begin{figure}[htbp]
\epsfig{file=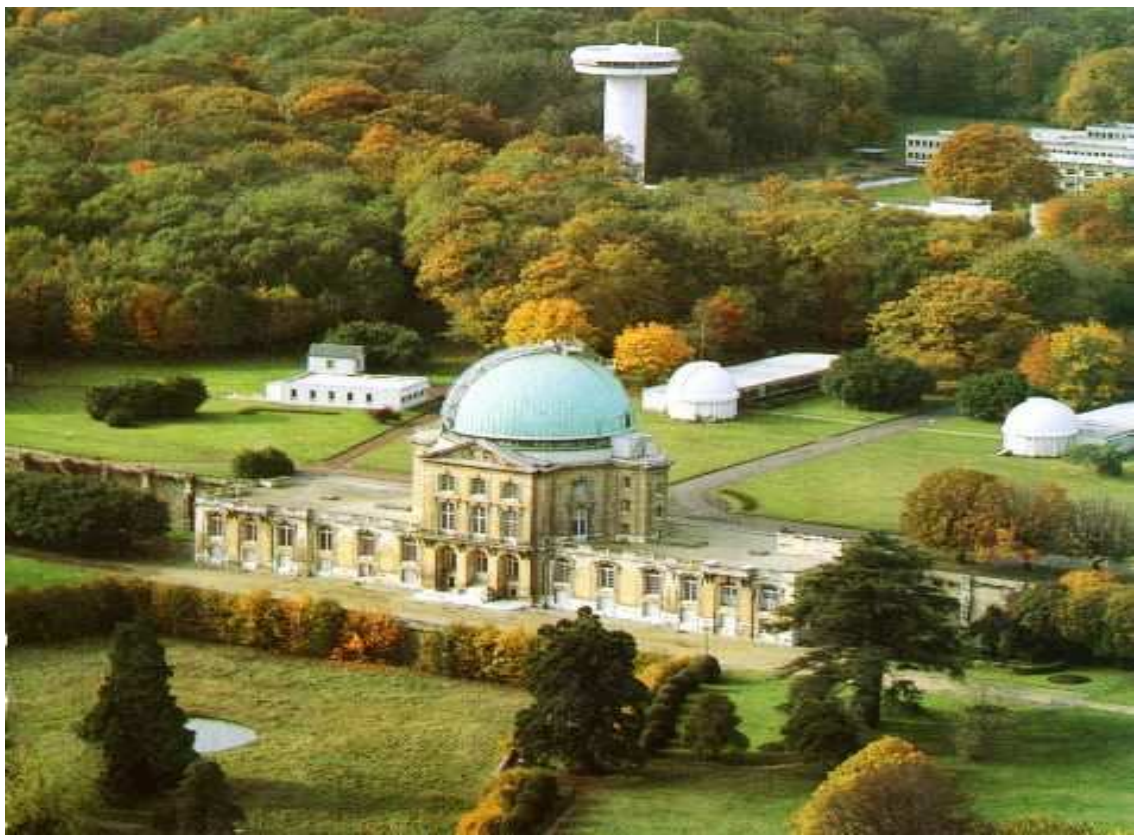,width=15cm,height=11cm}
\vskip 0.5 cm
\epsfig{file=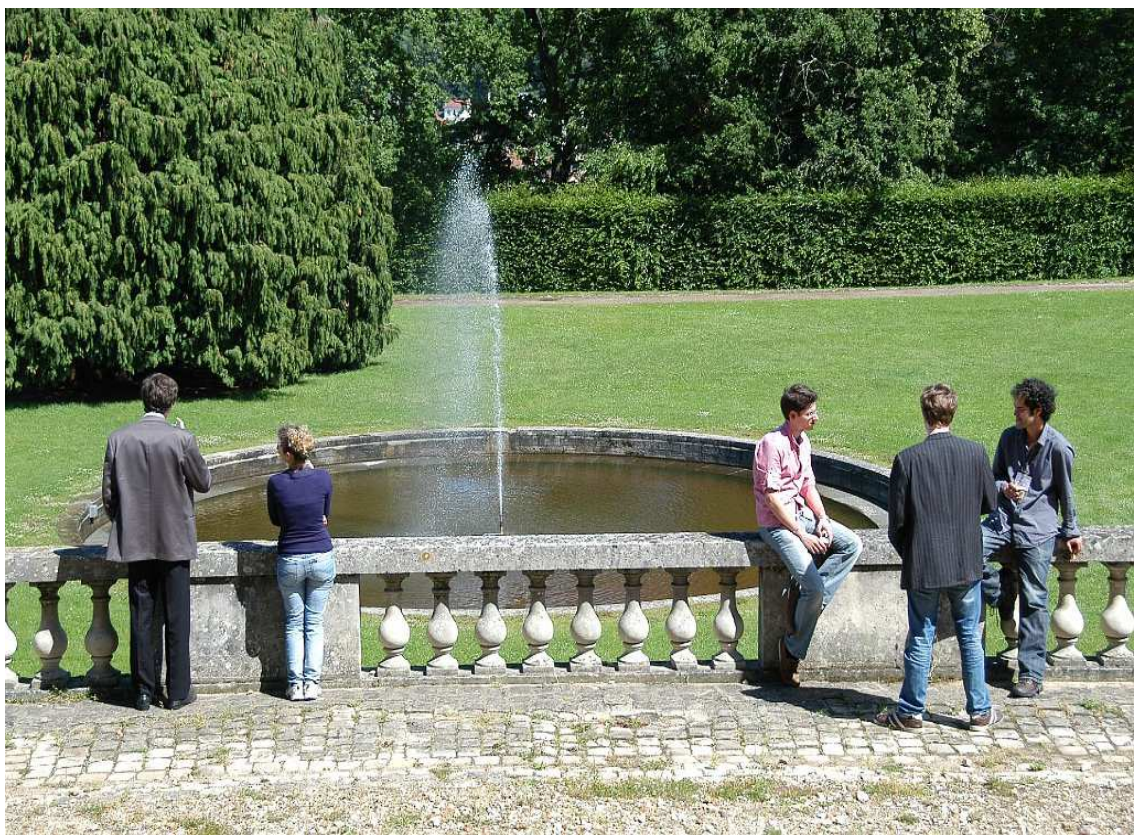,width=15cm,height=11cm}
\caption{Views of the Meudon Ch\^ateau}
\end{figure}

\end{document}